\pdfoutput=1
\documentclass[conference]{IEEEtran}
\IEEEoverridecommandlockouts
\usepackage{hyperref}
\usepackage{cite}
\usepackage{amsmath,amssymb,amsfonts}
\usepackage{graphicx}
\usepackage{textcomp}
\usepackage{xcolor}
\usepackage{algorithm}
\usepackage{algpseudocode}
\usepackage{mathtools}
\usepackage{breqn}
\usepackage{upgreek}
\usepackage{arydshln}
\usepackage{caption,subcaption}
\usepackage[numbers]{natbib}
\usepackage{multirow}
\usepackage[]{footmisc}

\usepackage{float}

\def\BibTeX{{\rm B\kern-.05em{\sc i\kern-.025em b}\kern-.08em
		T\kern-.1667em\lower.7ex\hbox{E}\kern-.125emX}}

\newtheorem{asum}{Assumption}

\begin{document}
	
	\title{Dynamic Circular  Formation Of Multi-Agent Systems With Obstacle Avoidance And Size Scaling: A Flocking Approach\\
	}
	\author{\IEEEauthorblockN{Seyed Mohammad Mahdi Seyed Sajadi}
		\IEEEauthorblockA{\textit{Department of Electrical Engineering} \\
			\textit{Amirkabir University of Technology (Tehran Polytechnic)}\\
			Tehran, Iran \\
			smmss@aut.ac.ir}
		\and
		\IEEEauthorblockN{Hajar Atrianfar}
		\IEEEauthorblockA{\textit{Department of Electrical Engineering} \\
			\textit{Amirkabir University of
				Technology (Tehran Polytechnic)}\\
			Tehran, Iran \\
			atrianfar@aut.ac.ir}
	}
	\maketitle
	\begin{abstract}
		Formation control with the flocking approach is an efficient method that can reach the formation without determining the agent's position. This paper focuses on reaching the circular formation around the leader or target with a specific geometric pattern for the second-order multi-agent system. 
		This means that the polygon formation is formed with arbitrary initial conditions. To create the circular formation, two potential function terms have been used for agent-agent and leader-agent interaction.
		In our approach, if some faults occur during the circular formation and some agents fail, the regular polygon formation will still form with fewer agents. Obstacle avoidance for a single-circle formation and collision-free motion is guaranteed. A circular formation with size scaling is proposed to better maneuver and pass through obstacles. Also, several circles with the desired radius can be reached with changes in the agent-leader potential function. In this work, optimization algorithms with different scenarios are compared to calculate the parameters of our algorithm.
		% The distance between the agents is adjusted so that a regular polygon formation is formed.
	\end{abstract}
	
	
	\section{Introduction}
	Flocking is a form of collective behavior of many agents inspired by nature, such as the cooperative movement of birds. First, Reynolds \cite{reynolds1987flocks} introduced the flocking approach in the form of three laws: 
	\begin{enumerate}
		\item “Flock Centering: attempt to stay close to nearby flockmates,”
		\item “Obstacle Avoidance: avoid collisions with nearby flockmates,”
		\item “Velocity Matching: attempt to match velocity with nearby flockmates.”
	\end{enumerate}
	Next, Olfati-Saber \cite{olfati2006flocking} proposed the mathematical equivalent of the flocking approach using the appropriate potential function. Olfati-Saber introduced three algorithms; in the first algorithm, only the problem of non-collision between agents and maintenance of flocking was investigated. Due to the finite interaction range between agents, fragmentation occurs in the flock. The second algorithm adds the leader term, and because of this, the flock becomes coherent and does not fragment.  Finally, the third algorithm adds obstacle avoidance to the flock using the potential function. 
	
	Formation control is another interesting approach in multi-agent systems used in many applications. Among the military applications, we can mention chase and pursuit, encirclement, escort, etc. Chen \textit{et al}. \cite{chen2018multitarget} have investigated the multi-target consensus pursuit in a circle formation and with a flocking approach. The methods of encircling a target  \cite{manzoor2017coordinated,ma2018cooperative} and rotating targets have been studied \cite{zhang2019distributed}. 
	
	Additionally, the surrounding control of targets in finite time and circle formation for escort of the UAV group have been proposed in \cite{ma2017finite} and \cite{wu2021autonomous}.
	Also, Brust \textit{et al.} \cite{brust2017defending} have offered  a UAV defense system that forms a formation around the malicious UAV and escorts it out of the flight zone. 
	\\
	One type of formation is circular, which was even used in some mentioned military fields.  Wang \textit{et al.} \cite{wang2020circle} introduced the circle formation control problem of mobile agents for second-order dynamics constrained to move on a circle. Song \textit{et al.} \cite{song2018circle} investigated circle formation for a limited interaction range with distributed switching control laws. Also, Wang \& Xie \cite{wang2017limit} studied limit-cycle-based decoupled design with collision avoidance among agents. First, the agents converge on a circle around the target, then adjust their distance from one another. The agents that rotate around the moving target of the circular formation are investigated in the fourth scenario \cite{jin2018circular}. The distributed event-trigger method with first- and second-order dynamics reduces communication between agents \cite{wen2018asynchronous,wen2019distributed,yang2022decoupled,yu2018event}. The circular formation has been developed for various applications, such as UAVs \cite{muslimov2020adaptive,chen2019circular} and fish-inspired robots \cite{berlinger2021implicit}.
	
	In this work, we deal with single- and multi-circle formations
	of the second-order multi-agent systems. The single-circle case is discussed as regular polygon formations in the presence of failed agents and obstacle avoidance with and without size scaling. Han \textit{et al.} \cite{han2015formation} proposed formation control with the size scaling method that shows formation to pass through narrow corridor shrinks. The novel size scaling method proposed in this paper is extended in order to make formation expand against the big obstacle and shrink between two obstacles.
	\\Section \ref{section2} defines the problems and presents our approach, algorithm, and flocking model. Section \ref{section3} covers our problem simulations\footnote{The simulation results can be viewed at https://youtu.be/h-orcIXugsc},  and  section \ref{section4} provides a summary and conclusion.
	\section{Problem Formulation and algorithm}\label{section2}
	The main idea of this paper is the dynamic circular  formation around the leader or target, in which agents can be formed in single- or multi-circles around the leader.
	The radius of the circle and the number of agents in the circle are determined by potential function terms between the agent-leader and the agent-agent. Whether the leader is moving or stationary, the circular formation is maintained. It is also guaranteed that the agents do not collide with the obstacles and with each other. The main features of the proposed algorithm in this paper are as follows:
	\begin{itemize}
		\item Maintaining regular polygon formation against failure and loss of agents
		\item Obstacle avoidance with fixed formation and size scaling of the formation
		\item Multi-circle formation 
		\item Obtaining the optimal parameters of (\ref{eq4}) with optimization algorithms.
	\end{itemize}
	\subsection{Graphs and Nets}
	In this algorithm, the interactions between the agents are spherical, and the graph is undirected. A graph G is a pair $ (\mathcal{V}, \mathcal{E})$ consisting of vertices  $ \mathcal{V}$ and a set of edges $ \mathcal{E}$. The quantities $|\mathcal{V}|$ and $|\mathcal{E}|$ are, respectively, called the order and size of the graph.\\
	The adjacency matrix in the flocking approach is zero when the distance between two agents is greater than the $r$ and one if it is less than the $r$.
	The interaction range between two agents is denoted by $r$.
	The set of spatial neighbors of agent $i$, which is placed in its radius $r$, is indicated  by:
	\begin{equation}\label{eq1}
		N_i=\left\{j \in \mathcal{V}:\left\|q_j-q_i\right\|<r\right\},
	\end{equation}
	and the set of edges for the spatially induced graph is denoted by:
	\begin{equation}\label{eq2}
		\mathcal{E}(q)=\left\{(i, j) \in \mathcal{V} \times \mathcal{V}:\left\|q_j-q_i\right\|<r, i \neq j\right\}.
	\end{equation}
	
	\subsection{Flocking Model}
	The flocking model algorithm consists of three potential functions. The first sets the distance between the agents and handles collision avoidance. The second is used for obstacle avoidance, and the third is used to set the distance between the agent and the leader. Two potential functions between agent-agent and agent-leader are the main cause of forming a circular formation. The appropriate parameters must be set to establish the trade-off between these tasks. \\
	The second-order dynamics of the agents are as follows:
	\begin{equation}\label{eq3}  
		\begin{array}{cc}
			\Dot{q}_i &= p_i \\
			\Dot{p}_i &= u_i.
		\end{array}
	\end{equation}
	The flocking algorithm that we use is given as the following:
	\begin{equation}\label{eq4}
		\resizebox{1.03\hsize}{!}{$
			\begin{array}{lll}
				u_{i}&=& c_1^{\alpha} \sum\limits_{j \in N_{i}^{\alpha}} \phi_{\alpha}\left(\left\|q_{j}-q_{i}\right\|_{\sigma}\right) \mathbf{n}_{i,j}+c_2^{\alpha} \sum\limits_{j \in N_{i}^{\alpha}} a_{i j}(q)\left(p_{j}-p_{i}\right) \\ &+ &c_1^{\beta} \sum\limits_{k \in N_{i}^{\beta}} \phi_{\beta}\left(\left\|\hat{q}_{i,k}-q_{i}\right\|_{\sigma}\right) \hat{\mathbf{n}}_{i,k }+c_2^{\beta} \sum\limits_{j \in N_{i}^{\beta}} b_{i,k}(q)\left(\hat{p}_{i,k}-p_{i}\right) \\&-  &c_1^{\gamma}\phi_{\alpha_L}\left(\left\|q_{i}-q_{r}\right\|_{\sigma}\right) \mathbf{n}_{i,r} - c_2^{\gamma} \left(p_i - p_r \right),
			\end{array}$}
	\end{equation}
	% 	\begin{equation}\label{eq4}
		% 	\resizebox{.5\textwidth}{!}{$
			% 	\begin{array}{lll}
				% 		u_{i}&=& c_1^{\alpha} \sum\limits_{j \in N_{i}^{\alpha}} \phi_{\alpha}\left(\left\|q_{j}-q_{i}\right\|_{\sigma}\right) \mathbf{n}_{i,j}+c_2^{\alpha} \sum\limits_{j \in N_{i}^{\alpha}} a_{i j}(q)\left(p_{j}-p_{i}\right) \\ &+ &c_1^{\beta} \sum\limits_{k \in N_{i}^{\beta}} \phi_{\beta}\left(\left\|\hat{q}_{i,k}-q_{i}\right\|_{\sigma}\right) \hat{\mathbf{n}}_{i,k }+c_2^{\beta} \sum\limits_{j \in N_{i}^{\beta}} b_{i,k}(q)\left(\hat{p}_{i,k}-p_{i}\right) \\&-  &c_1^{\gamma}\phi_{\alpha_L}\left(\left\|q_{i}-q_{r}\right\|_{\sigma}\right) \mathbf{n}_{i,r} - c_2^{\gamma} \left(p_i - p_r \right)
				% 	\end{array}$}
		% \end{equation}
	% 	\begin{equation}\label{eq4}
		% 	\begin{array}{lll}
			% 		u_{i}&=& c_1^{\alpha} \sum\limits_{j \in N_{i}^{\alpha}} \phi_{\alpha}\left(\left\|q_{j}-q_{i}\right\|_{\sigma}\right) \mathbf{n}_{i,j}+c_2^{\alpha} \sum\limits_{j \in N_{i}^{\alpha}} a_{i j}(q)\left(p_{j}-p_{i}\right) \\ &+ &c_1^{\beta} \sum\limits_{k \in N_{i}^{\beta}} \phi_{\beta}\left(\left\|\hat{q}_{i,k}-q_{i}\right\|_{\sigma}\right) \hat{\mathbf{n}}_{i,k }+c_2^{\beta} \sum\limits_{j \in N_{i}^{\beta}} b_{i,k}(q)\left(\hat{p}_{i,k}-p_{i}\right) \\&-  &c_1^{\gamma}\phi_{\alpha_L}\left(\left\|q_{i}-q_{r}\right\|_{\sigma}\right) \mathbf{n}_{i,r} - c_2^{\gamma} \left(p_i - p_r \right)
			% 	\end{array}
		% \end{equation}
	
	where $ c_1^{\alpha},c_2^{\alpha},c_1^{\beta},c_2^{\beta},c_1^{\gamma},c_2^{\gamma}>0$. Equation (\ref{eq4}) consists of $\alpha$-agent and virtual agents associated with $\alpha$-agents that are called $\beta$- and $\gamma$-agents. The $\beta$-agents are actually on the surface of the obstacles, and when it is in the $\alpha$-agent interaction range, the repulsive force is activated. $\gamma$-agent is also a representation of a leader with an infinitive interaction range with $\alpha$-agents. More details are given in \cite{olfati2006flocking}.
	
	In (\ref{eq4}), the terms $ a_{i j}$, and $ b_{i,k}$ are spatial and heterogeneous adjacency elements respectively, $\phi_{\alpha}(z)$ and $ \phi_{\alpha L}(z)$ are action functions between agent-agent and agent-leader respectively, and $ \phi_{\beta}$ is a repulsive action function between agent-obstacle. Also, $n_{i j}$ is a vector along the line connecting $q_i$ to $q_j$, and $\sigma$-norm is used to construct smooth collective potential functions and is differentiable everywhere. All the terms mentioned are defined as follows:
	\begin{equation}\label{eq5}
		 \begin{array}{cll}
			\phi_{\alpha}(z) &=&\rho_{h}(z/r_{\alpha})\phi(z-d_{\alpha}) \\
			\phi(z) &=& \frac{1}{2} \left[\left(a+b\right)\sigma_1 \left(z+c\right)+(a-b)\right] \\
			\phi_{\alpha L}(z) &=&\rho_{h}(z/r_{\alpha L  })\phi_L(z-d_{\alpha L }) \\
			\phi_L(z) &=& \frac{1}{2} \left[\left(a_L+b_L\right)\sigma_L \left(z+c\right)+(a_L-b_L)\right] \\
			\phi_{\beta}(z) &=&\rho_{h}(z/d_{\beta})(\sigma_1(z-d_{\beta})-1) \\
			\mathbf{n}_{i j} &=& \dfrac{q_j - q_i}{\sqrt{1+\epsilon\|q_j - q_i\|^2}}\\
			a_{i j}(q) &=& \rho_{h}(\|q_j-q_i)\|_{\sigma}/r_{\alpha} \in [0 , 1 ] \\
			b_{i,k}(q) &=& \rho_{h}(\|\hat{q}_{i,k}-q_i)\|_{\sigma}/d_{\alpha} \in [0 , 1 ] \\
			\sigma_1(z) &=& \dfrac{z}{\sqrt{1+z^2}} \\
			c &=& \dfrac{|a-b|}{\sqrt{4ab}} \\
			\|z\|_{\sigma} &=&\dfrac{1}{\epsilon}\left[\sqrt{1+\epsilon\|z\|^2}-1 \right], \\
		\end{array} 
	\end{equation}
	and $ \rho_h(z) $ is as follows:
	\begin{equation}\label{eq6}
		\rho_h (z) = \left\{ \begin{array}{lr}
			1,  & z\in [0,h) \\
			\frac{1}{2}[1+\cos{(\pi \frac{(z-h)}{(1-h)})}],  & z\in [h,1] \\
			0 & \text{otherwise}
		\end{array} \right.
	\end{equation}
	
	where $ \phi(z)$ and $ \phi_L(z)$ are uneven sigmoidal functions with parameters $0<a\leq b$ and $0<a_L\leq b_L$ to guarantee $ \phi(0)=0$ and $ \phi_L(0)=0$. 
	 The rest of the parameters can be expressed as:
	\begin{equation*}
		 \begin{array}{@{}cc@{}}
			d_{\alpha} = \|d\|_{\sigma}, &  r_{\alpha} = \|r\|_{\sigma} \\
			d_{\beta} = \|d^{\prime}\|_{\sigma}, &  r_{\beta} = \|r^\prime\|_{\sigma}  \\
			d_{\alpha L}=\|d_L\|_{\sigma}, &  r_{\alpha L} = \|r_L\|_{\sigma} \\
			0<\epsilon<1  ,&  0<h<1  \\
			0<\epsilon_L<1  ,&  0<h_L<1,  \\
		\end{array}
	\end{equation*}
	in which $d,\: d^{\prime}$, and $ d_L$ are the desired distance between agent-agent, agent-obstacle, and agent-leader, respectively.
	Also, the interaction range between agent-agent, agent-obstacle, and agent-leader are denoted by $r,\: r^{\prime}$, and $ r_L$, respectively, and $r$ is defined as $r=\kappa d$ with $1< \kappa \ll 2$.\\
	Also, the pairwise potential function is defined as:
	\begin{equation}\label{eq7}
		\psi_\alpha(z)=\int_{d_\alpha}^z \phi_\alpha(s) d s.
	\end{equation}
	According to (\ref{eq6}), $\rho_{h}(z/r_{\alpha})$ is zero at distance $ r$ (interaction range) and more, so it makes the action function also zero. Hence, $\rho_h (z)$ can be called a cut-off for the action function.
	With the introduced leader action function, all agents everywhere are attracted to the leader; therefore, no cut-off is considered for the action function. \\
	\begin{asum}\label{asum:1} 
		The action function of the leader is assumed without cut-off, so the $\rho_{h}(z/r_{\alpha 1}) $ term of the leader vanishes for single circle formation.
	\end{asum}
	
\begin{figure}[ht]
		\centering % <-- added
		\begin{subfigure}{0.24\textwidth}
			\includegraphics[width=\linewidth]{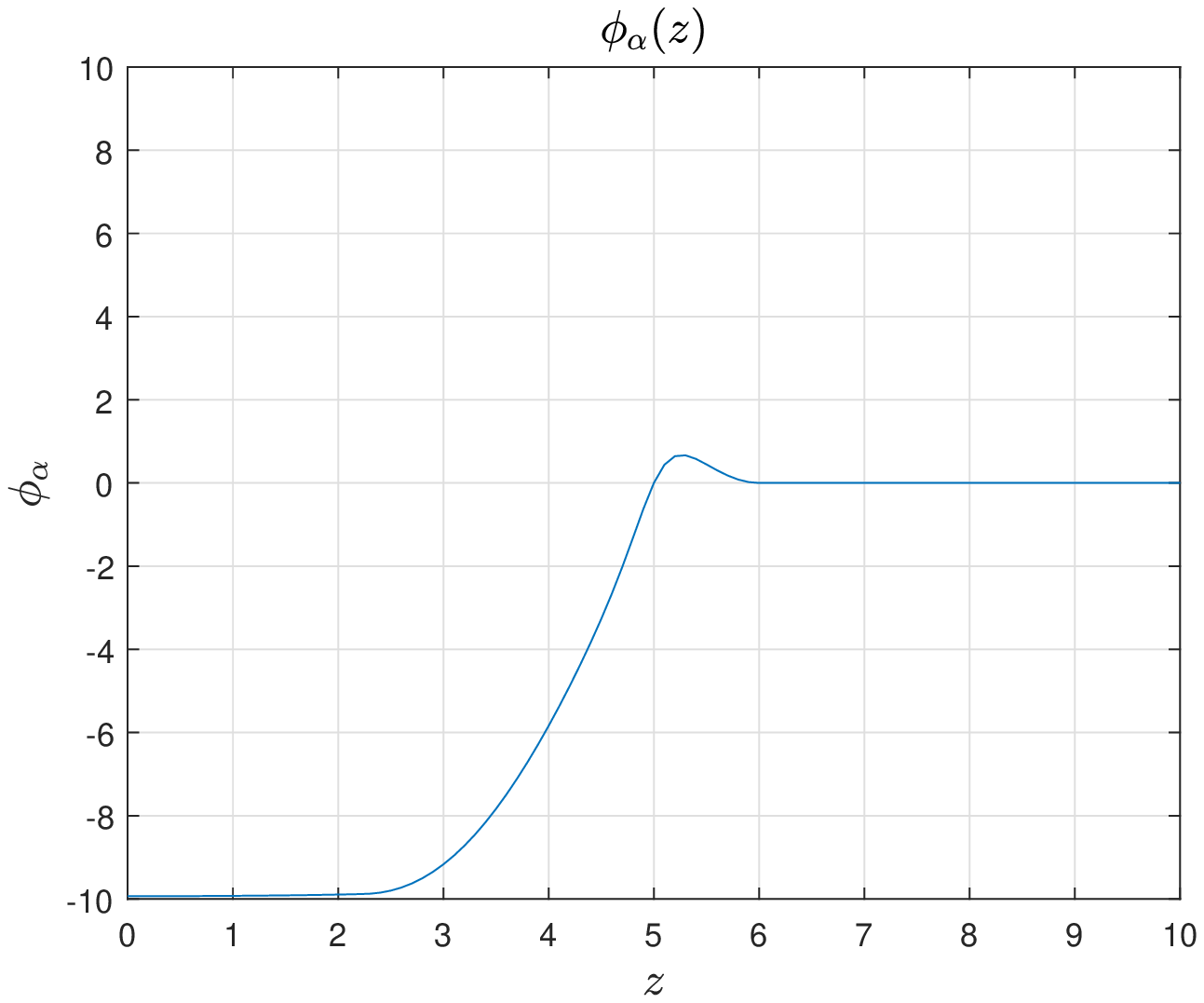}
			\caption{$ \phi_\alpha$ with cut-off}
			\label{fig:phi_alpha}
		\end{subfigure}\hfil % <-- added
		\begin{subfigure}{0.24\textwidth}
			\includegraphics[width=\linewidth]{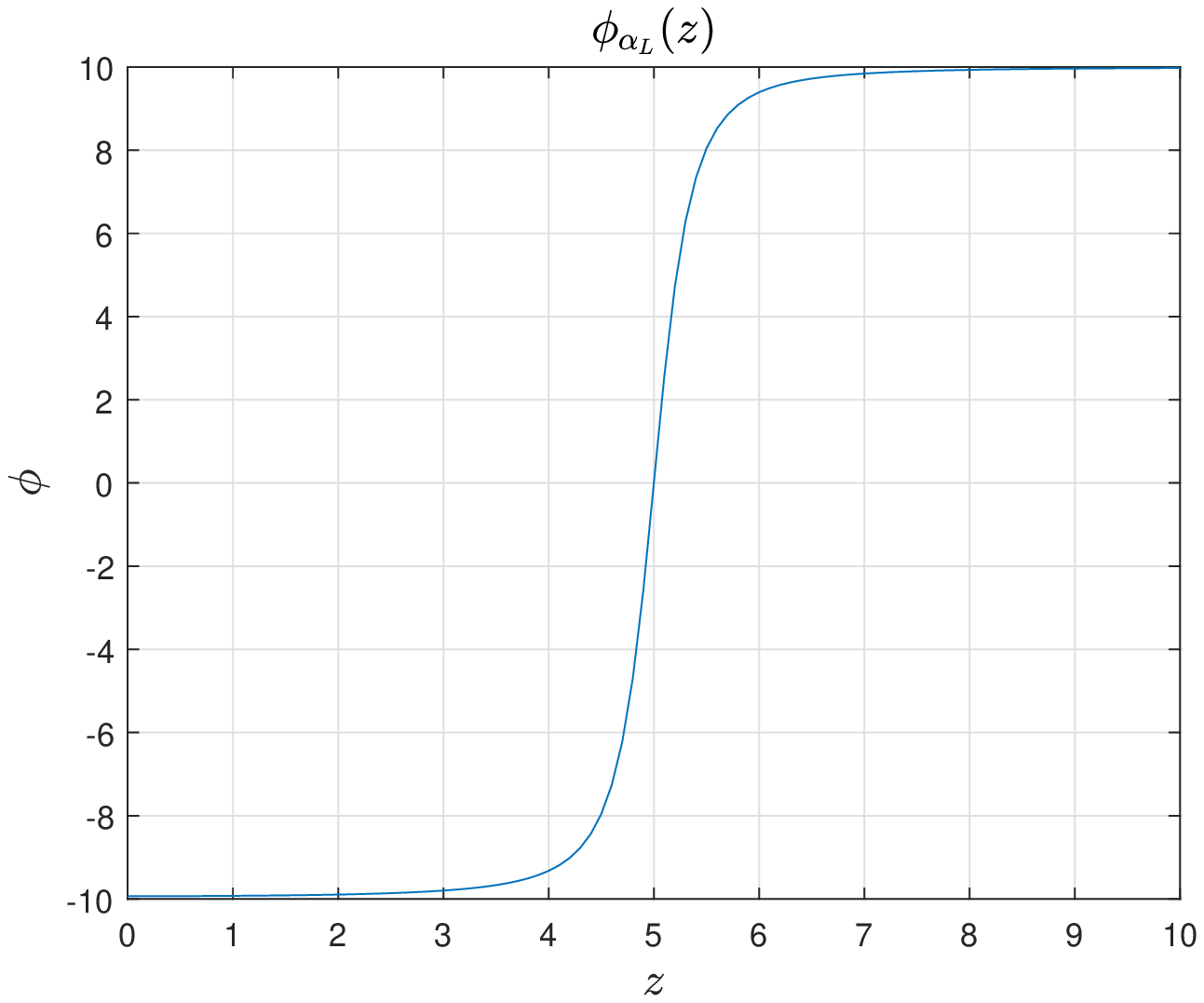}
			\caption{$\phi_\alpha$ without cut-off}
			\label{fig:phi}
		\end{subfigure} % <-- added
		\caption{Action function with $ d=5 $}
		\label{fig:fig1}
	\end{figure}

\begin{figure}[ht]
		\centering % <-- added
		\begin{subfigure}{0.24\textwidth}
			\includegraphics[width=\linewidth]{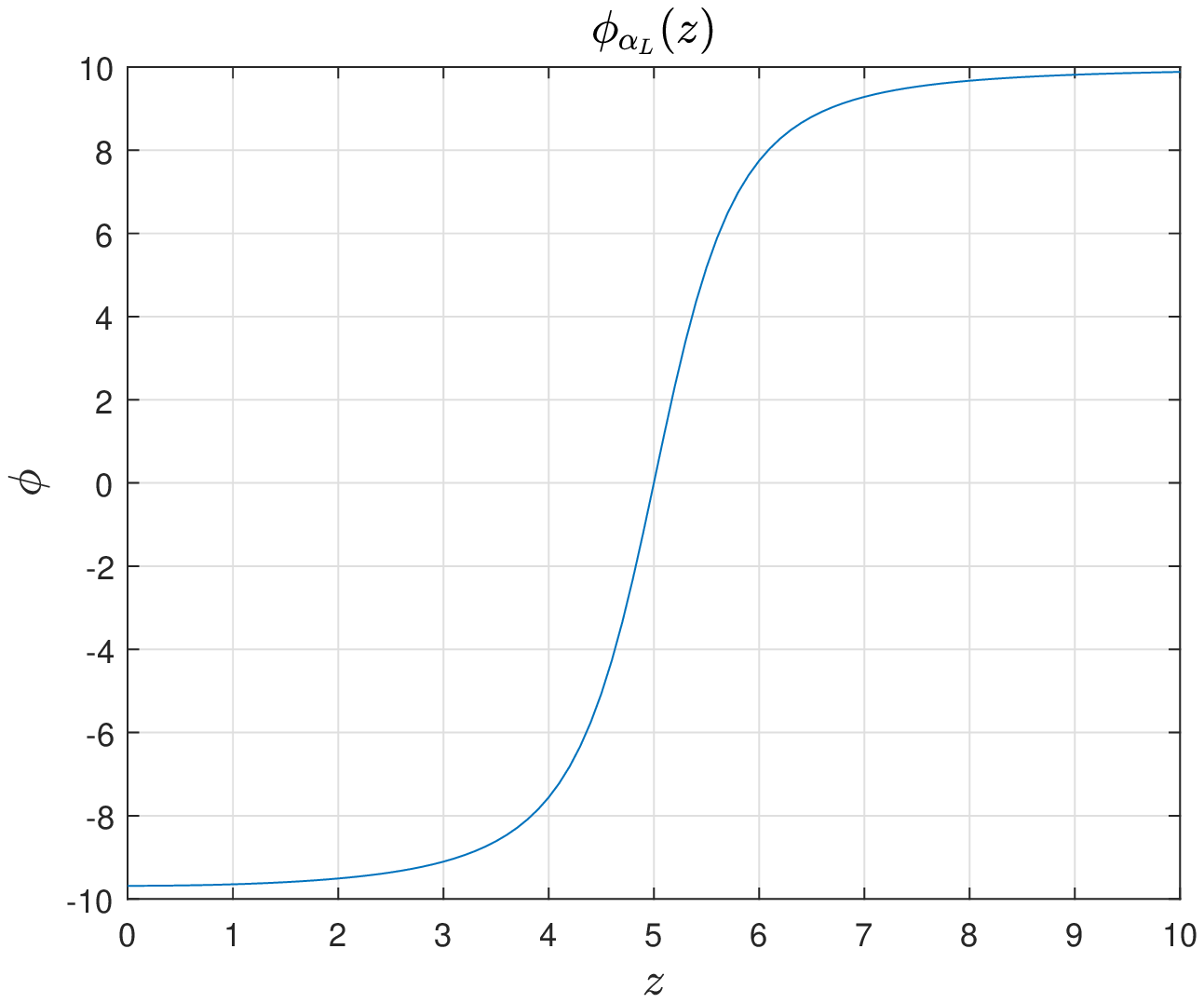}
			\caption{$ \phi_\alpha$ without cut-off}
			\label{fig:phi_alpha_lf}
		\end{subfigure}\hfil % <-- added
		\begin{subfigure}{0.24\textwidth}
			\includegraphics[width=\linewidth]{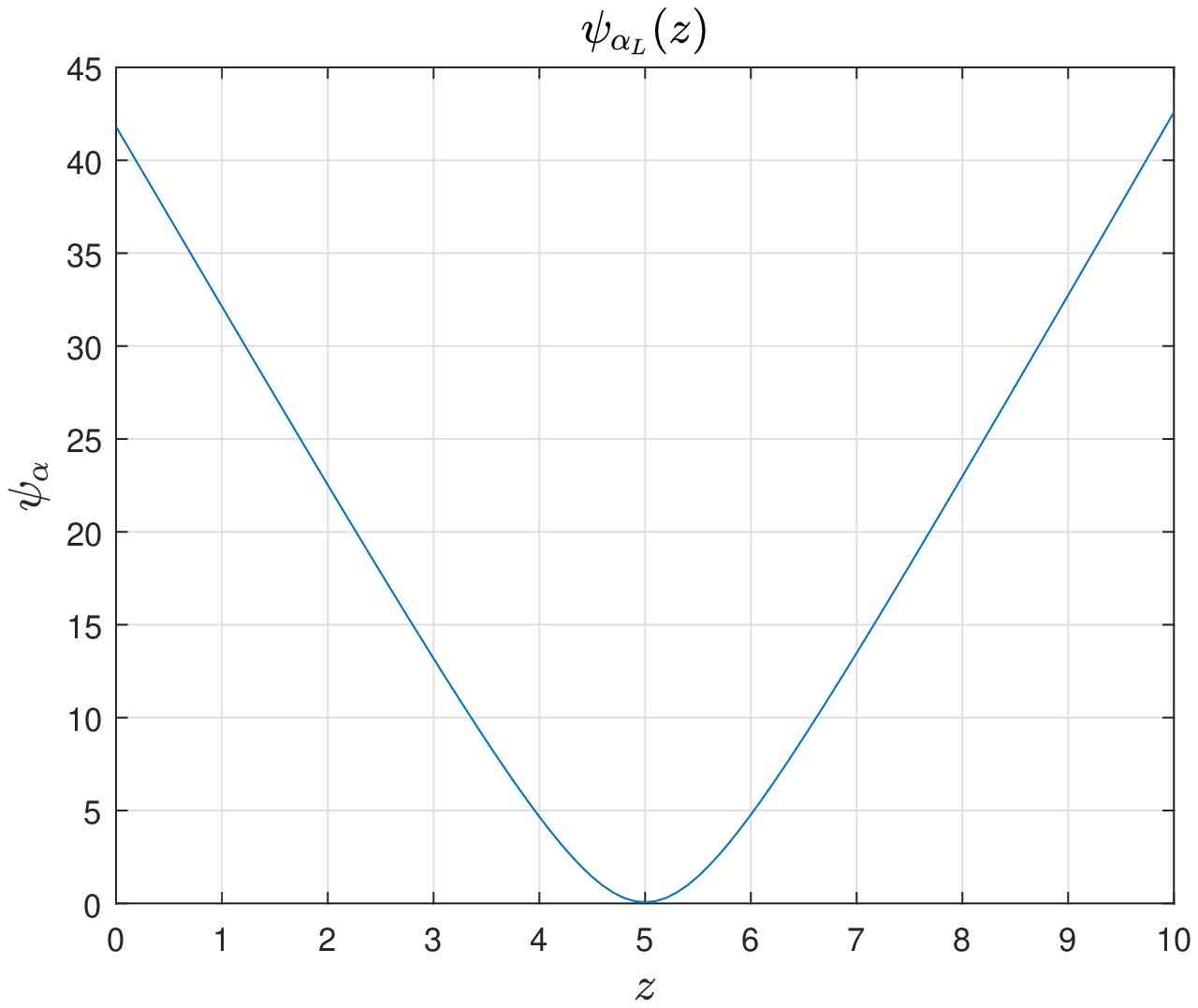}
			\caption{$\psi_\alpha$ without cut-off}
			\label{fig:psi}
		\end{subfigure} % <-- added
		\caption{Action and potential function}
		\label{fig:fig2}
	\end{figure}

	The difference between the action function with and without cut-off is shown in Fig. \ref{fig:fig1}.
	As seen in Fig. \ref{fig:phi_alpha} desired distance $d=5$ is considered, and according to $ \kappa = 1.2$, the interaction range $r=6$ is obtained, so for $z=6$ and more, the action function is zero. In Fig. \ref{fig:phi}, the slope of the action function is steep and may cause problems such as oscillation in the desired distance; therefore, the slope can be reduced by Assumption \ref{asum:2}.
	\begin{asum}\label{asum:2}
		$ \sigma_L$ is modified as follows:\\
		% \begin{equation}\label{eq8}
			%  \sigma_L(z) =\frac{z}{\sqrt{a_{\sigma_L}\times d+z^2}}, \quad a_{\sigma_L}\geq 1  
			% \end{equation}
		\begin{equation*}
			\sigma_L(\frac{a_{\sigma_L}}{d_{\alpha L}} z+c)
		\end{equation*}
		\\
		The slope of the potential function in Fig. \ref{fig:phi} decreases with the sigma in the above equation, and for each $d$, an almost identical slope is obtained. The slope of the potential function can be increased or decreased with parameter $a_{\sigma_L}$. As $a_{\sigma_L}$ increases, the slope decreases; conversely,  as $a_{\sigma_L}$ decreases, the slope rises. In this work, $a_{\sigma_L}=1$ is considered.
	\end{asum}
	Finally, the diagram of leader action and potential function with $ a_{\sigma_L} =1$ is depicted in Fig. \ref{fig:fig2}.
	According to potential and action functions in Fig. \ref{fig:fig2}, all agents move and are placed around the virtual leader at desired distance $d_L=5$.
	\subsection{Polygon formation}
	The virtual leader action function causes the agents to be placed on the circle of the leader with the desired radius around the leader.
	Also, the distance between the agents must be adjusted to form a polygon formation on the circle of the leader. The distance between agents depends on the number of agents placed on the circle of the leader.
	\begin{figure}
		\centering
		\includegraphics[width=0.75\linewidth]{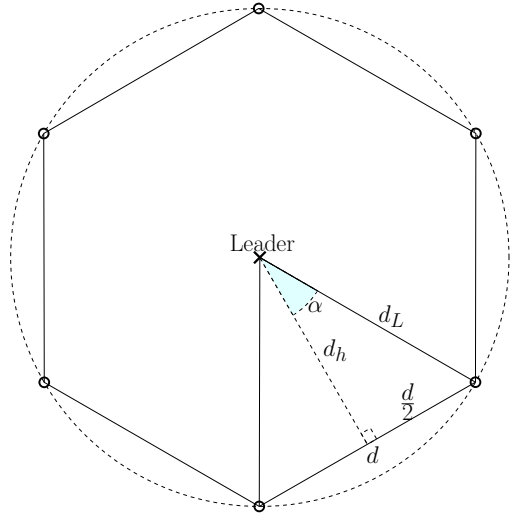}
		\caption{An example of polygon formation for 6 agents}
		\label{fig:fig3}
	\end{figure}
	Fig. \ref{fig:fig3} shows a polygon formation for $N = 6$ on the circle of the leader with a desired radius of $d_L$. As seen in Fig. \ref{fig:fig3}, $d$ is the distance between the agents, $\alpha$ is the half angle between two neighbor agents, $d_L$ is the distance between the agent with the leader, and $d_h$ is the distance orthogonal to $d$. According to Fig. \ref{fig:fig3}, the distance between agents is calculated as follows:
	\begin{equation}\label{eq8}
		d=2\sin{\frac{\pi}{N}}d_L  , \quad \textit{for d}\geq1,
	\end{equation}
	where $N$ is the number of agents. In principle, one of the reasons to form a polygon is the repulsive force between agents. \\
	%Equation (\ref{eq8}) is derived from Figure \ref{fig:fig3}.
	If one or some agents fail, the polygon formation will not be formed correctly during the polygon formation. To solve this problem, Assumption \ref{asum:3} is proposed to form polygon formation in the presence of failure of agents.
	\begin{asum}\label{asum:3}
		If the fault occurs and some of the agents fail, the time of fault should be detected, and then (8) should be updated with the new number of agents.
	\end{asum}
	It is guaranteed that the polygon formation is formed by Assumption \ref{asum:3}. In other words, when an agent is failed, the desired distance between agents is updated and increased, so the repulsive force between the agents causes the polygon formation to be formed with fewer agents. 
	\subsection{Obstacle avoidance and size scaling}
	The second term of (\ref{eq4}) alone guarantees obstacle avoidance, but sometimes it is better to resize the formation to better pass through obstacles. Obstacles in this work are considered as circles.  The idea of size scaling of the formation is in such a way that when facing a circle obstacle with a radius equal to or larger than our formation,  the circle formation will expand. When facing two side obstacles, the radius of the circular formation will shrink.
	% (The idea of changing the scale of the formation is that when facing a circle with a radius equal to or greater than our formation,  the circle formation is expanding and when facing two side obstacles, the circular formation shrinks). or The idea of size scaling of the formation is to increase the radius of the circular formation when facing a circular obstacle with a radius equal to or larger than our formation, and when facing two side obstacles,  the circular formation shrinks.).
	To resize the circular formation, we use the interaction range between agent-obstacle and agent-agent, which is presented as follows:
	% Algorithm \ref{alg1}.
	%\begin{algorithm}[ht]
	%		\caption{$\mathrm{Size \: scaling \: formation \:  \: }$}\label{alg1}
	%		\begin{algorithmic}[1]
		%            \While {$t<t_{final}$}
		%            \For{$ i=1$ to $ N$ }
		%            	\State $ B = b_{i,k}, A = a_{i,j}$ 
		%            \EndFor
		%                \If {$ \text{numel}(B) \geq N/4$}
		%                    %\State $ $
		%                    \If {$ (a_{i,j}) \leq 0$, in neighborhood}
		%                    \State $ d_L$ increase
		%                    \State Update Eq (\ref{eq8})
		%                    \Else 
		%                    \State $ d_L$ decrease
		%                    \State Update Eq (\ref{eq8})
		%                    \EndIf
		%                \EndIf
		%            \EndWhile
		%		\end{algorithmic}
	%	\end{algorithm}
%	\\
%Algorithm \ref{alg1} is such that 
%If at least a quarter of the agents is in the interaction range with the obstacle and the same agents are located in each other's neighborhood, (or in other words, they are in the interaction range with each other), the circle formation expands. But if they are not in each other's neighborhood, the formation shrinks. Also, Equation \ref{eq8} should be updated after the circular formation expands or shrinks. The amount of expansion and shrink (ie $d_L$) of circular formation is arbitrary. 
\begin{enumerate}
	\item If at least a quarter of the agents is in the interaction range with the obstacle:
	\begin{enumerate}
		\item If those agents are located in each other's neighborhood, or in other words, they are in the interaction range with each other, the circle formation expands.
		\item Else, the circle formation shrinks.
	\end{enumerate}
	\item Update equation (\ref{eq8})
	\item After several time steps, the radius of the circular formation returns to its initial condition, and the previous steps are re-checked. 
\end{enumerate}
Note that the interaction range between agent-obstacle and agent-agent is obtained from $a_{i,j}$, and $b_{i,k}$ of (\ref{eq5}), respectively. Also, the radius of the circle and the duration is arbitrary for expansion or shrinkage.
\subsection{Multi-circle}
The multi-circle formation requires a large number of agents to form. 
However, due to the limitation on the distance between agents in (\ref{eq8}), only a limited number of agents can be placed on a circle formation.
After the leader's circle formation is completed, extra agents are placed outside the leader's circle to create a larger circle due to the potential function between the agents. These extra agents push the front agents towards the leader, decreasing the distance between the front agents and the leader. However, due to the uncertainty of the exact placement of these agents, the multi-circle formation may become irregular. This makes it difficult to control the circle formation, as the radius of the circles cannot be accurately calculated. This problem is due to the simultaneous existence of two potential functions, one between agents and the other between agents and the leader  ($\alpha$- and $\gamma$-agent). To address this issue, the piecewise action function is proposed: 
\begin{equation}\label{eq9}
	\phi_{\alpha_L}= \left\{ \begin{array}{cc}
		a_1\rho_h\left(z / r_{\alpha L_{1}}\right) \phi\left(z-d_{\alpha L_{1}}\right) & \text { if } z \leq r_{\alpha L_{1}} \\
		a_2\rho_h\left(z / r_{\alpha L_{2}}\right) \phi\left(z-d_{\alpha L_{2}}\right) & \text { elseif } z \leq r_{\alpha L_{2}} \\
		\vdots & \vdots \\
		\phi\left(z-d_{\alpha L_{n}}\right) & \text { else } 
	\end{array}\right.
\end{equation}
where $n$ is the number of circles around the leader and $r_{\alpha L_{i}} $ is the cut-off of the $i$th circle. Also, only the action function of the last circle is without a cut-off. Note that by determining the radius of the circles and the distance between the agents, the total number of agents can be obtained from (\ref{eq8}). 
\begin{figure}[ht]
	\centering % <-- added
	\begin{subfigure}{0.24\textwidth}
		\includegraphics[width=\linewidth]{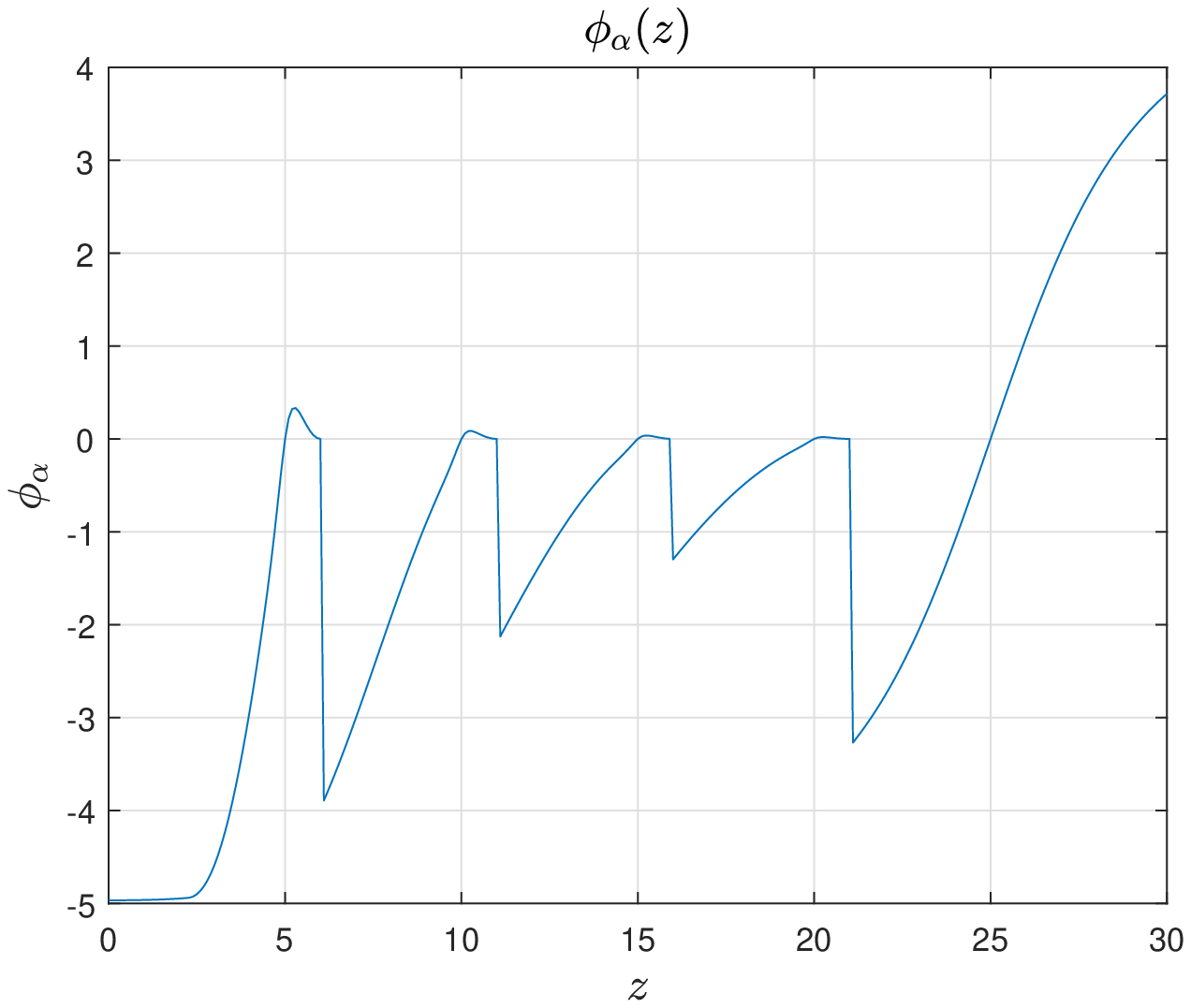}
		\caption{Action function $\phi_\alpha$}
		\label{fig:phi_alpha_m}
	\end{subfigure}\hfil % <-- added
	\begin{subfigure}{0.24\textwidth}
		\includegraphics[width=\linewidth]{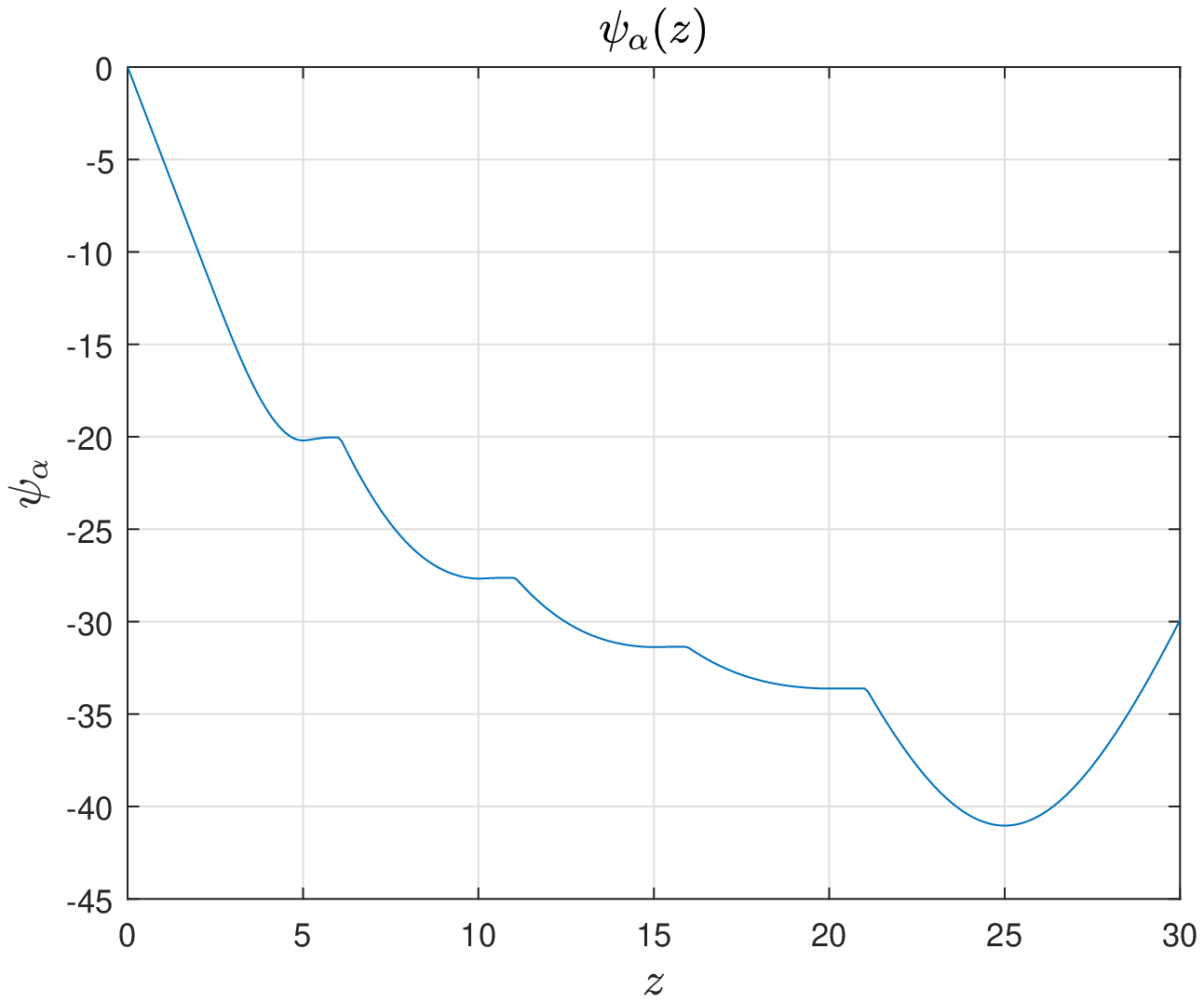}
		\caption{Potential function $\psi_\alpha$}
		\label{fig:psi_m}
	\end{subfigure} % <-- added
	\caption{An example of piecewise action and potential functions for five circles}
	\label{fig:fig4}
\end{figure}
The new piecewise action and potential functions are shown in Fig. \ref{fig:fig4}.
As seen in Fig. \ref{fig:fig4}, the radius of the circles is 5, 10, 15, 20, and 25. The first four circles have a cut-off with $\kappa=1.2$.
Also, in Fig. \ref{fig:fig4}, the action function amplitude  for long distances decreases. However, adjusting $ a_1, a_2,\cdots,a_{n-1}$ of (\ref{eq9}) can compensate for this decrease in the action function.
Equation (\ref{eq9}) is dependent on the initial condition of agents. Therefore, two scenarios can be considered for it. First, if the initial positions of agents are in the first circle,  the repulsive force of the leader pushes the agents to the first radius of the circle. The cut-off of the first circle causes extra agents to be repelled and go to the second circle, and in the same way, the extra agents of each circle go to the next circle and finally to the last circle. Nevertheless, due to the high density of agents in the first circle, agents may collide with each other in the initial moments.
Second, suppose the initial conditions of the agents are random everywhere. In that case, the completion of the circles at any radius depends on the initial conditions of the agents, so some circles may not be completed. \\
To address these issues, a switching action function algorithm is proposed. The idea behind the switching is to form the circles in an outward order from the inside, respectively. In other words, the first circle is formed first, followed by the second circle, and so on, until the last circle is finished. The switching action function algorithm is as follows: \\
\begin{equation}\label{eq10}
		\phi_{\alpha_L}=
	\resizebox{0.9\hsize}{!}{$
		\left\{\begin{array}{cc}
			\phi(z-d_{\alpha L_{1}}) &  \text{$0 \leq t  \leq ts$}\\ \\
			\left\{ \begin{array}{cc}
				a_1\rho_h\left(z / r_{\alpha L_{1}}\right) \phi\left(z-d_{\alpha L_{1}}\right) & \text { if } z \leq r_{\alpha L_{1}} \\
				\phi\left(z-d_{\alpha L_{2}}\right) & \text { else } 
			\end{array}\right. & \text{$ ts < t \leq 2ts$} \\ \\
			\left\{\begin{array}{cc}
				a_1\rho_h\left(z / r_{\alpha L_{1}}\right) \phi\left(z-d_{\alpha L_{1}}\right) & \text { if } z \leq r_{\alpha L_{1}} \\
				a_2\rho_h\left(z / r_{\alpha L_{2}}\right)\phi\left(z-d_{\alpha L_{2}}\right) & \text { elseif }  z \leq r_{\alpha L_{2}} \\
				\phi\left(z-d_{\alpha L_{3}}\right) & \text { else } 
			\end{array}\right. & \text{$ 2ts < t \leq 3ts$}\\ \\
			\vdots & \vdots \\ \\
			\left\{  \begin{array}{cc}
				a_1\rho_h\left(z / r_{\alpha L_{1}}\right) \phi\left(z-d_{\alpha L_{1}}\right) & \text { if } z \leq r_{\alpha L_{1}} \\
				a_2\rho_h\left(z / r_{\alpha L_{2}}\right) \phi\left(z-d_{\alpha L_{2}}\right) & \text { elseif } z \leq r_{\alpha L_{2}} \\
				\vdots & \vdots \\
				\phi\left(z-d_{\alpha L_{n}}\right) & \text { else } 
			\end{array}\right. &  \text{$ nts < t  \leq t_f $} \\ \\
		\end{array}\right.
		$}
\end{equation}
where $ts$ and $t_f$ are switching and final simulation times.
As seen in  (\ref{eq10}), in the first switch, i.e., $0 \leq t  \leq t_1$, all agents are attracted to the first circle. In the second switch, by considering the cut-off for the first circle, the remaining agents are attracted to the second circle, and extra agents of the first circle are repelled and go to the second circle. Moreover, the final switch is defined in the same way.
Switching   ($ts$) depends on agent velocity and algorithm parameters and is defined practically.
% \begin{equation}
	% \resizebox{.5\textwidth}{!}{$
		% \left\{\begin{array}{cc}
			%  \phi(z-d_{\alpha L_{1}}) &  \text{$t  \leq t_1$}\\ \\
			% \left\{ \begin{array}{cc}
				%  a_1\rho_h\left(z / r_{\alpha L_{1}}\right) \phi\left(z-d_{\alpha L_{1}}\right) & \text { if } z \leq r_{\alpha L_{1}} \\
				%  \phi\left(z-d_{\alpha L_{2}}\right) & \text { else } 
				% \end{array}\right\} & \text{$ t_1 \leq t \leq t_2$} \\ \\
			% \left\{\begin{array}{cc}
				% a_1\rho_h\left(z / r_{\alpha L_{1}}\right) \phi\left(z-d_{\alpha L_{1}}\right) & \text { if } z \leq r_{\alpha L_{1}} \\
				% a_2\rho_h\left(z / r_{\alpha L_{2}}\right)\phi\left(z-d_{\alpha L_{2}}\right) & \text { elseif }  z \leq r_{\alpha L_{2}} \\
				% \phi\left(z-d_{\alpha L_{3}}\right) & \text { else } 
				% \end{array}\right\} & \text{$ t_2 \leq t \leq t_3$}\\ \\
			% \vdots & \vdots \\ \\
			% \left\{  \begin{array}{cc}
				% a_1\rho_h\left(z / r_{\alpha L_{1}}\right) \phi\left(z-d_{\alpha L_{1}}\right) & \text { if } z \leq r_{\alpha L_{1}} \\
				% a_2\rho_h\left(z / r_{\alpha L_{2}}\right) \phi\left(z-d_{\alpha L_{2}}\right) & \text { elseif } z \leq r_{\alpha L_{2}} \\
				% \vdots & \vdots \\
				% \phi\left(z-d_{\alpha L_{n}}\right) & \text { else } 
				% \end{array}\right\} &  \text{$t  > t_n$} \\ \\
			% \end{array}\right.
		% $}
	% \end{equation}
\\The leader cut-off value ($ r_{ L}$) is another critical parameter that should be adjusted so that the extra agents on the circle are repelled and go to the next outer circle. According to Fig. \ref{fig:fig3}, \ref{fig:fig5} and \ref{eq9}, the distance between the leader and the extra agent can be obtained as follows:
\begin{figure}
	\centering
	\includegraphics[width=\linewidth]{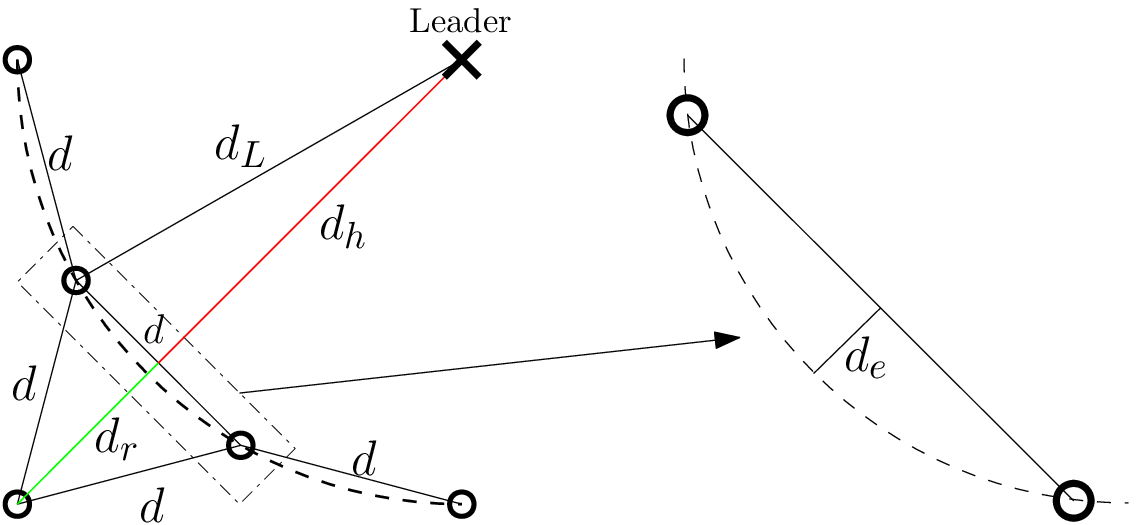}
	\caption{Extra agents on circle}
	\label{fig:fig5}
\end{figure}
\begin{equation}\label{eq11}
	\left\{ \begin{array}{ccl}
		d_{h_i} & = & d_{L_i}\cos{\alpha}, \quad \alpha_i = \dfrac{\pi}{N_i} \\
		d_{e_i} & = & d_{L_i} - d_{h_i} \\
		d_r & = & \dfrac{\sqrt{3}}{2}d  \\
		d_{Le_i}& = & d_{L_i} + d_r - d_{e_i} =  d_{h_i} + d_r,
	\end{array} \right.
\end{equation}
where $d_{Le}$ is the distance between the leader with the extra agent, and $i$ represents the $i$th circle. 
Now, to cut the extra agents, the cut-off of the circles must be less than the corresponding $d_{Le_i}$.
\begin{equation}\label{eq12}
	r_{L_i} = d_{Le_i} - d_{\varepsilon},
\end{equation}
and $ d_{\varepsilon}>0$.\\
Another fundamental problem is that the interaction range between two agents equals or exceeds the distance between the two circles.
Therefore the interaction range between agents should also be adjusted according to the radius of circles.
% \begin{figure}
	%     \centering
	%     \includegraphics[width=\linewidth]{image/asum41.eps}
	%     \caption{Setting the interaction range between agents considering the leader's cut-off limit}
	%     \label{fig:fig6}
	% \end{figure}
\begin{figure}
	\centering
	\includegraphics[width=0.75\linewidth]{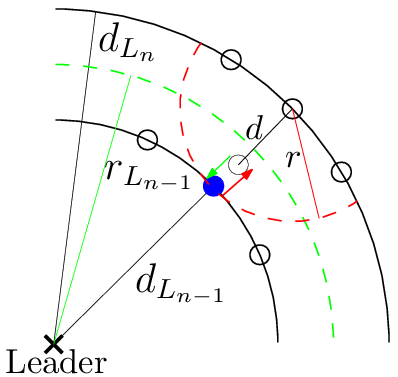}
	\caption{Setting the interaction range between agents considering the leader's cut-off limit}
	\label{fig:fig6}
\end{figure}
As shown in Fig. \ref{fig:fig6}, the blue agent is in the attraction range of the leader's first circle and the agent of the second circle, and two opposing forces affect the agent.
If two opposing forces are equal, the agent may remain stationary.
Alternatively, if the force between the agents is more substantial, the blue agent will move toward the second circle.
This problem causes an irregular circular formation, so Assumption \ref{asum:4} is given to solve this problem.
\begin{asum}\label{asum:4}
	The interaction range between the agents (i.e., $r$) should be less than the distance between every two circles.
	\begin{equation}\label{eq13}
		r < d_{ L_{n}} - d_{L_{n-1}}
	\end{equation}
\end{asum}
The agents of two adjacent circles do not exert opposing forces on each other by Assumption \ref{asum:4}.

				\subsection{Optimization of algorithm parameters}
				This section aims to obtain the parameters of alpha and gamma agents with different scenarios. 
				The interaction between the potential functions of the $\alpha$- and $\gamma$-agents makes it difficult to determine the optimal parameters manually.
				There are many optimization algorithms, but in this paper, GA \cite{mitchell1998introduction}, PSO \cite{kennedy1995particle}, and GWO \cite{mirjalili2014grey} algorithms are used to obtain the optimal parameters. The parameters of the algorithm that are calculated from optimization are $ (c_1^{\alpha}, c_2^{\alpha}, c_1^{\gamma}, c_2^{\gamma} )$ in (\ref{eq4}), and $ (a, b, a_L, b_L, \epsilon, h, \epsilon_L)$ in (\ref{eq5}, \ref{eq6}). For optimization, single circle formation and the maximum number of agents on the circle, according to (\ref{eq8}), are considered.
				The fitness function to be optimized includes the distance between adjacent agents and between agents with the leader. \\
									\begin{figure*}[ht]
					\centering % <-- added
					\begin{subfigure}{0.25\textwidth}
						\includegraphics[width=\linewidth, height=0.2\textheight]{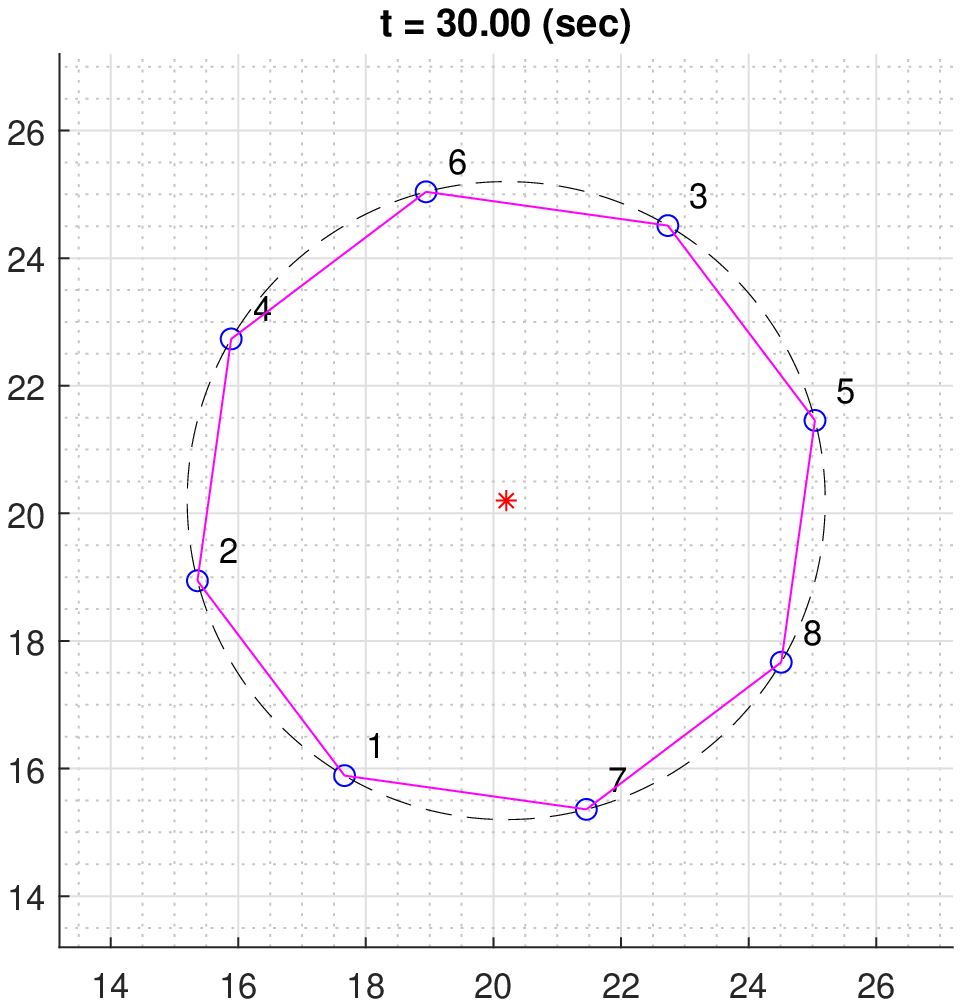}
						\caption{}
						\label{fig:flt1}
					\end{subfigure}\hfil % <-- added
					\begin{subfigure}{0.25\textwidth}
						\includegraphics[width=\linewidth, height=0.2\textheight]{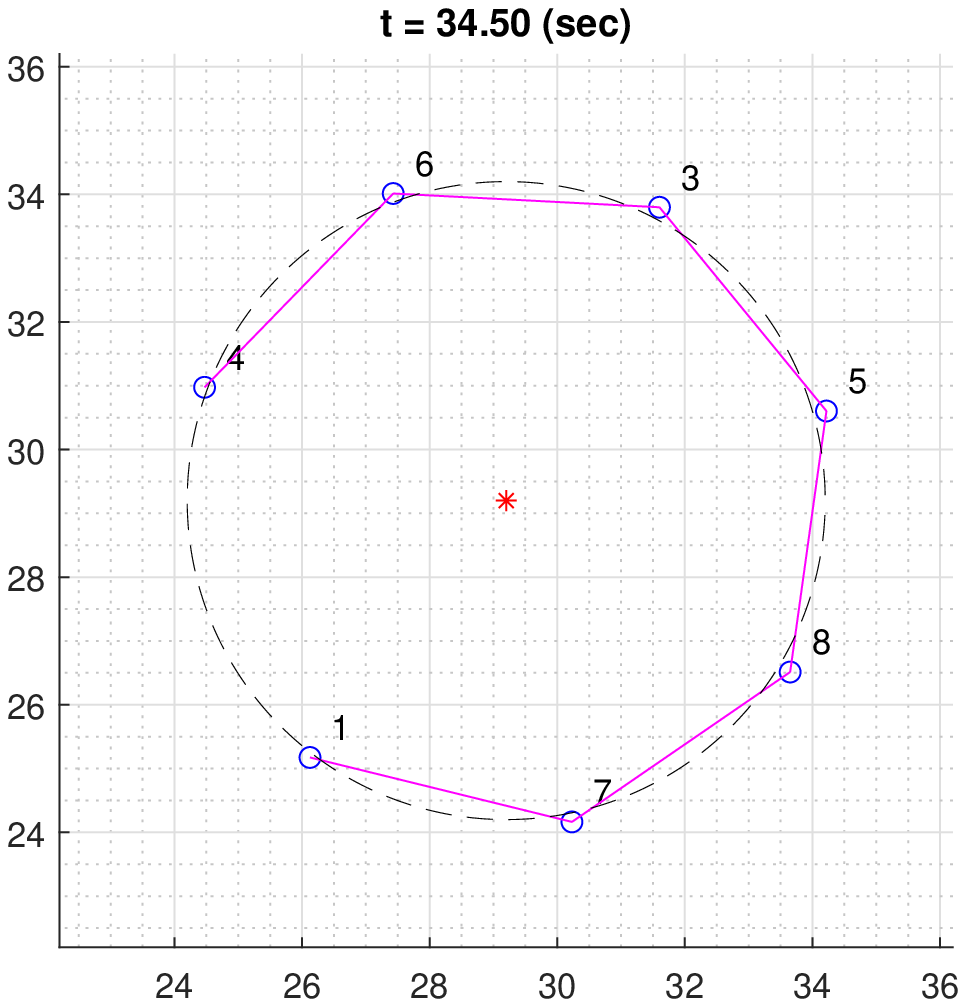}
						\caption{}
						\label{fig:flt2}
					\end{subfigure}\hfil % <-- added
					\begin{subfigure}{0.25\textwidth}
						\includegraphics[width=\linewidth, height=0.2\textheight]{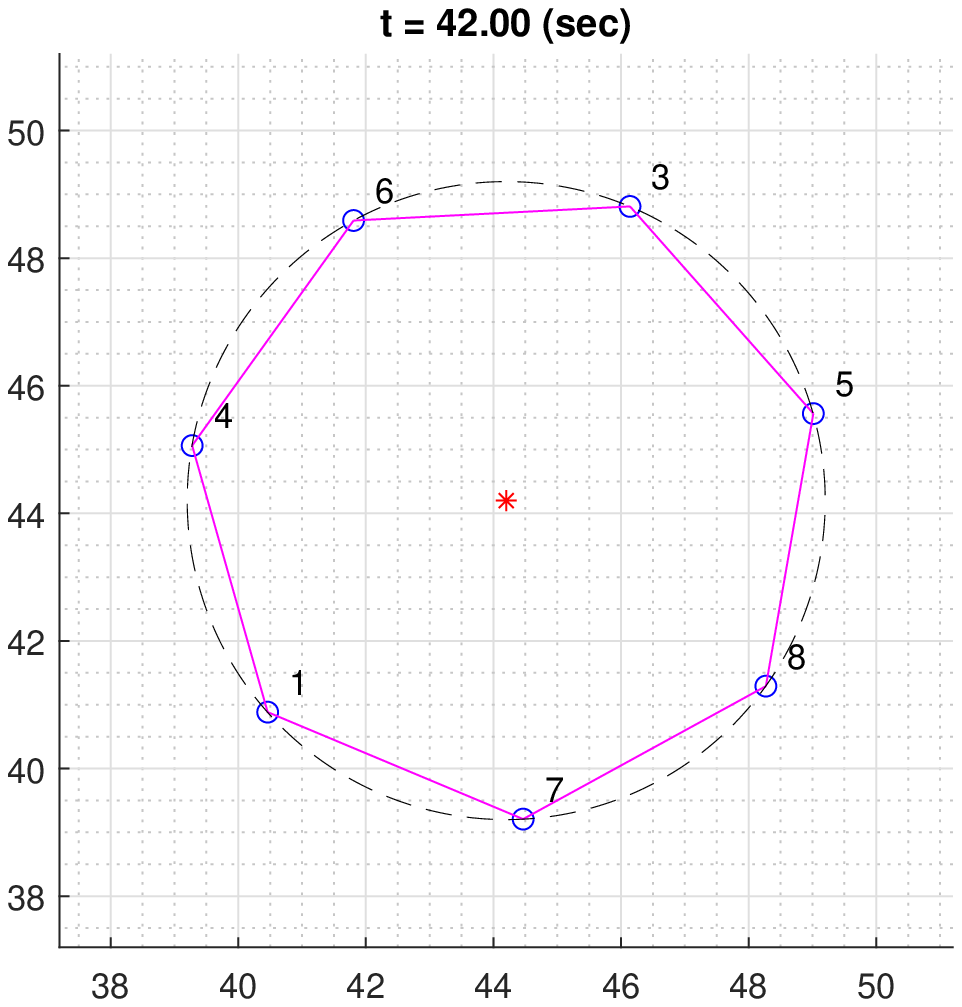}
						\caption{}
						\label{fig:flt3}
					\end{subfigure}%\hfil % <-- added
					\begin{subfigure}{0.25\textwidth}
						\includegraphics[width=\linewidth, height=0.2\textheight]{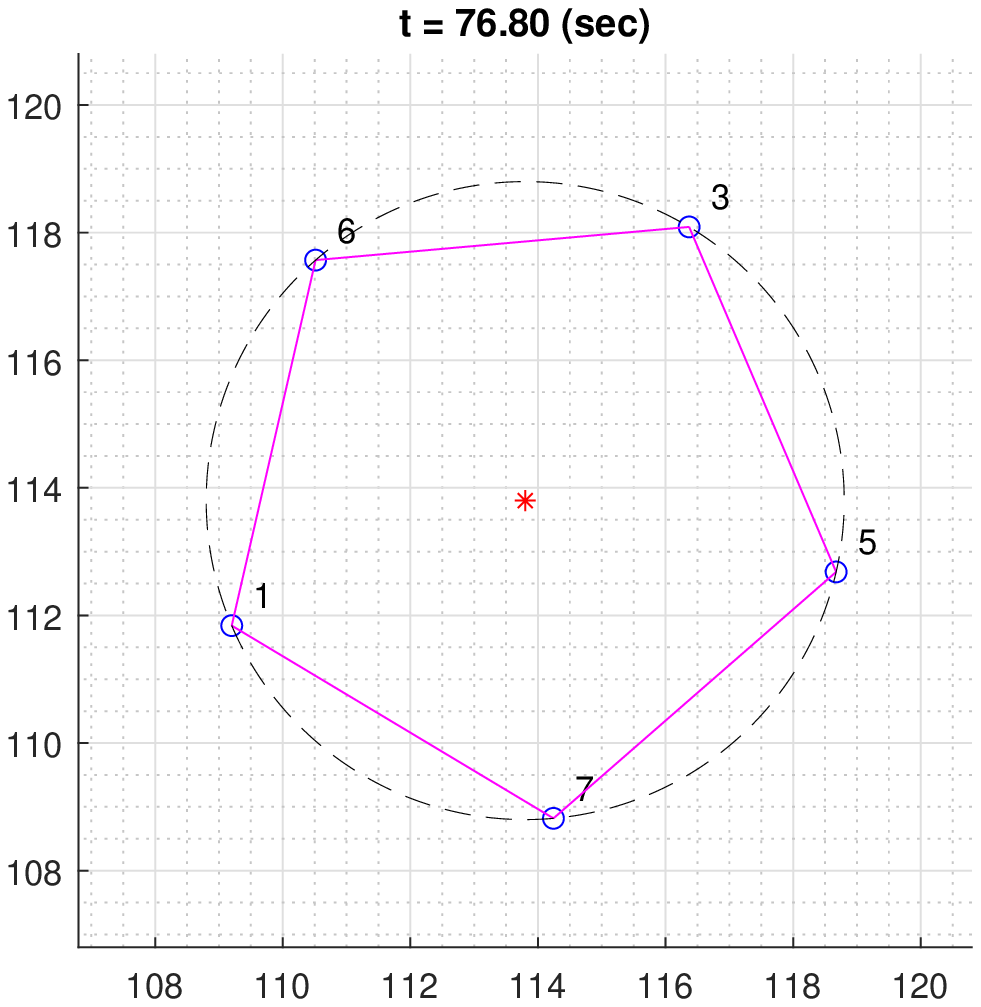}
						\caption{}
						\label{fig:flt6}
					\end{subfigure}\\ %\hfil % <-- added
					\begin{subfigure}{0.25\textwidth}
						\includegraphics[width=\linewidth, height=0.2\textheight]{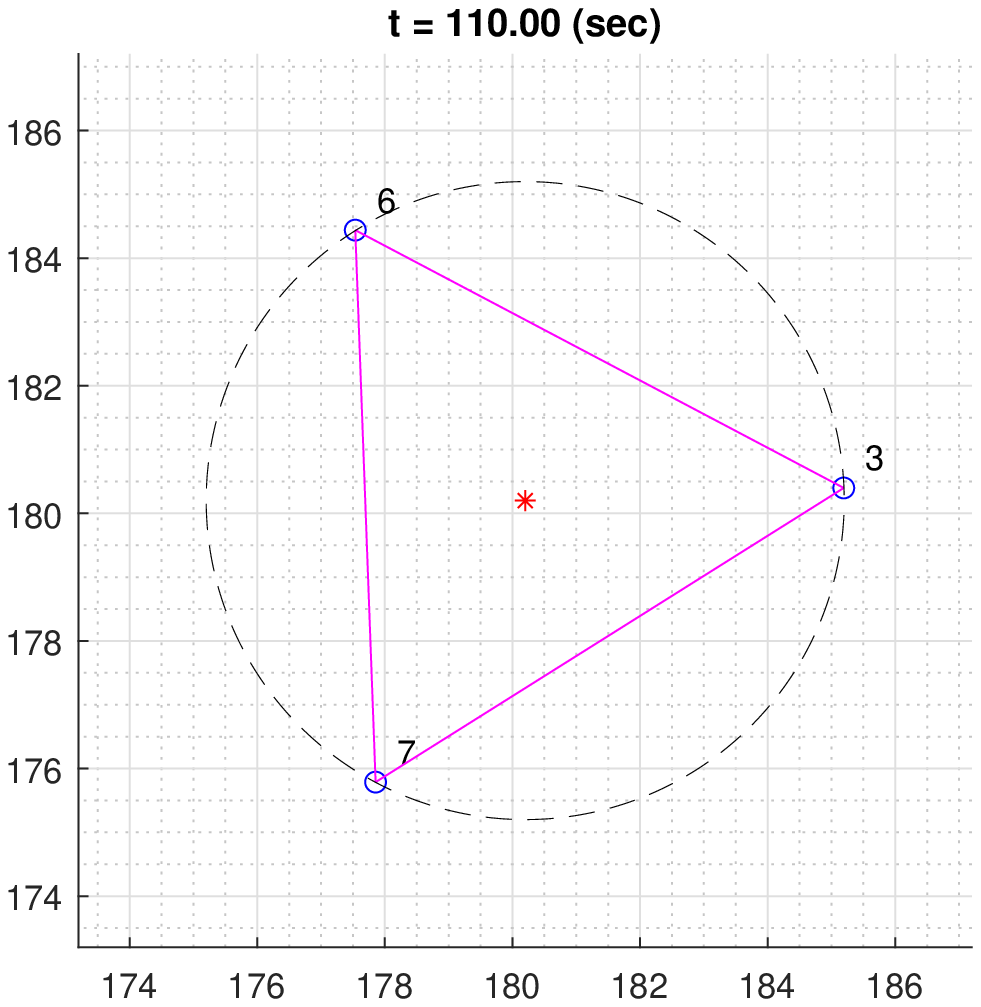}
						\caption{}
						\label{fig:flt7}
					\end{subfigure}\hfil % <-- added
					\begin{subfigure}{0.375\textwidth}
						\includegraphics[width=\linewidth, height=0.2\textheight]{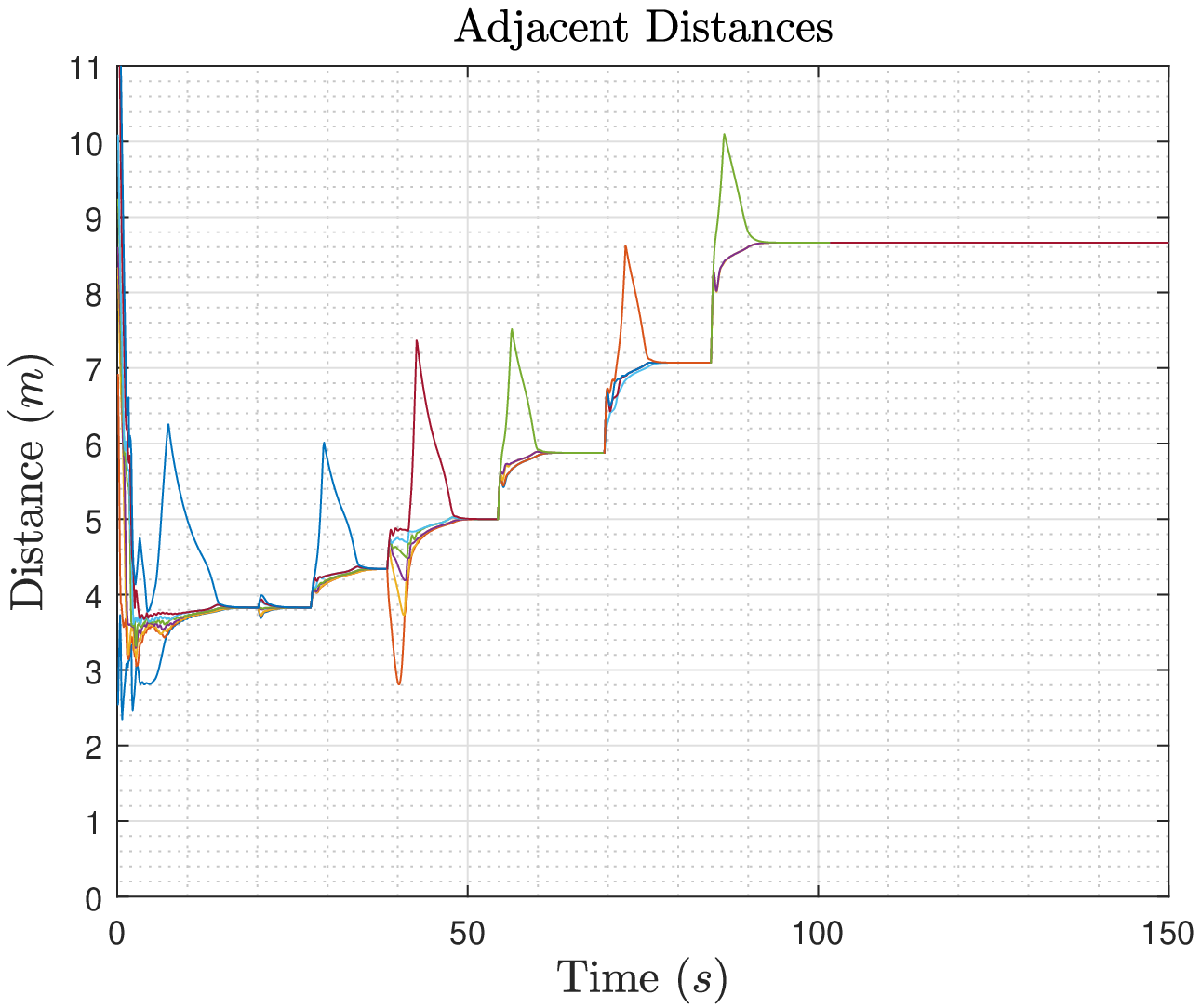}
						\caption{}
						\label{fig:flt8}
					\end{subfigure}\hfil % <-- added				
					\begin{subfigure}{0.375\textwidth}
						\includegraphics[width=\linewidth, height=0.2\textheight]{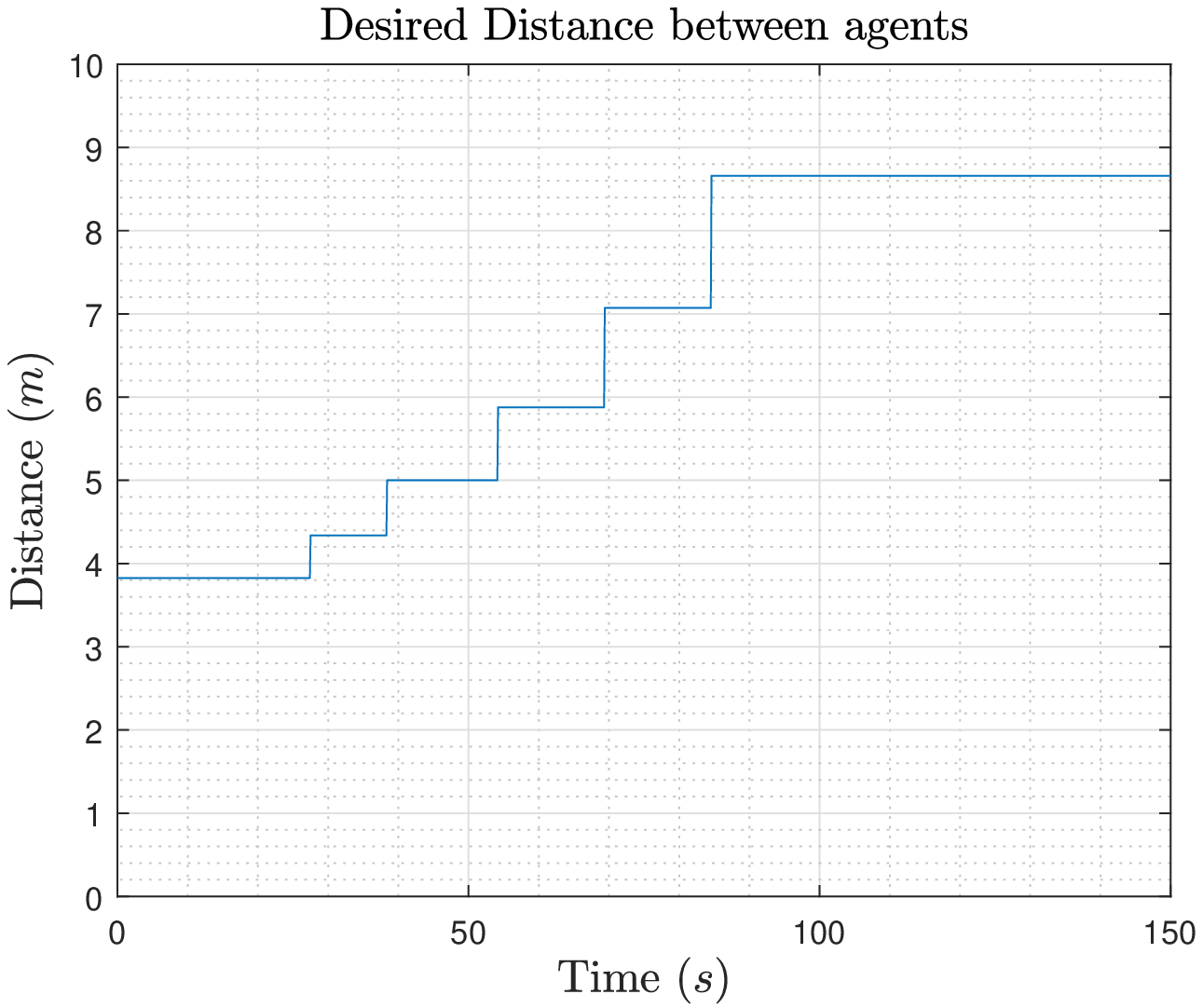}
						\caption{}%The desired distance and the distance between agents}
						\label{fig:flt9}
					\end{subfigure}\\
					\caption{Dynamic polygon formation with failed agents and moving leader for $N=8$ }
					\label{fig:fig7}
				\end{figure*}
				The following relation calculates the total number of distances between the agents:
				\begin{equation}\label{eq14}
					D = \frac{1}{2}N(N-1).
				\end{equation}	
				The agents must eventually be placed on a circle of the leader. Hence they are neighbors to only two adjacent agents.
				From the total number of inter-agent distances, the shortest possible distances between agents are selected based on the number of agents used in the simulation for the purpose of optimization. However, these distances may not be adjacent during the simulation. The fitness function used in this optimization is as follows:
				%\begin{equation}\label{eq1}
				%   F = \left\| X - X_d \right\|_f
				%\end{equation}
				\begin{equation}\label{eq15}
					\left\|J\right\|_F=\sqrt{\sum\limits_{i=1}^{N}\sum\limits_{j=1}^{m} \left((d_{i t_j}-d )^2 + (d_{L_{i t_j}}-d_L )^2 \right)},
				\end{equation}
				where $ \left\|.\right\|_F$ is the Frobenius norm,  $N$ is the number of agents, and $m$ is the final time.
				Also, $d_{i t_j}$ is the $ i$-th distance of two adjacent agents at time $t_j $, and  $d_{L_{i t_j}}$ is the $i$-th distance of the agent with the leader at time $t_j$.
				$d$ is the desired distance between agents, and $d_L$ is the desired distance between the agent and the leader.
				%\begin{equation*}
				%    X = \left[
				%\begin{array}{cccc:cccc}
				%d_{1 t_1} &  d_{1 t_2} & \cdots & d_{1 t_m} & d_{L_1 t_1} &  d_{L_1 t_2} & \cdots & d_{L1 t_m}\\ 
				%d_{2 t_1} &  d_{2 t_2} & \cdots & d_{2 t_m} & d_{L_2 t_1} &  d_{L_2 t_2} & \cdots & d_{L_2 t_m} \\
				%\vdots & \vdots & \ddots & \vdots & \vdots & \vdots & \ddots & \vdots \\ 
				%d_{N t_1} &  d_{N t_2} & \cdots & d_{N t_m} & d_{L_N t_1} &  d_{L_N t_2} & \cdots & d_{L_N t_m} \\
				%\end{array} \right] 
				%\end{equation*}
				
				%\begin{equation*}
				%    X_d = \left[
				%\begin{array}{cccc:cccc}
				%d &  d & \cdots & d & d_L &  d_L & \cdots & d_L\\ 
				%d &  d & \cdots & d & d_L &  d_L & \cdots & d_L \\
				%\vdots & \vdots & \ddots & \vdots & \vdots & \vdots & \ddots & \vdots \\ 
				%d &  d & \cdots & d & d_{L} &  d_{L} & \cdots & d_{L} \\
				%\end{array} \right] 
				%\end{equation*}
				Finally, (\ref{eq15}) is optimized by different algorithms with different scenarios, the obtained fitness function values are compared, and their minimum is selected.
				\section{Simulation }\label{section3}
					\begin{figure*}[ht]
					\centering % <-- added
						\begin{subfigure}{0.25\textwidth}
							\includegraphics[width=\linewidth, height=0.2\textheight]{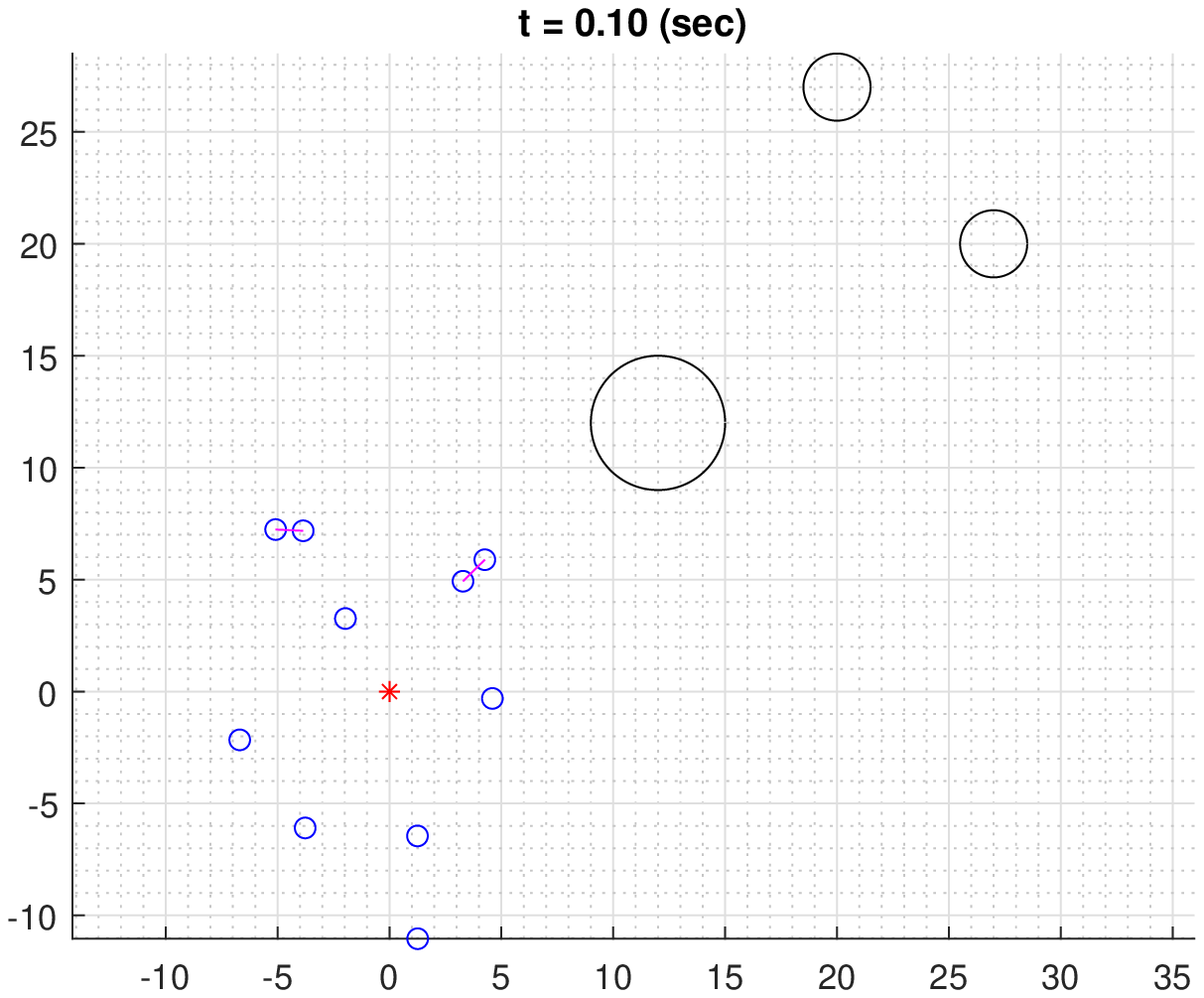}
							\label{fig:S1_obs1}
							\caption{}
						\end{subfigure}\hfil % <-- added
						\begin{subfigure}{0.25\textwidth}
							\includegraphics[width=\linewidth, height=0.2\textheight]{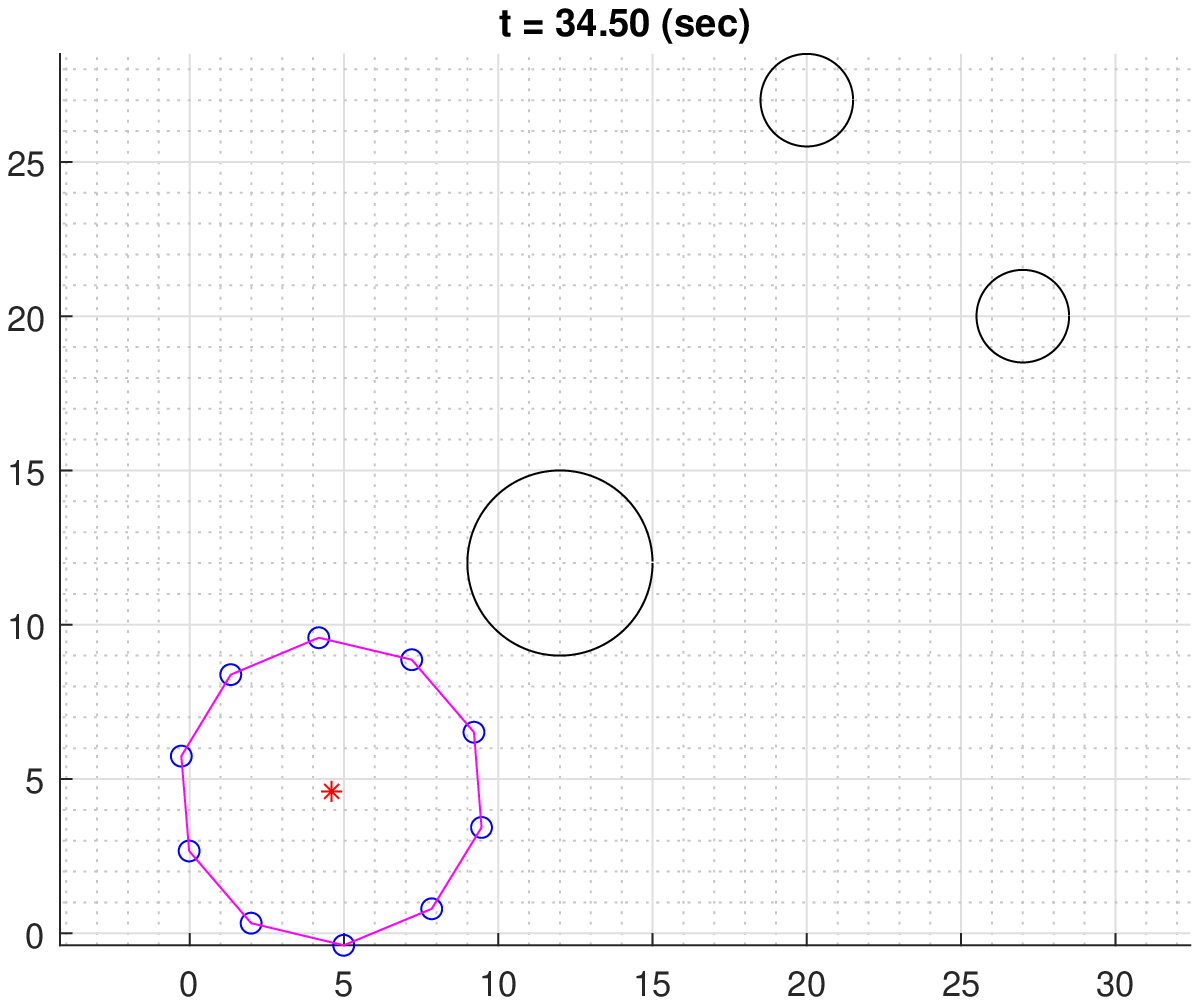}
							\label{fig:S1_obs2}
							\caption{}
						\end{subfigure}\hfil
						\begin{subfigure}{0.25\textwidth}
							\includegraphics[width=\linewidth, height=0.2\textheight]{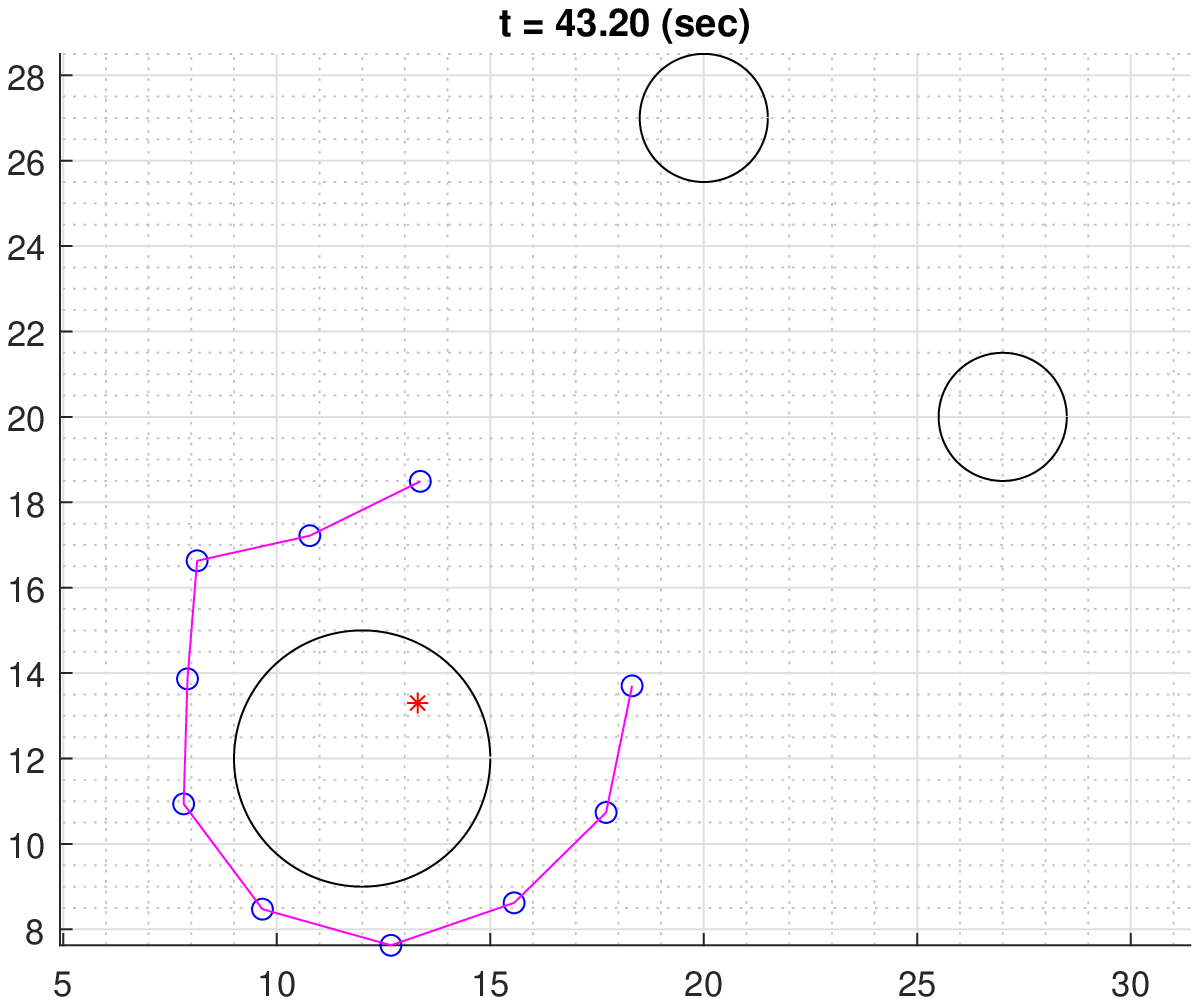}
							\label{fig:S1_obs3}
							\caption{}
						\end{subfigure}\hfil % <-- added
						\begin{subfigure}{0.25\textwidth}
							\includegraphics[width=\linewidth, height=0.2\textheight]{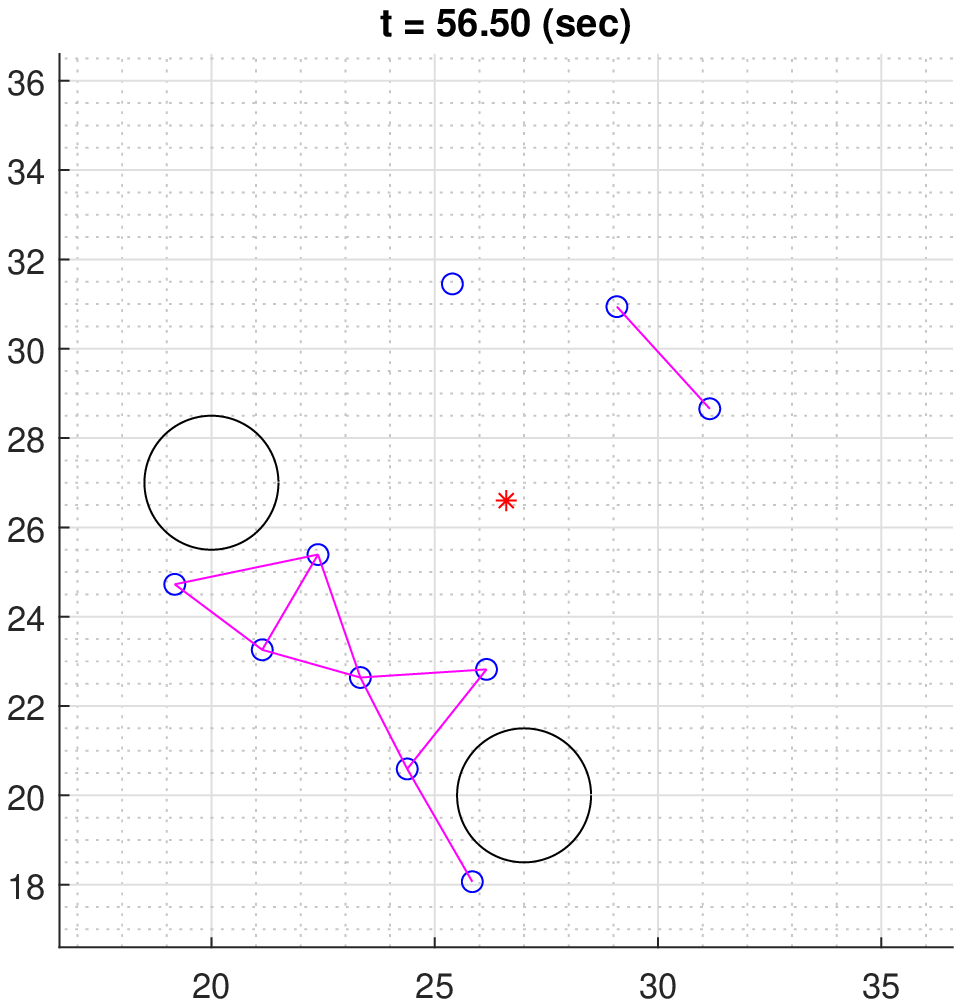}
							\label{fig:S1_obs4}
							\caption{}
						\end{subfigure}\\ % <-- added
						\begin{subfigure}{0.25\textwidth}
							\includegraphics[width=\linewidth, height=0.2\textheight]{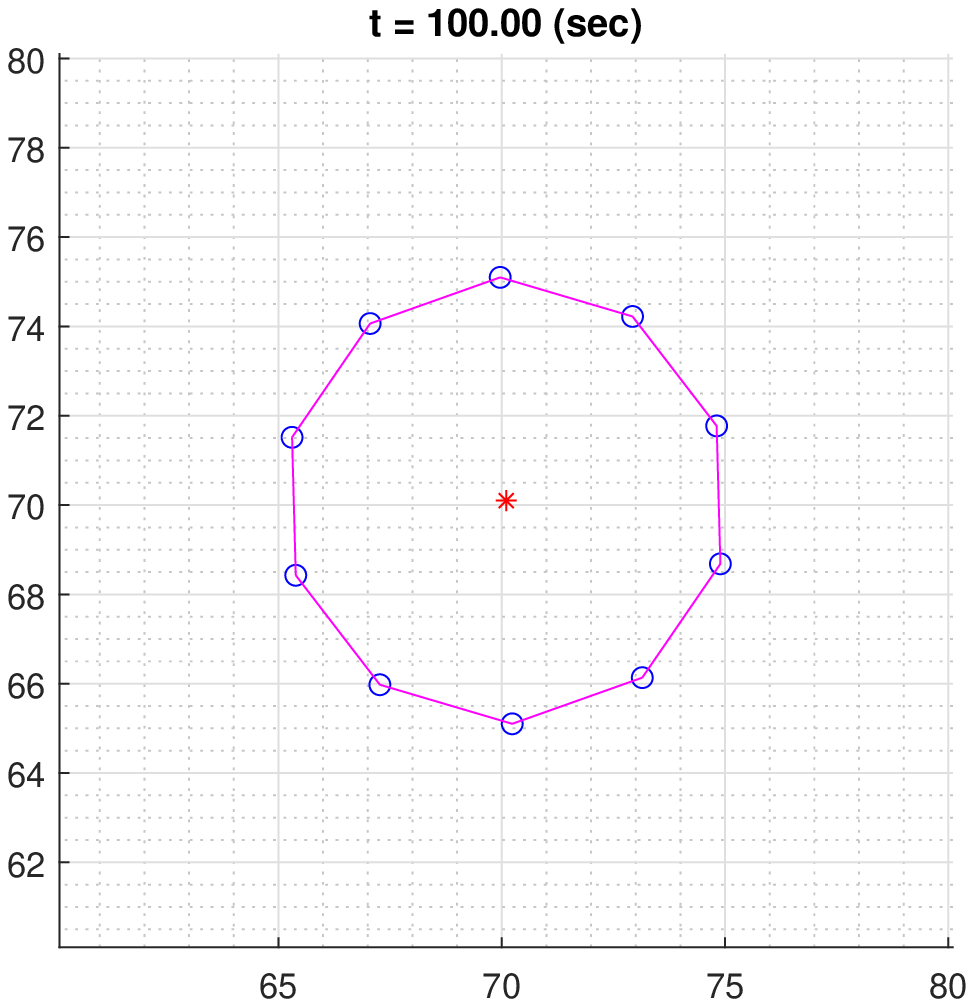}
							\caption{}
							\label{fig:S1_obs5}
						\end{subfigure}\hfil % <-- added
						\begin{subfigure}{0.375\textwidth}
							\includegraphics[width=\linewidth, height=0.2\textheight]{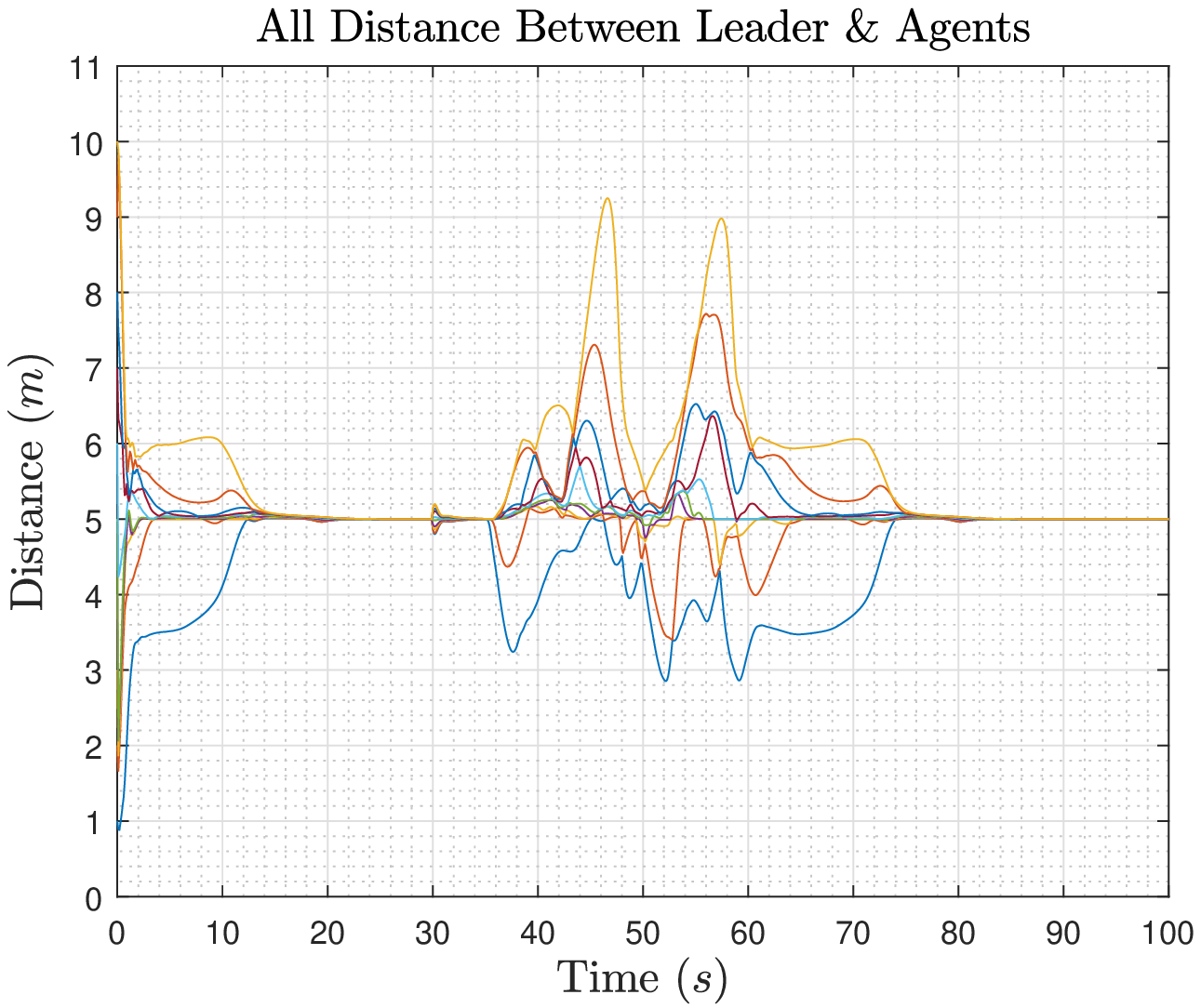}
							\caption{}
							\label{fig:S1_obs6}
						\end{subfigure}\hfil
						\begin{subfigure}{0.375\textwidth}
							\includegraphics[width=\linewidth, height=0.2\textheight]{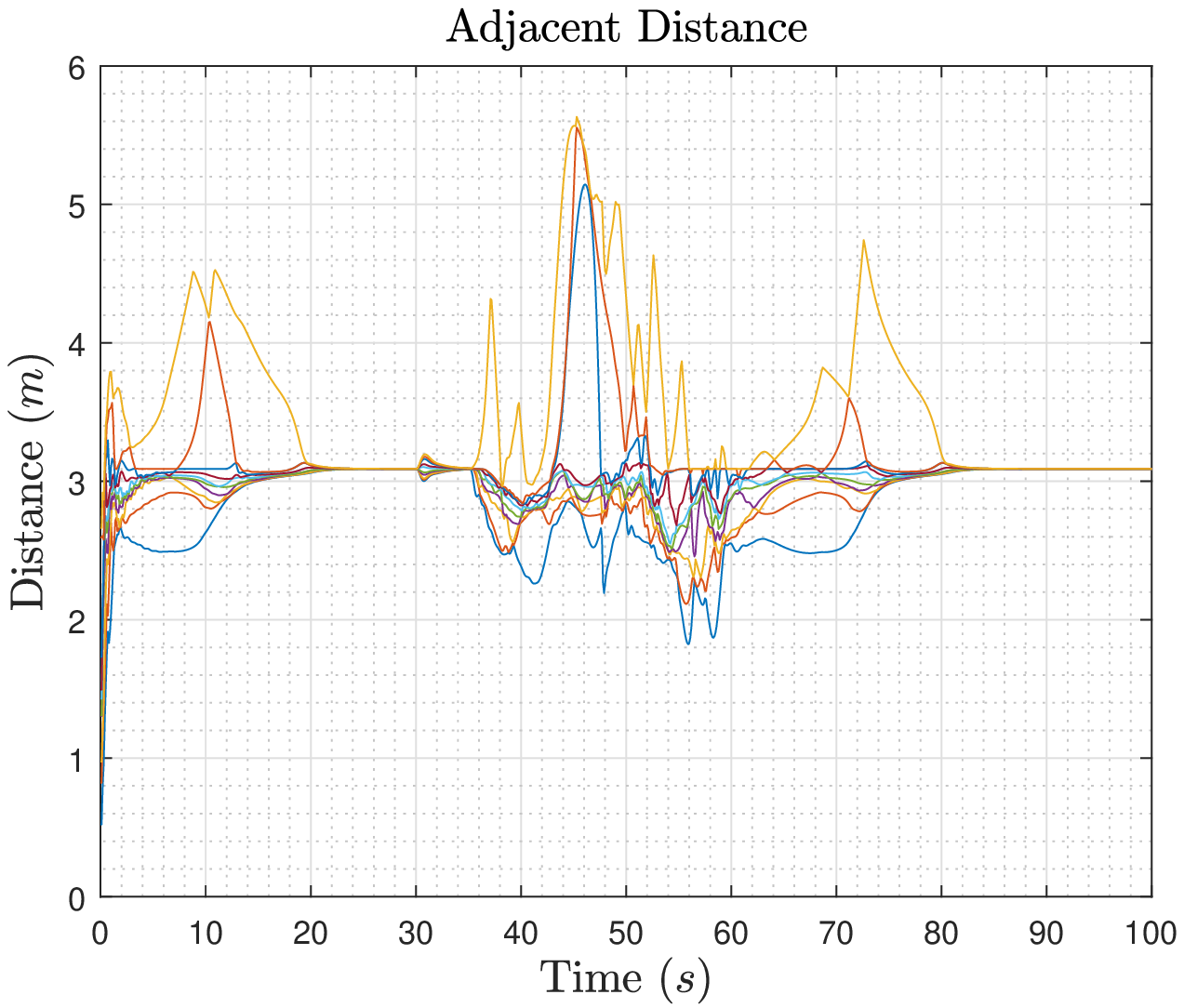}
							\caption{}
							\label{fig:S1_obs7}
						\end{subfigure}
						\caption{The first scenario of obstacle avoidance with the moving leader and $N=10$}
						\label{fig:fig8}
					\end{figure*}
				\begin{figure*}[ht]
					\centering % <-- added
					\begin{subfigure}{0.25\textwidth}
						\includegraphics[width=\linewidth, height=0.2\textheight]{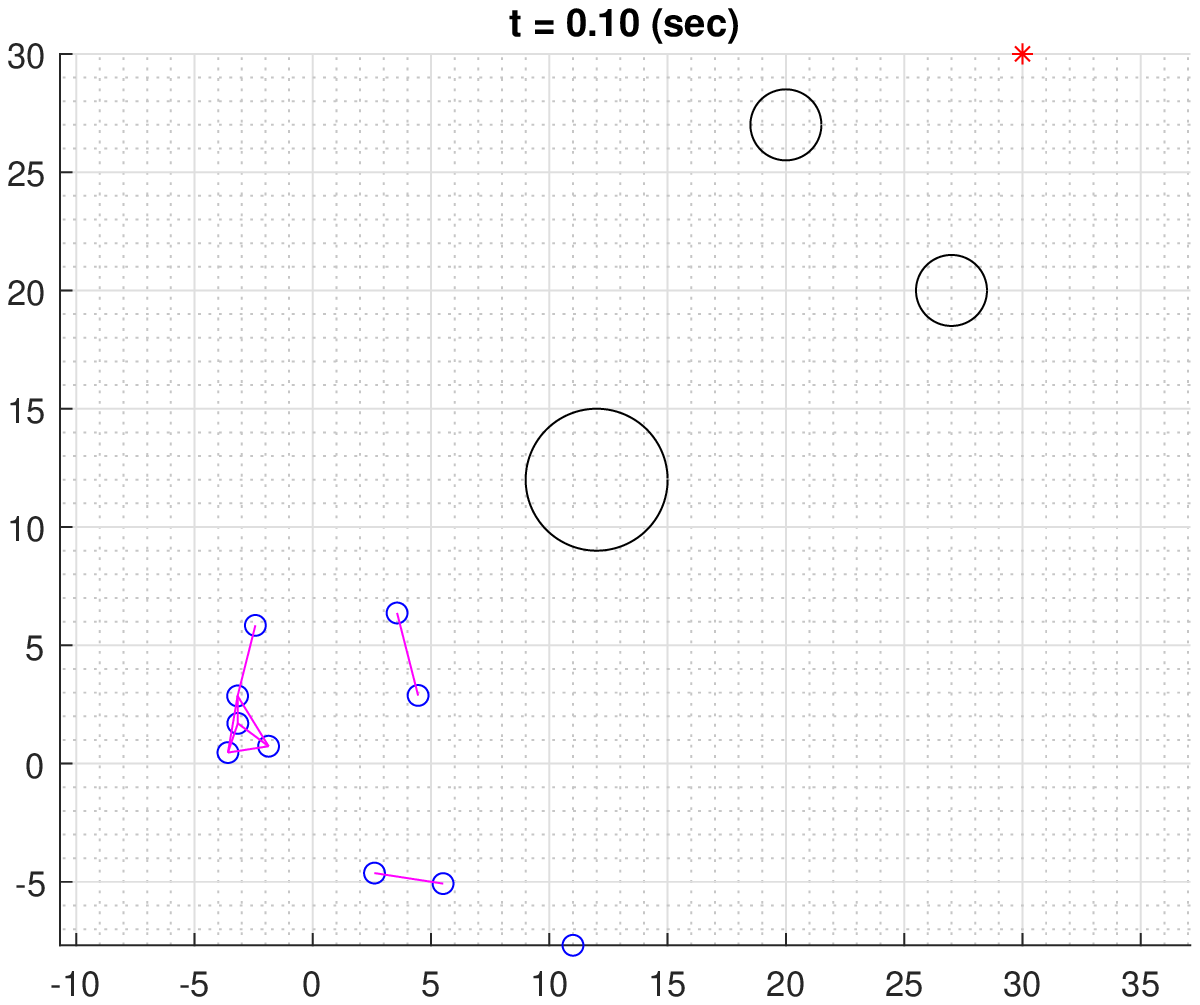}
						\label{fig:S2_obs1}
						\caption{}
					\end{subfigure}\hfil % <-- added
					\begin{subfigure}{0.25\textwidth}
						\includegraphics[width=\linewidth, height=0.2\textheight]{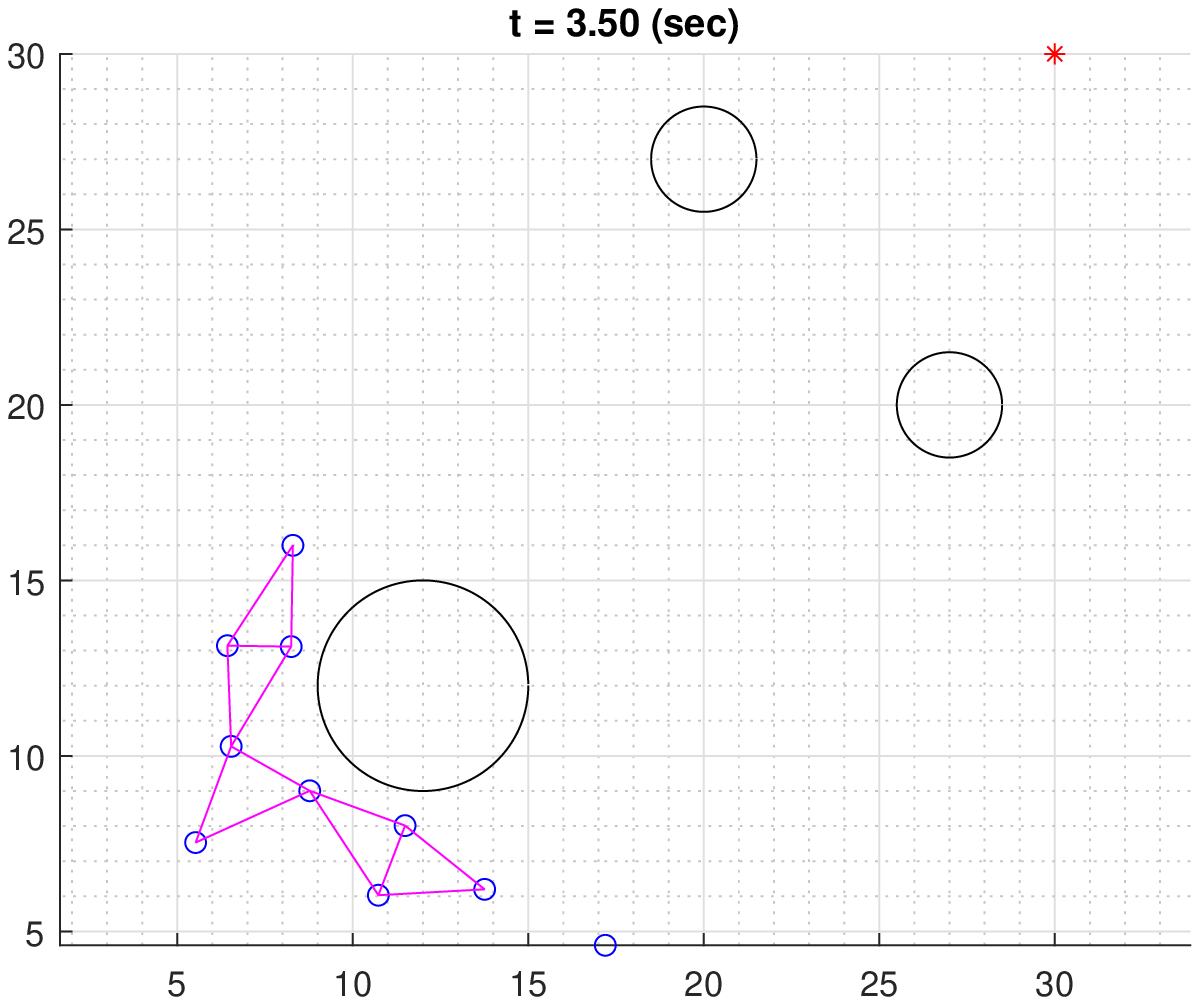}
						\label{fig:S2_obs2}
						\caption{}
					\end{subfigure}\hfil % <-- added
					\begin{subfigure}{0.25\textwidth}
						\includegraphics[width=\linewidth, height=0.2\textheight]{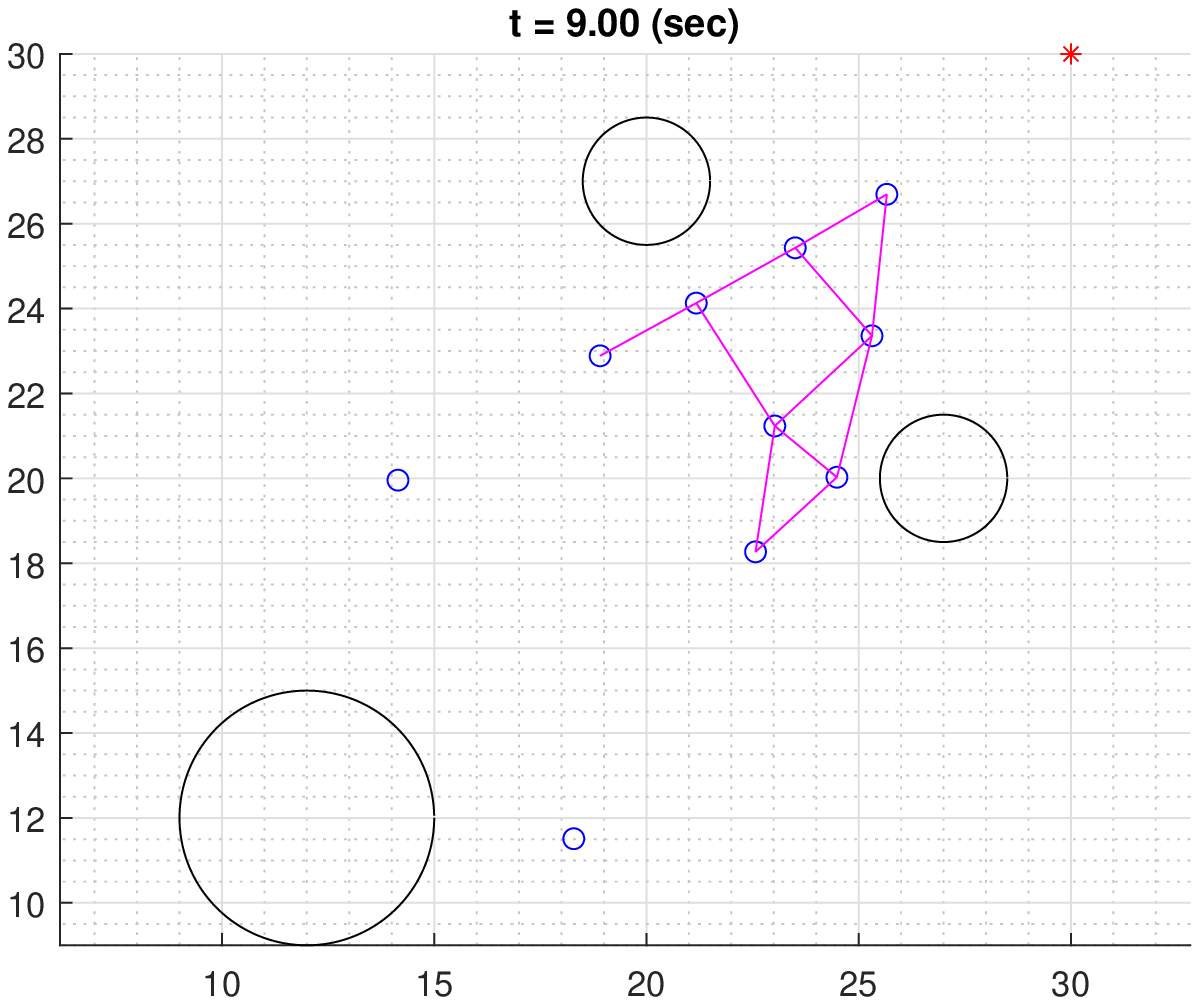}
						\label{fig:S2_obs3}
						\caption{}
					\end{subfigure}\hfil % <-- added
					\begin{subfigure}{0.25\textwidth}
						\includegraphics[width=\linewidth, height=0.2\textheight]{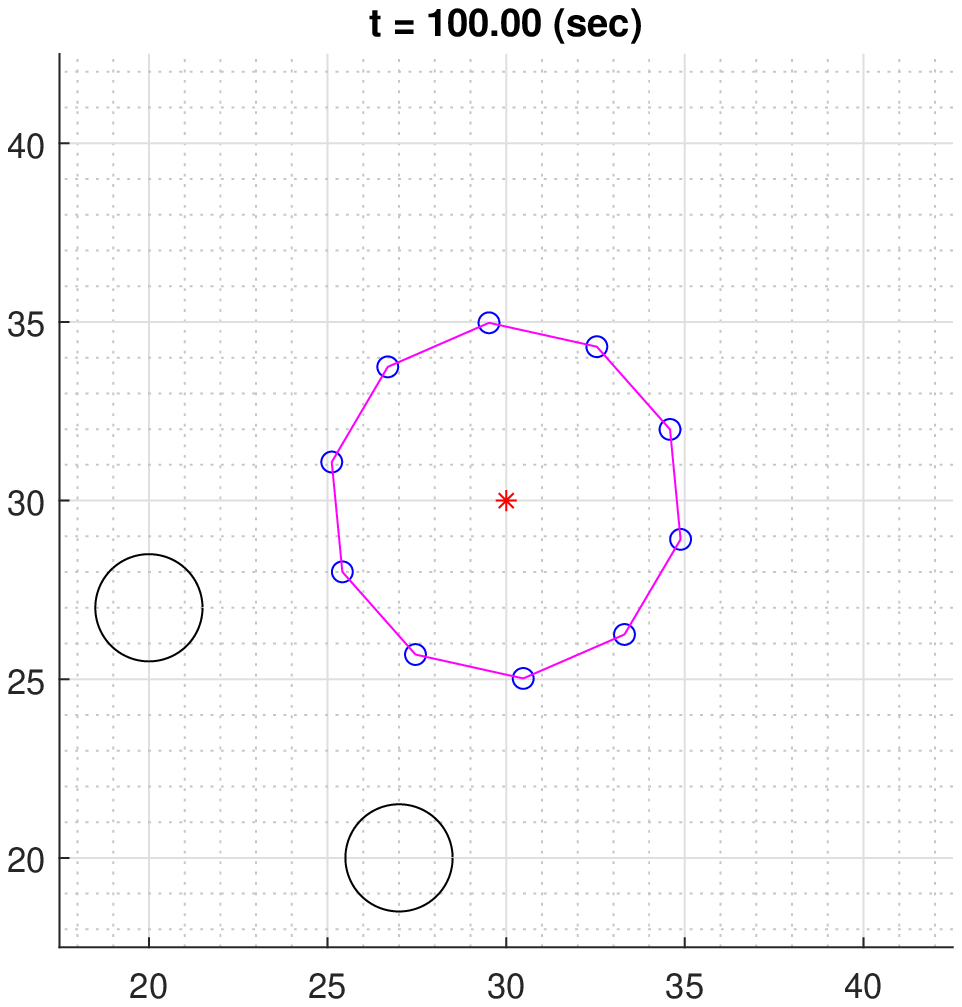}
						\label{fig:S2_obs4}
						\caption{}
					\end{subfigure}\\ % <-- added
					\begin{subfigure}{0.4\textwidth}
						\includegraphics[width=\linewidth, height=0.2\textheight]{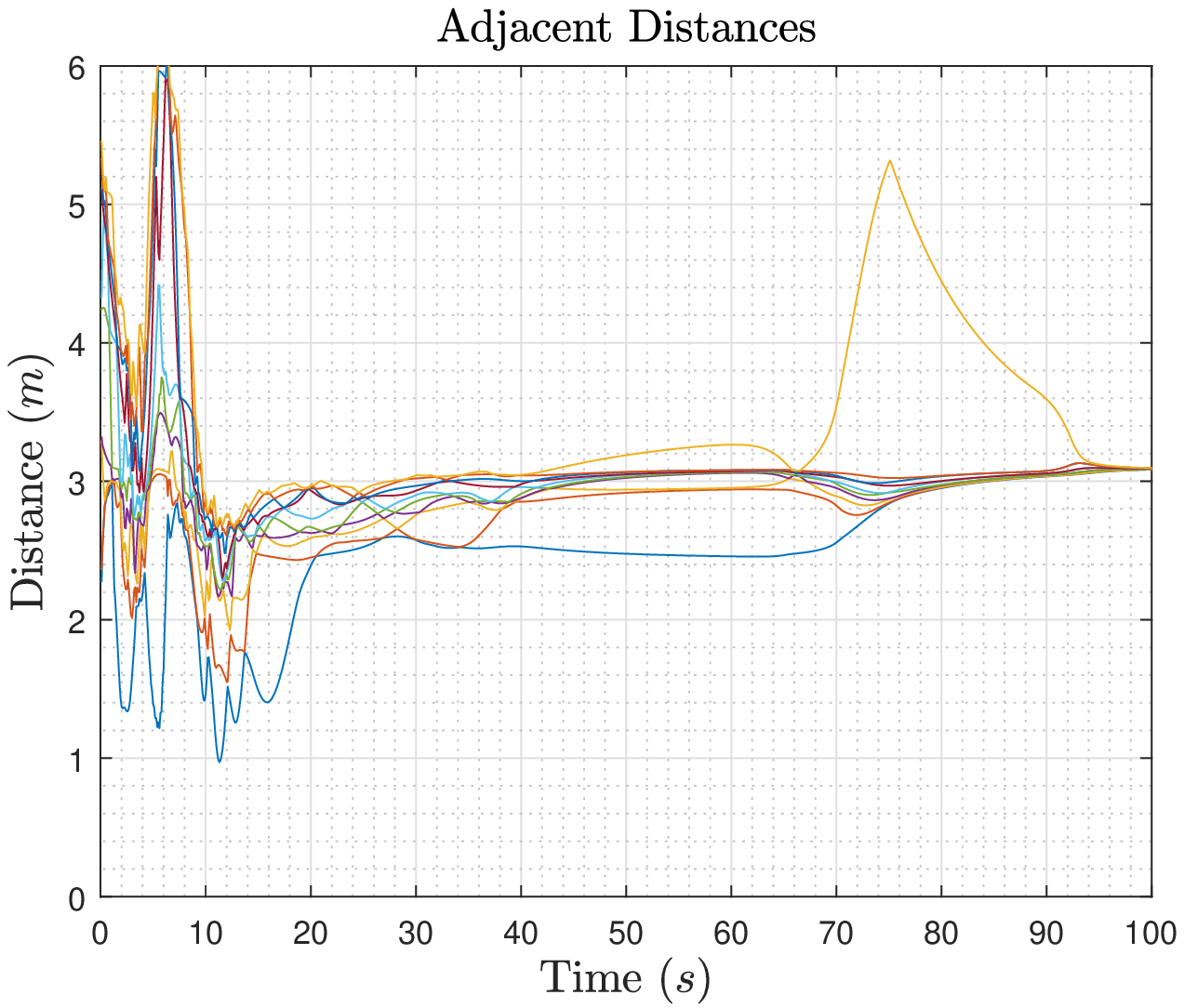}
						\label{fig:S2_obs5}
						\caption{}
					\end{subfigure}\hfil
					\begin{subfigure}{0.4\textwidth}
						\includegraphics[width=\linewidth, height=0.2\textheight]{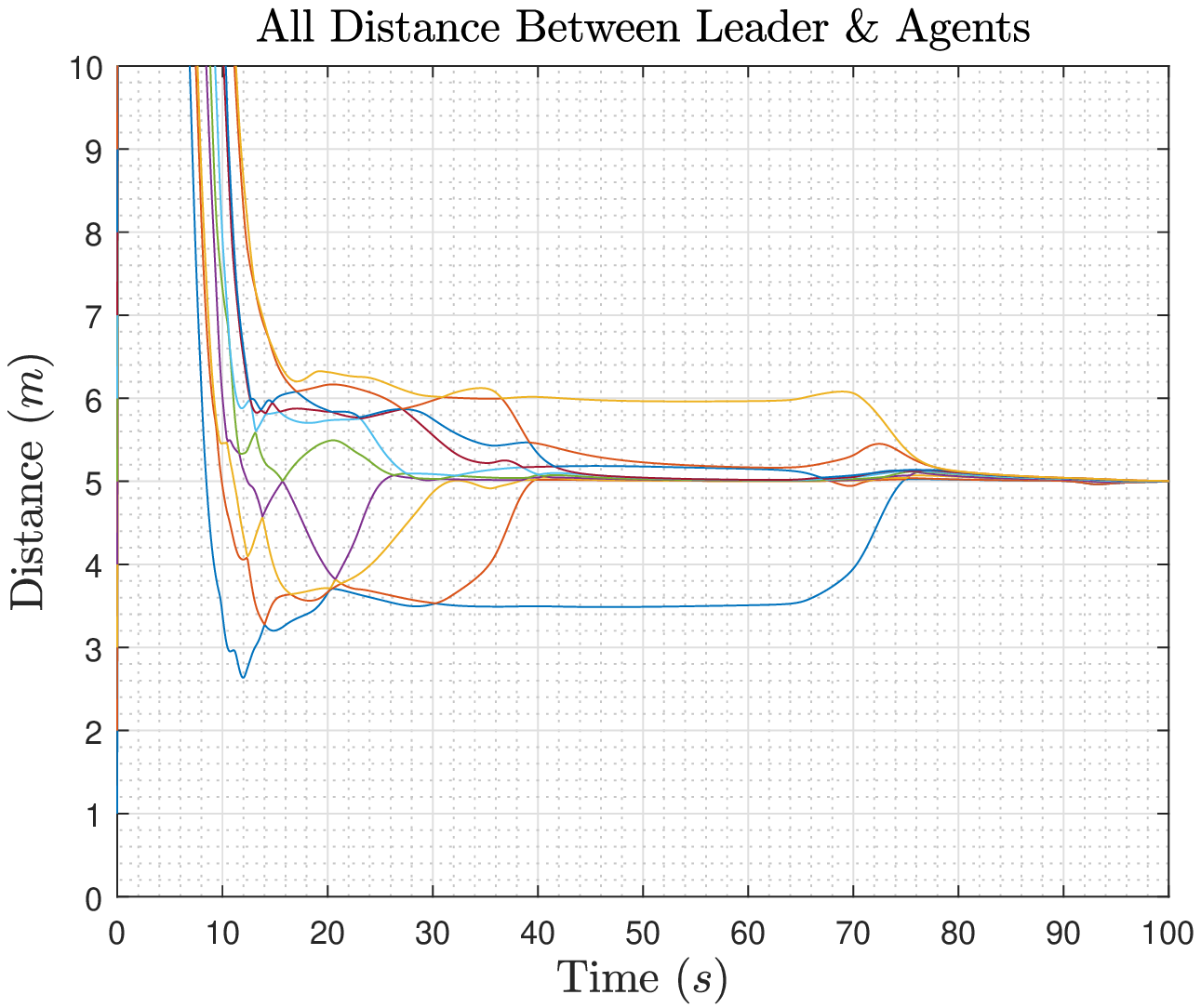}
						\label{fig:S2_obs6}
						\caption{}
					\end{subfigure}\hfil
					\caption{Obstacle avoidance with a stationary leader for $N=10$}
					\label{fig:fig9}
				\end{figure*}
				First, the minimum fitness value obtained from different scenarios of optimization algorithms is selected, then it is used to obtain the optimal parameters of the following sections.
				The sampling time during simulation 0.1 is considered, and the initial position of the virtual leader is (0,0).
				\subsection{Optimization}
				As mentioned in this section, single-circle formation without obstacles is considered. The initial conditions of the agents are considered as normally distributed random numbers with a mean of zero and a standard deviation of 10. 
				The initial settings for optimization are $d_L=5$, $Time=50 $, $Population =30$, $Iteration =500 $, $N=31$, and according to $N$, the distance between agents $d=1.0117$    is obtained from (\ref{eq8}). Four scenarios are considered to obtain optimal parameters. In the first scenario, 11 parameters are used for optimization, including $ c_1^{\alpha}, c_2^{\alpha}, c_1^{\gamma}, c_2^{\gamma}, a, b, a_L, b_L, e, h, \; \text{and} \; e_L$. In the second scenario, the same previous parameters except for $ a=5, b=5, a_L=3,  \; \text{and} \; b_L=3$ are used for optimization. In the third scenario, the same parameters as the first scenario except for $ e=0.1,\; h=0.2, \; \text{and} \; e_L=0.1$ are used for optimization. In the fourth  scenario, $ c_1^{\alpha}, c_2^{\alpha}, c_1^{\gamma}, \; \text{and} \; c_2^{\gamma}$ parameters are used for the optimization.
				% 	\begin{table}[ht]
					% 		\centering
					% 		\caption{Comparison of optimization algorithms}
					%      \begin{tabular}{|c|c|c|c|c|c|c|c|c|c|c|c|c|} \hline 
						% 	 Algorithm & \multicolumn{4}{c|}{GWO}  & \multicolumn{4}{|c|}{PSO}  & \multicolumn{4}{|c|}{GA} \\  
						% 	\hline
						% 	 Method & 1 & 2 & 3 & 4 & 1 & 2 & 3 & 4 & 1 & 2 & 3 & 4      \\\hline
						% % 	 Population & \multicolumn{3}{c|}{} &  \multicolumn{3}{|c|}{}  & \multicolumn{3}{|c|}{} \\\hline
						% % 	 Iteration  &  \multicolumn{3}{c|}{} &  \multicolumn{3}{|c|}{}  & \multicolumn{3}{|c|}{}    \\\hline
						%       \# parameters    & 11 & 8 & 7 & 4 & 11 & 7 & 8 & 4 & 11 & 7 & 8 & 4 \\\hline
						% 	 Fitness    & 2.4241 &  & 24.7733 & 1.61788 & 169.331 & &  &  & 59.88 & & & \\\hline
						% \end{tabular}
					% 	\end{table}
				\begin{table}[ht]
					\centering
					\caption{Comparison of optimization algorithms}
					\begin{tabular}{|c|c|c|c|} \hline 
						\label{Table1}
						& Scenario & \# of parameters  & Cost\\  
						\hline
						\multirow{4}{*}{GWO} & 1 & 11 & 2.4241     \\
						& 2 & 8 & 9.1389            \\
						& 3 & 7 & 6.3343     \\
						& 4 & 4 & 1.61788     \\\hline
						\multirow{4}{*}{PSO} & 1 & 11 & 20.5801 \\
						& 2 & 8 & 10.1402\\
						& 3 & 7 & 3.83659 \\
						& 4 & 4 & 4.86521\\\hline
						\multirow{4}{*}{GA}  & 1 & 11 & 34.8627\\
						& 2 & 8 & 33.7109\\
						& 3 & 7 & 26.4257\\
						& 4 & 4 & 25.6392\\\hline						
					\end{tabular}
				\end{table}
				\\
				According to table \ref{Table1}, the lowest cost function obtained is related to the GWO algorithm, scenario 4.
				
				\subsection{Dynamic polygon formation}
				This section shows that polygon formation will still be formed if some agents fail. To simulate this section, eight agents are considered, and it is assumed that five failures occur for the agents during the simulation. The leader is stationary in the beginning of the simulation until the agents around the leader form the circle formation, and then after 20 seconds, the leader starts moving. The parameters of  (\ref{eq4}) are $c_1^{\alpha}=8.1$, $c_2^{\alpha}=2.3$, $c_1^{\gamma}=6.5$, and $c_2^{\gamma}=8$.							
				 As seen in Fig. \ref{fig:fig7}, the circular formation with eight agents in the beginning is formed, and five agents randomly fail during the formation.
				 At the time of failure,  (\ref{eq8}) is updated (Fig. \ref{fig:flt9}), and after a short time, the distance between agents reaches the desired distance (Fig. \ref{fig:flt8}), and the polygon formation is formed. 
				\subsection{Obstacle avoidance}
				In this section, obstacle avoidance with and without size scaling is simulated and shown in Fig. \ref{fig:fig8}.
				To demonstrate obstacle avoidance, two scenarios are considered. Firstly, the circular formation starts moving with the virtual leader and passing through obstacles. Secondly, the agents reach the leader after passing through obstacles.
				For simulating this section, $N=10$, $d_L=5$, and according to  (\ref{eq8}), $d=3.0902$ is obtained. The initial condition is normally distributed random numbers with a mean of zero and a standard deviation of 5. 
				The radius of the first obstacle is $3$, and the rest is $1.5$. In the first scenario, in the beginning, the leader remains stationary until the polygon formation is formed, then the leader starts moving.
				Fig. \ref{fig:fig8} and Fig. \ref{fig:fig9} show the first and second scenarios, respectively. As seen in Fig. \ref{fig:fig8} and Fig. \ref{fig:fig9}, obstacle avoidance is demonstrated. Also, in Fig. \ref{fig:S1_obs7}, the smallest distance between agents is 2, so they do not collide with each other.			
				The circular formation is disarranged against an obstacle with a radius equal to or greater than it in the first scenario. hence, size scaling is used to change the radius of the formation to better passing through obstacles. The radius of the first obstacle is equal to the radius of the circular formation, i.e., it is considered to be 5. The number of agents is regarded as 12 because it can be divided by 4, and according to  (\ref{eq8}), $d=2.5882$ is obtained.
				\begin{figure*}[ht]
					\centering % <-- added
					\begin{subfigure}{0.25\textwidth}
						\includegraphics[width=\linewidth, height=0.2\textheight]{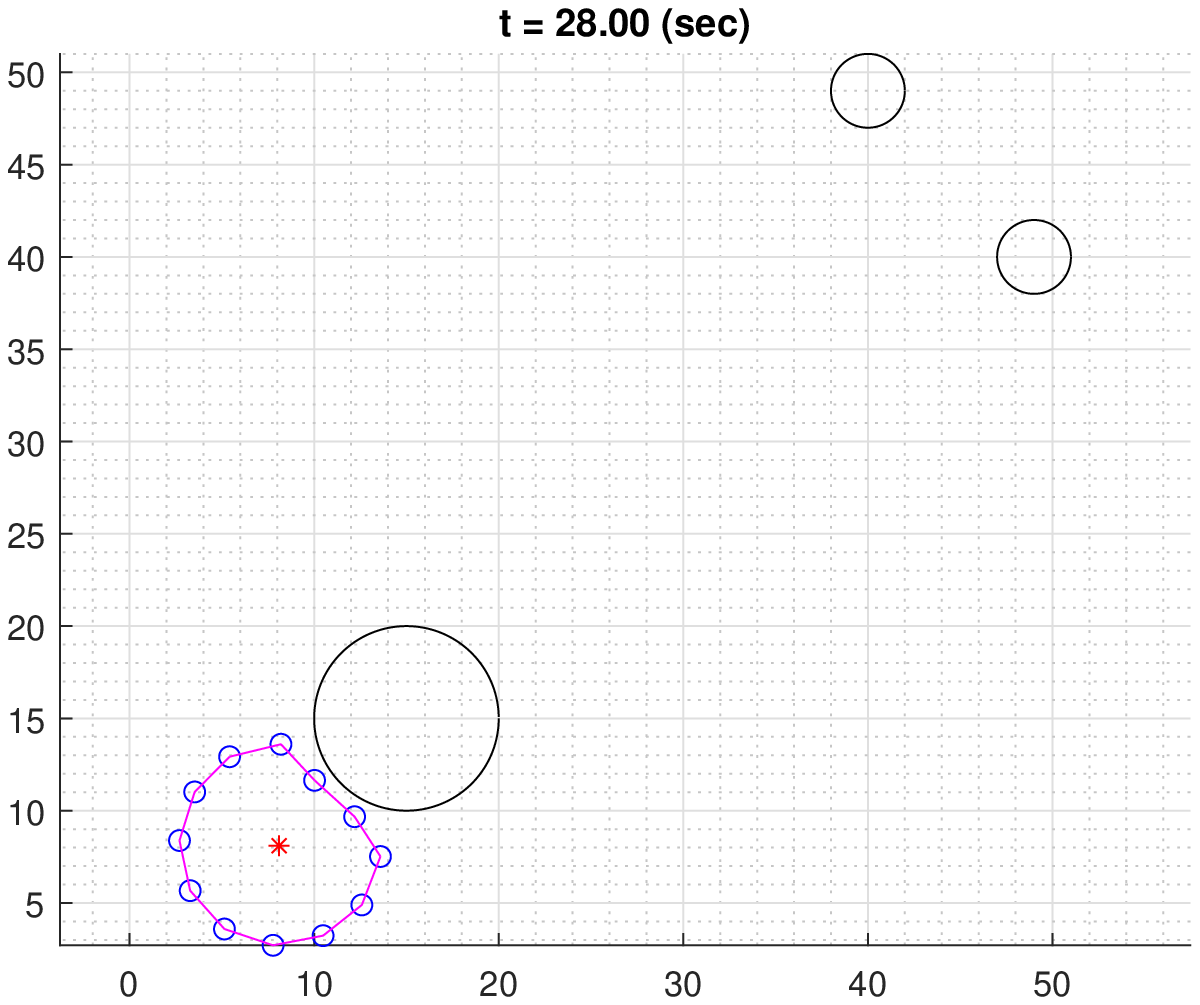}
						\caption{}
						\label{fig:scl1}
					\end{subfigure}\hfil % <-- added
					\begin{subfigure}{0.25\textwidth}
						\includegraphics[width=\linewidth, height=0.2\textheight]{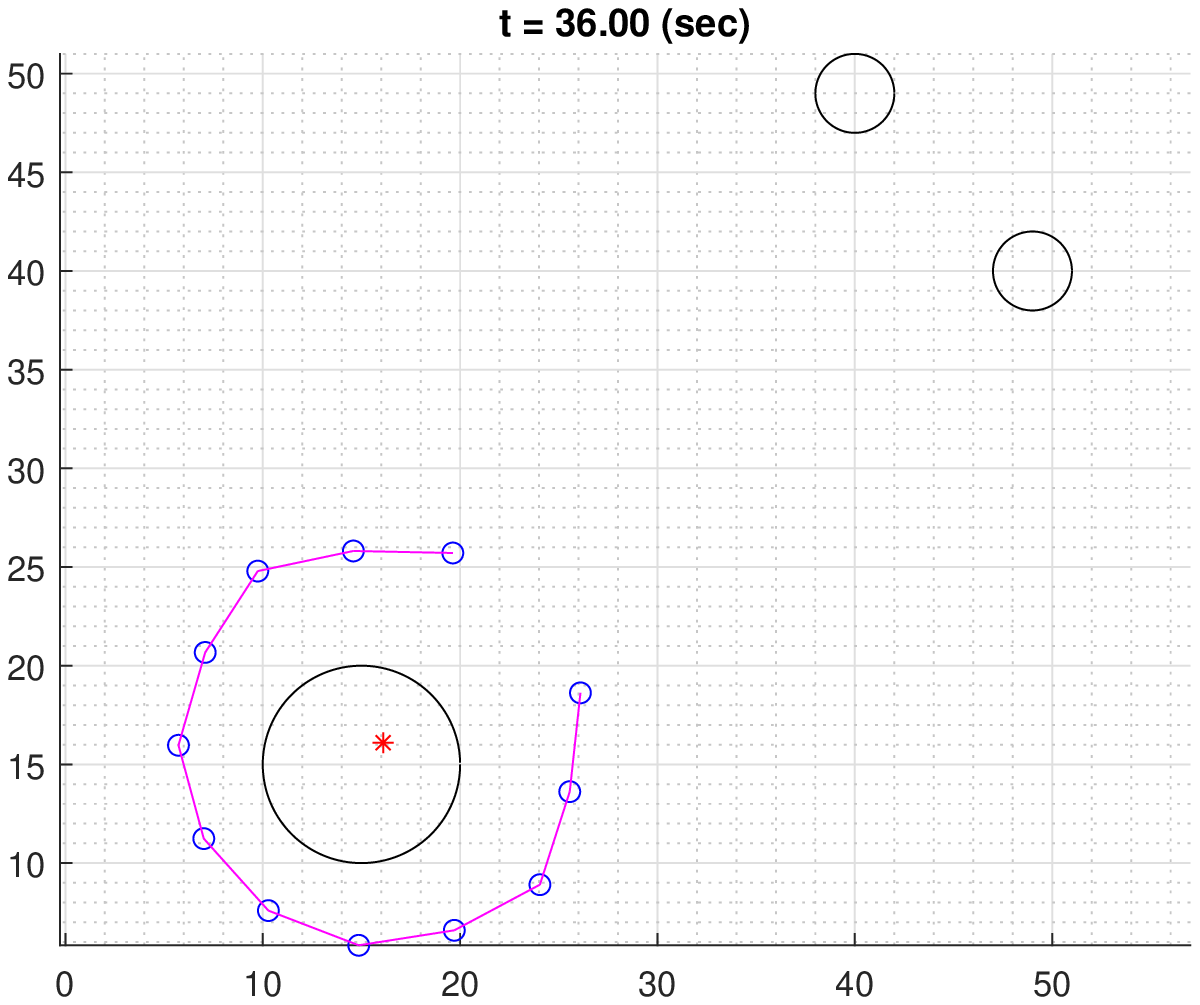}
						\caption{}
						\label{fig:scl2}
					\end{subfigure}\hfil % <-- added
					\begin{subfigure}{0.25\textwidth}
						\includegraphics[width=\linewidth, height=0.2\textheight]{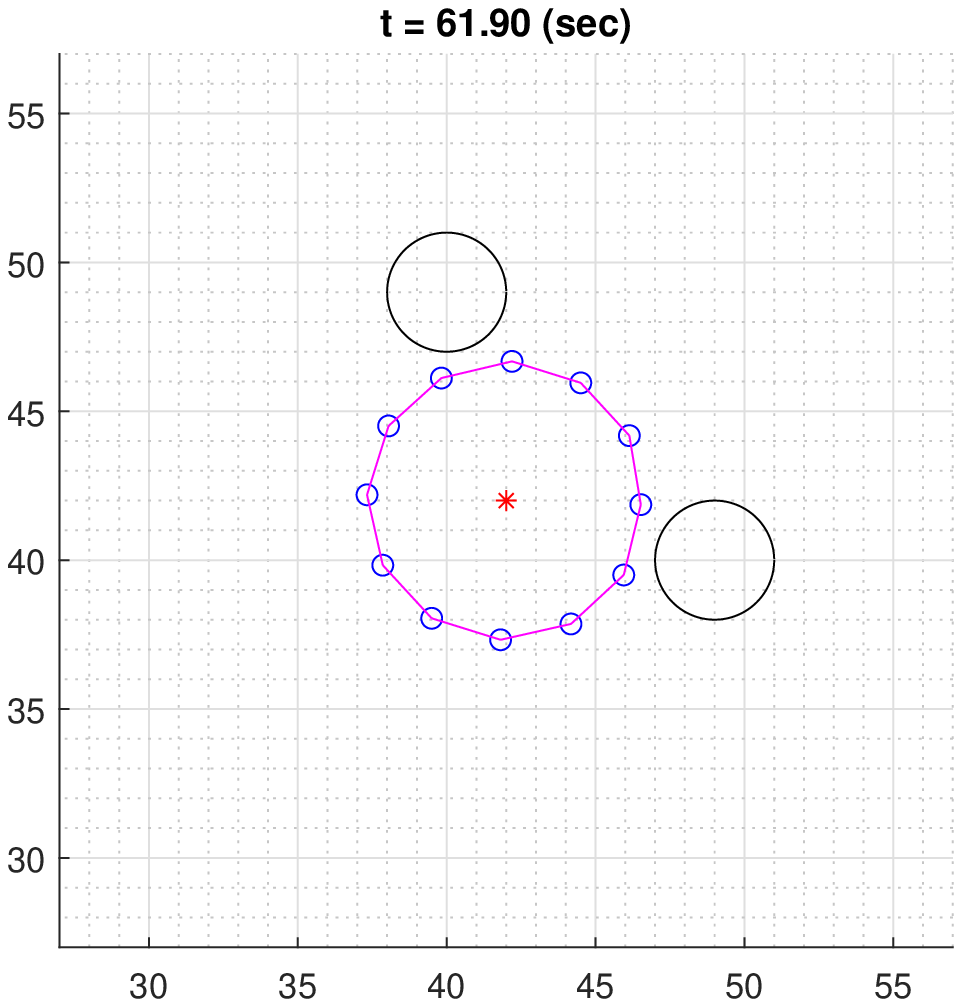}
						\caption{}
						\label{fig:scl3}
					\end{subfigure}\hfil 
					\begin{subfigure}{0.25\textwidth}
						\includegraphics[width=\linewidth, height=0.2\textheight]{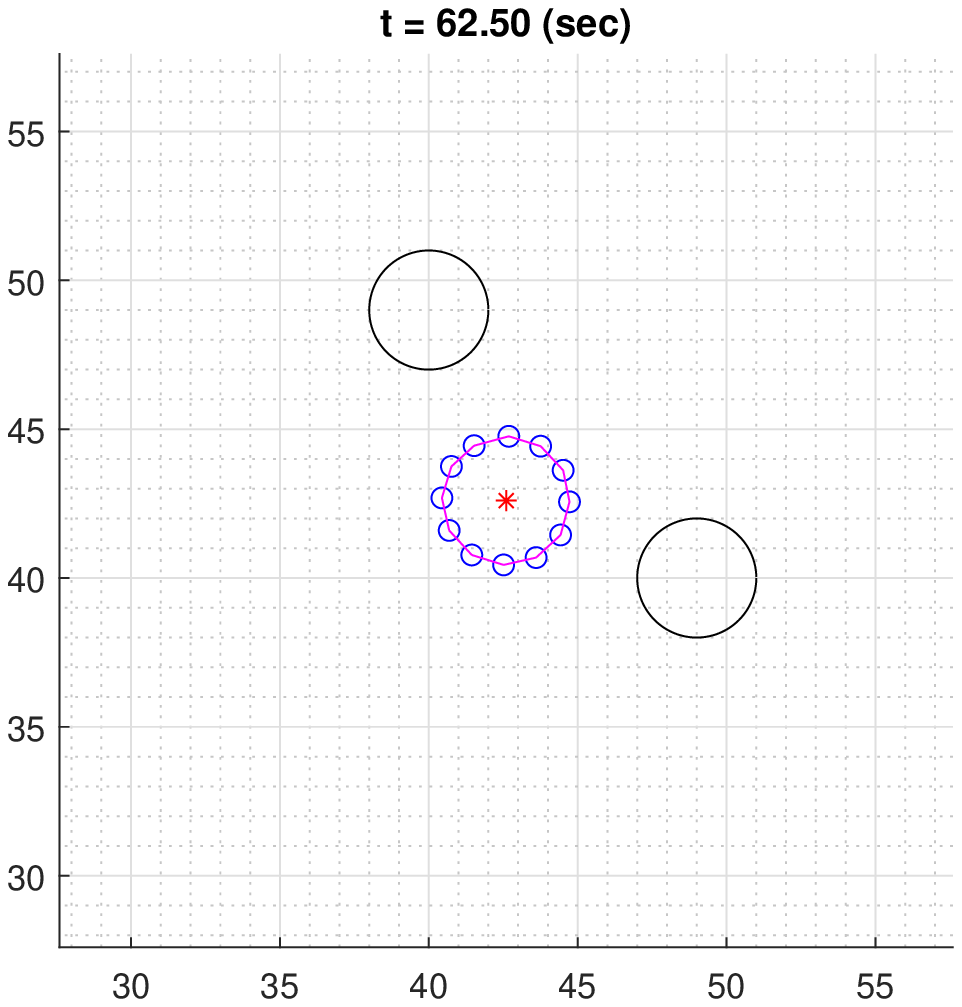}
						\caption{}
						\label{fig:scl4}
					\end{subfigure}\\
					\begin{subfigure}{0.25\textwidth}
						\includegraphics[width=\linewidth, height=0.2\textheight]{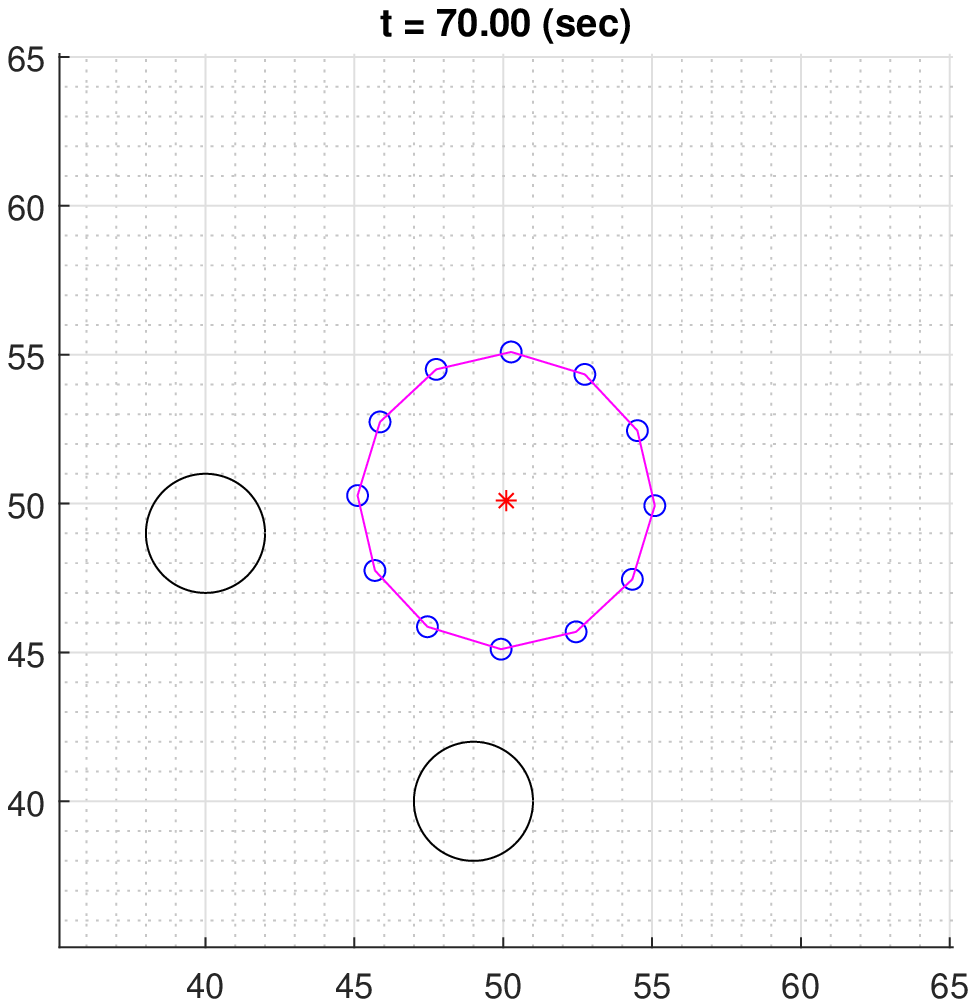}
						\caption{}
						\label{fig:scl5}
					\end{subfigure}\hfil % <-- added
					\begin{subfigure}{0.375\textwidth}
						\includegraphics[width=\linewidth, height=0.2\textheight]{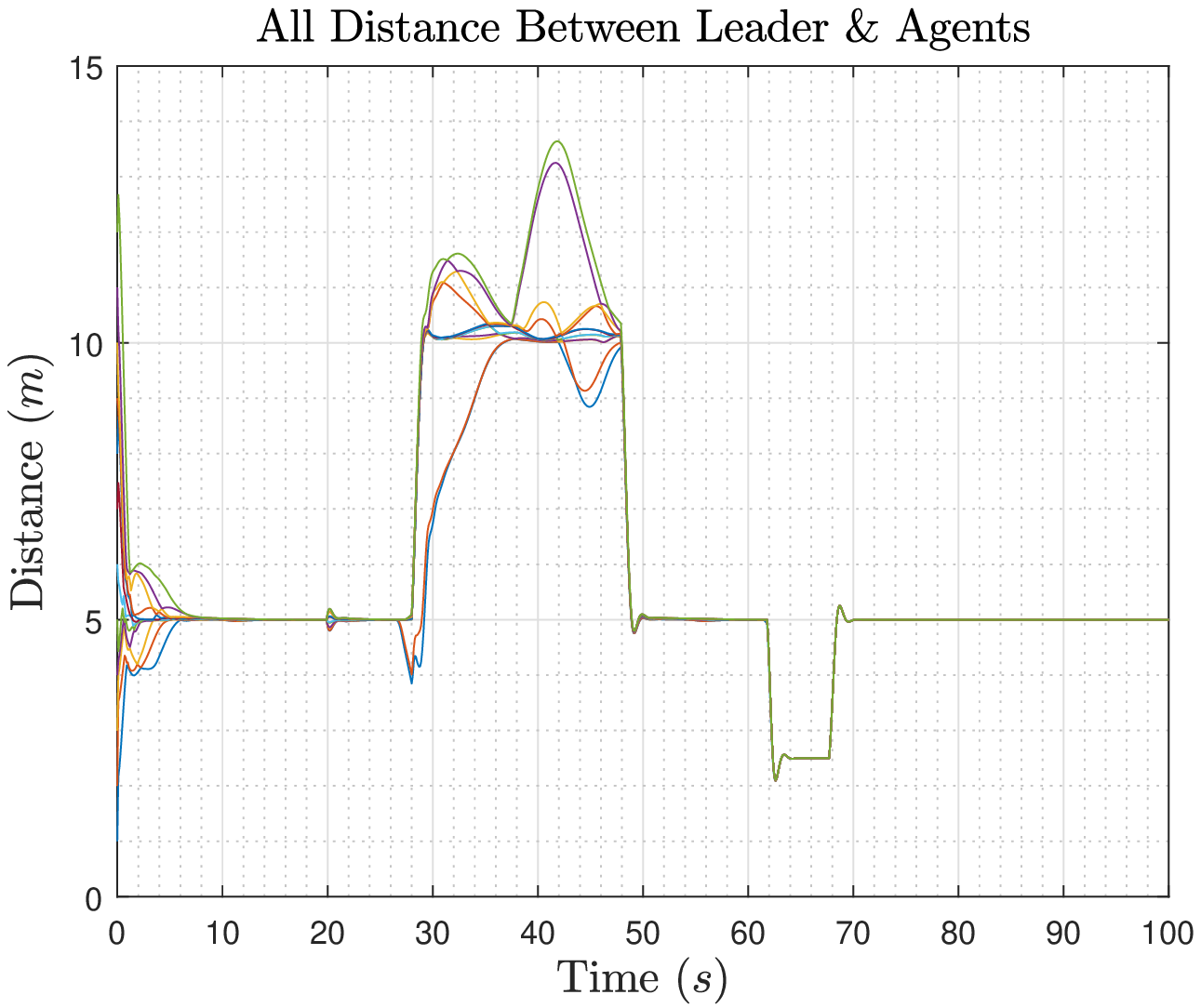}
						\caption{}
						\label{fig:scl6}
					\end{subfigure}\hfil % <-- added
					\begin{subfigure}{0.375\textwidth}
						\includegraphics[width=\linewidth, height=0.2\textheight]{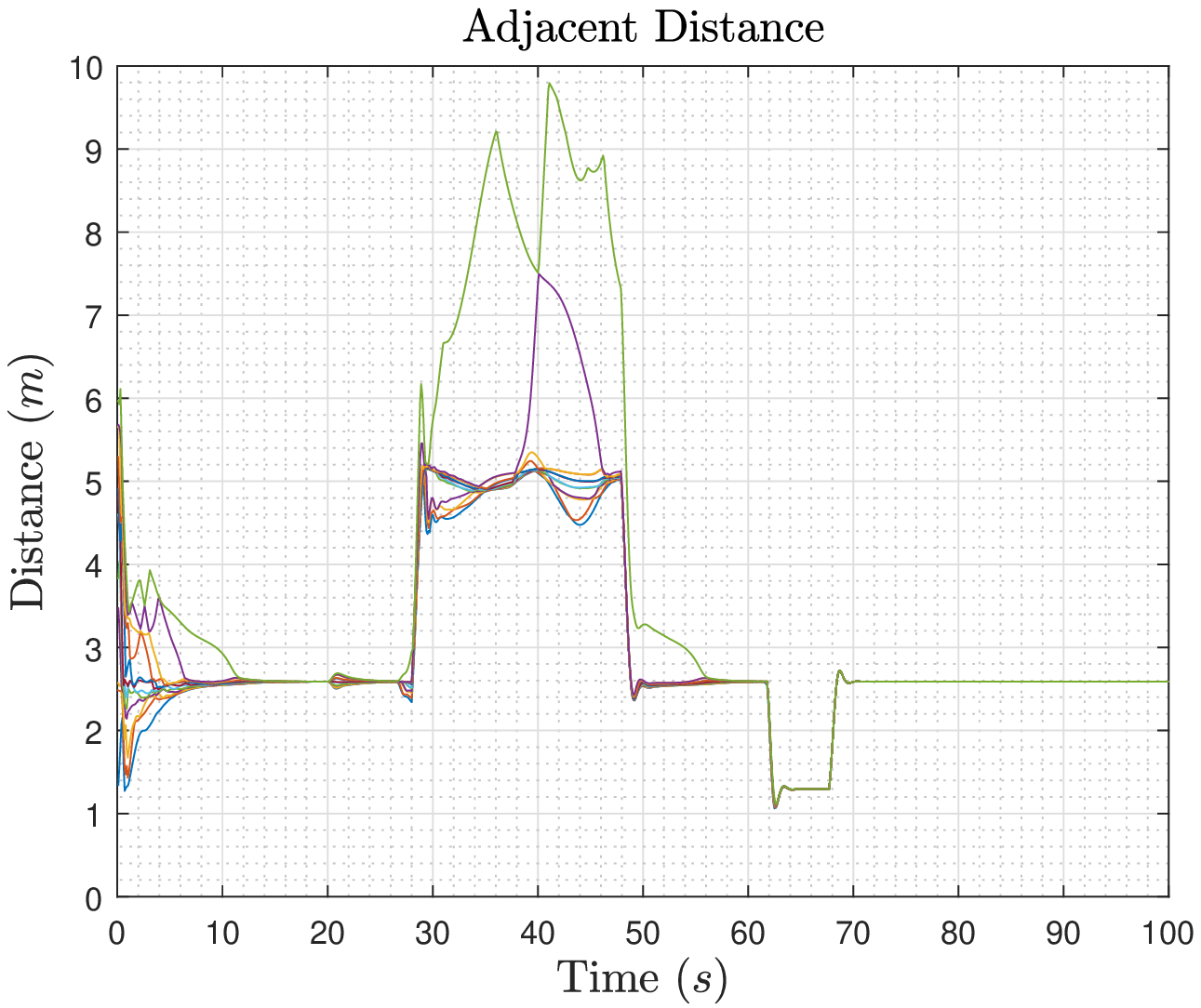}
						\caption{}
						\label{fig:scl7}
					\end{subfigure}%\hfil % <-- added
					\caption{Size scaling for $N=12$ }
					\label{fig:fig10}
				\end{figure*}
				As seen in Fig. \ref{fig:fig10}, the radius of the formation is doubled after the three agents are placed in the interaction range of the obstacle that they are also in the neighborhood of each other. After 20 seconds and passing through the obstacle, the radius of the formation returns to the first state. When agents reach side obstacles, the formation radius is halved. This is because the agents in the interaction range of the obstacles are not in the neighborhood of each other. After several time steps and passing through the obstacle, the radius returns to the first state. The parameters of  (\ref{eq4}) are as table \ref{Table2}.
				\begin{table}[ht]
					\centering
					\caption{Algorithm parameters}
					\begin{tabular}{|c|c|c|c|c|c|c|c|} \hline 
						\label{Table2}
						& $c_1^{\alpha}$ & $c_2^{\alpha}$ & $c_1^{\beta}$ & $c_2^{\beta}$ & $c_1^{\gamma}$ & $c_2^{\gamma}$ & $N$\\  
						\hline
						First scenario & 6.6 & 2.4 & 15 & 7 & 4.3 & 11.2 & 10  \\\hline
						Second scenario & 6.1 & 2.8 & 15 & 7 &4.9 & 12.7 & 10 \\\hline
						Size scaling & 5.9 & 2.3 & 15 & 7 & 5.2 & 13.3 & 12    \\\hline
					\end{tabular}
				\end{table} 
				\subsection{Multi-circles}
				Four circle formation is demonstrated in this section. The distance between agents is determined according to the number of agents in the first circle. The number of agents for three different types of formations in the first circle is 3, 5, and 6, which form triangles, pentagons, and hexagons, respectively.  The number of agents and the radius in the first circle determine the distance between agents.
				\begin{figure*}[ht]
					\centering % <-- added
					\begin{subfigure}{0.33\textwidth}
						\includegraphics[width=\linewidth, height=0.2\textheight]{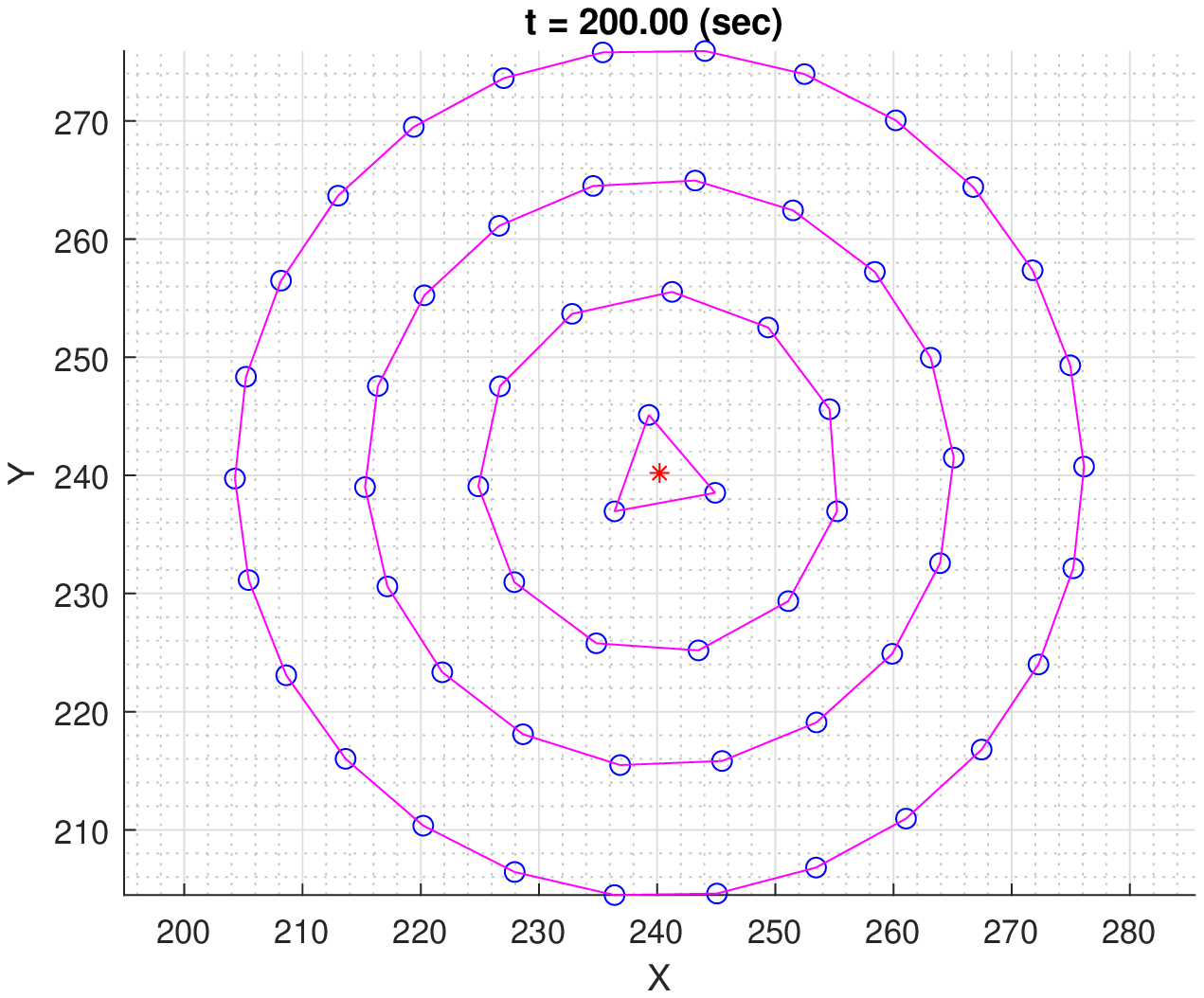}
						\caption{Triangle}
						\label{fig:mlt1}
					\end{subfigure}\hfil % <-- added
					\begin{subfigure}{0.33\textwidth}
						\includegraphics[width=\linewidth, height=0.2\textheight]{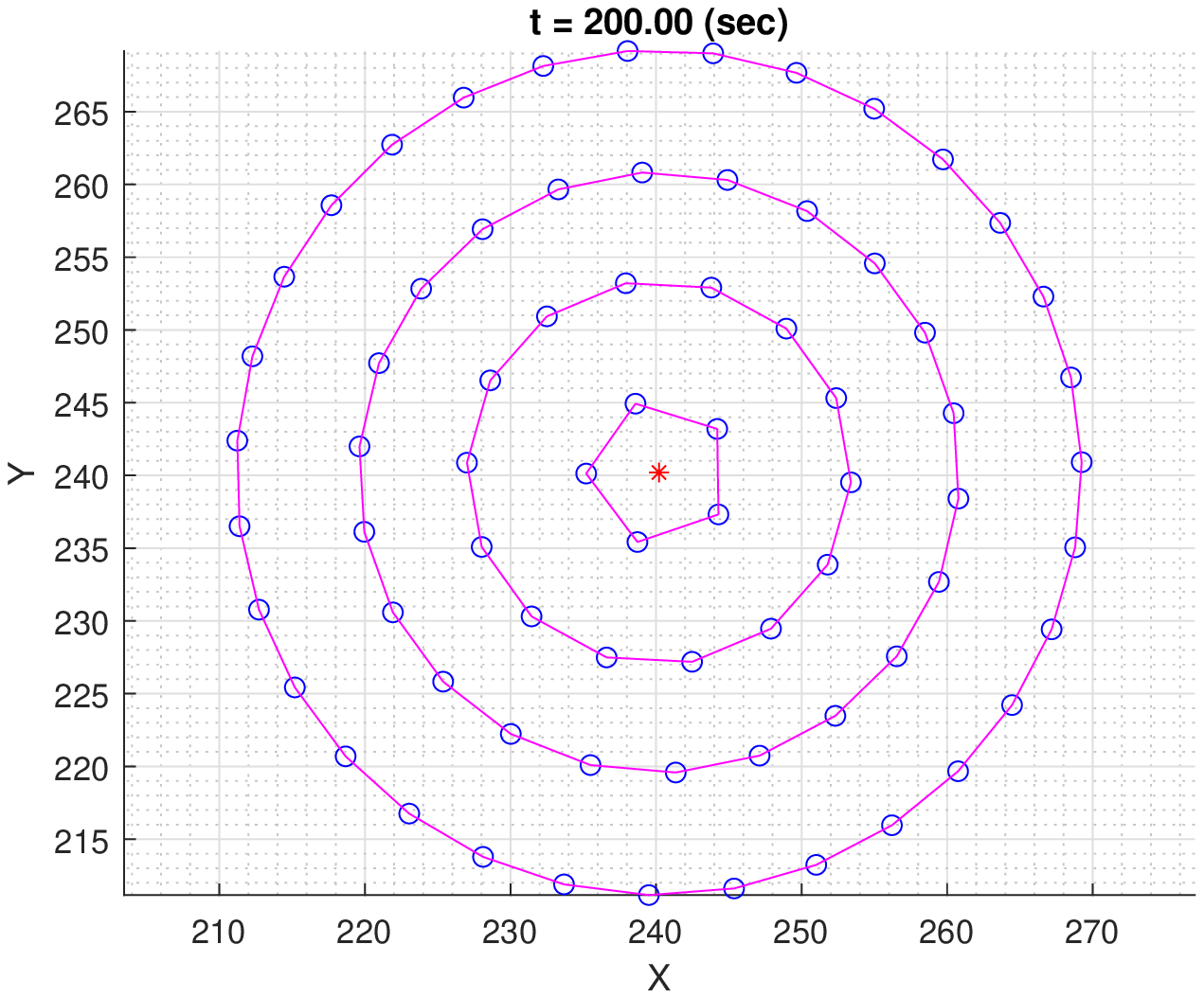}
						\caption{Pentagon}
						\label{fig:mlt2}
					\end{subfigure}\hfil % <-- added
					\begin{subfigure}{0.33\textwidth}
						\includegraphics[width=\linewidth, height=0.2\textheight]{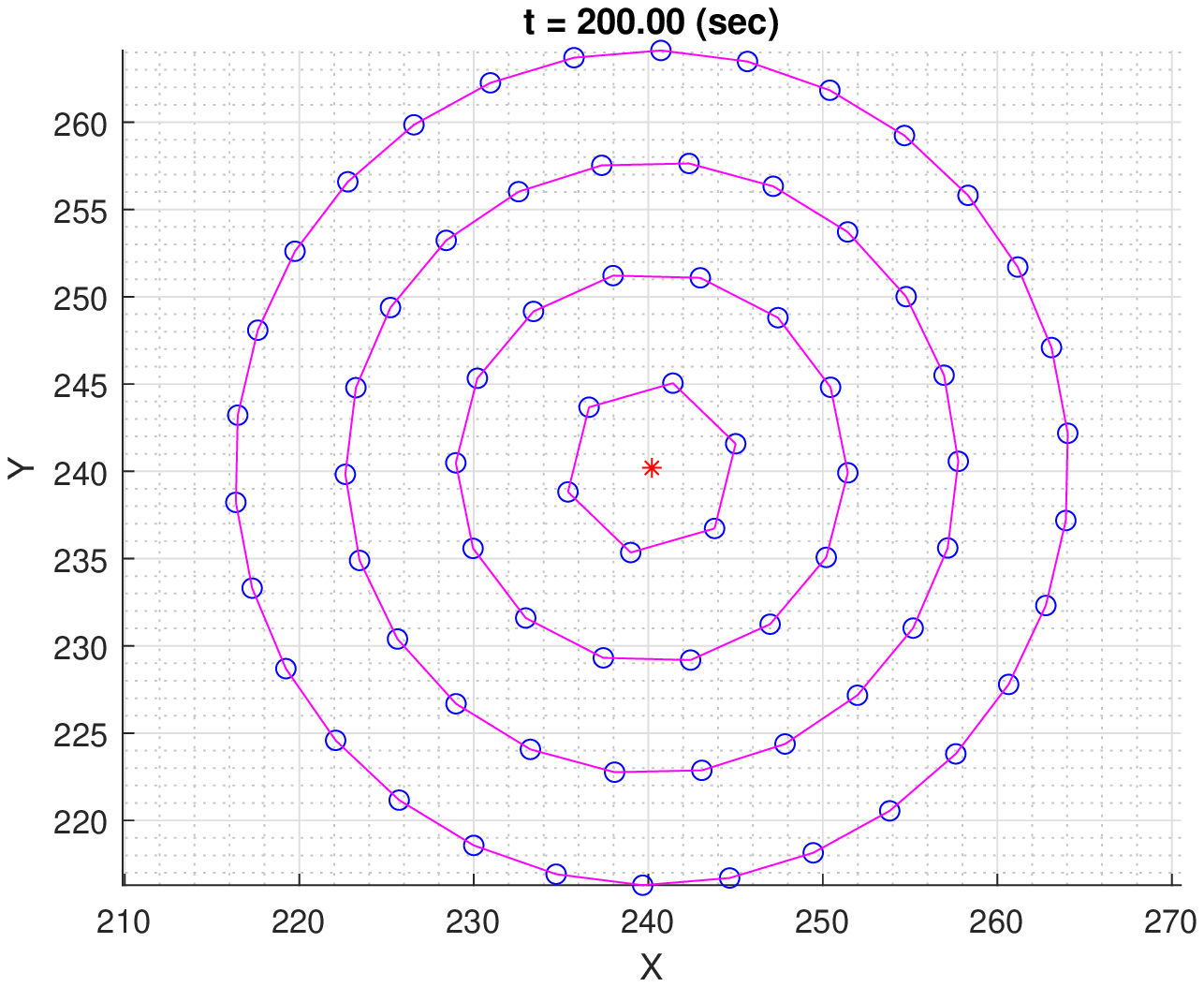}
						\caption{Hexagon}
						\label{fig:mlt3}
					\end{subfigure}\\%\hfil % <-- added
					\begin{subfigure}{0.33\textwidth}
						\includegraphics[width=\linewidth, height=0.2\textheight]{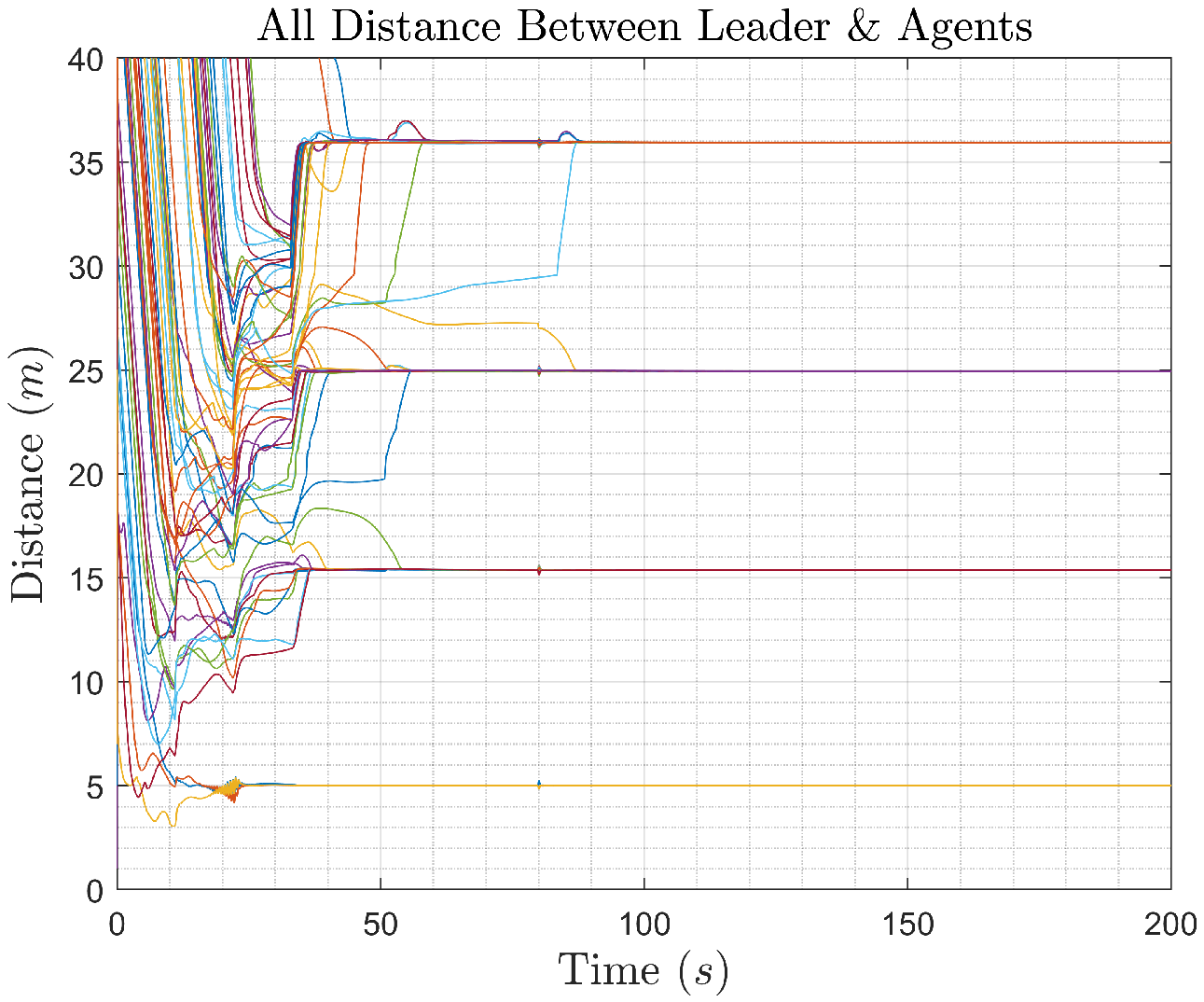}
						\caption{Triangle}
						\label{fig:mlt4}
					\end{subfigure}\hfil
					\begin{subfigure}{0.33\textwidth}
						\includegraphics[width=\linewidth, height=0.2\textheight]{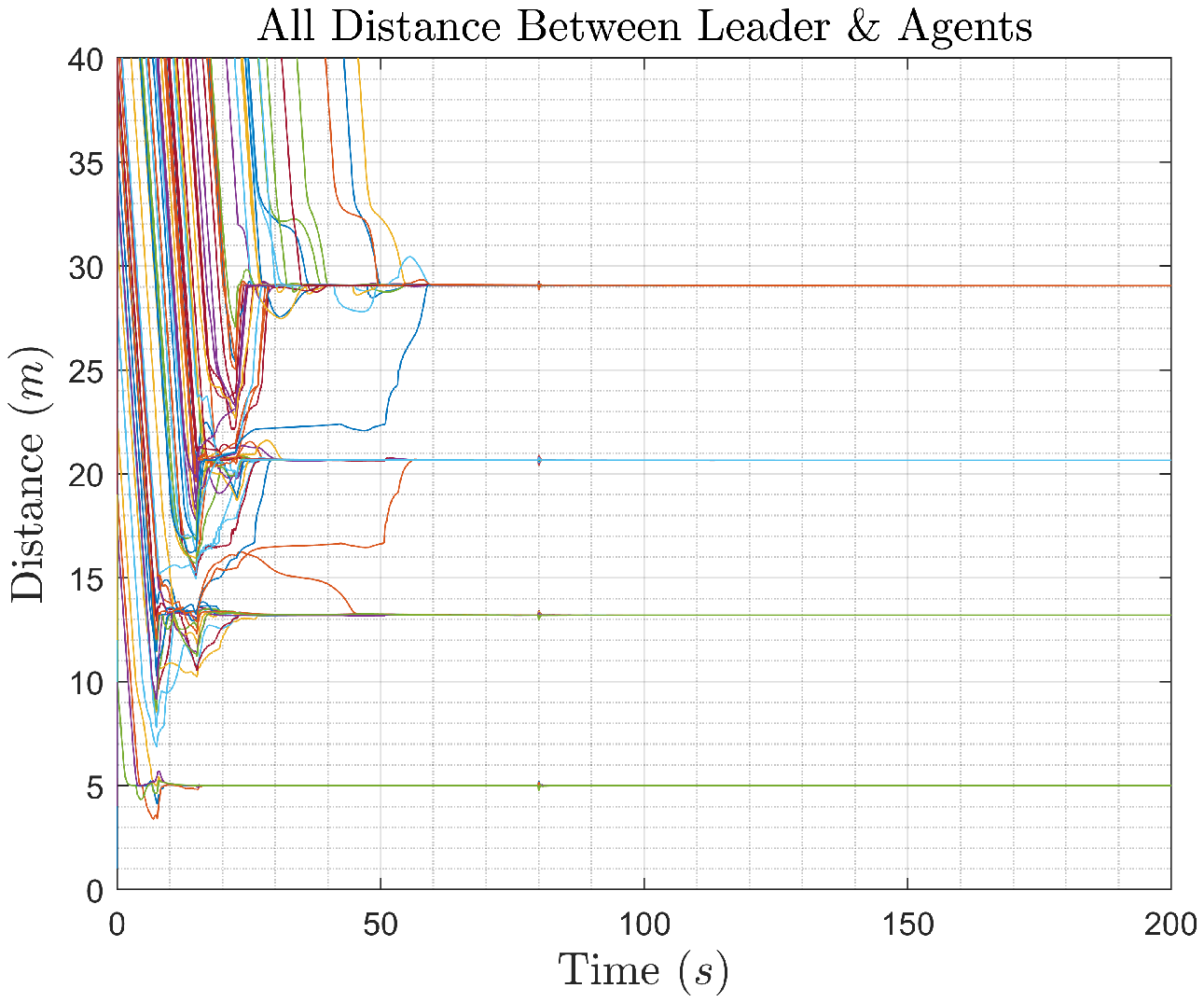}
						\caption{Pentagon}
						\label{fig:mlt5}
					\end{subfigure}\hfil % <-- added
					\begin{subfigure}{0.33\textwidth}
						\includegraphics[width=\linewidth, height=0.2\textheight]{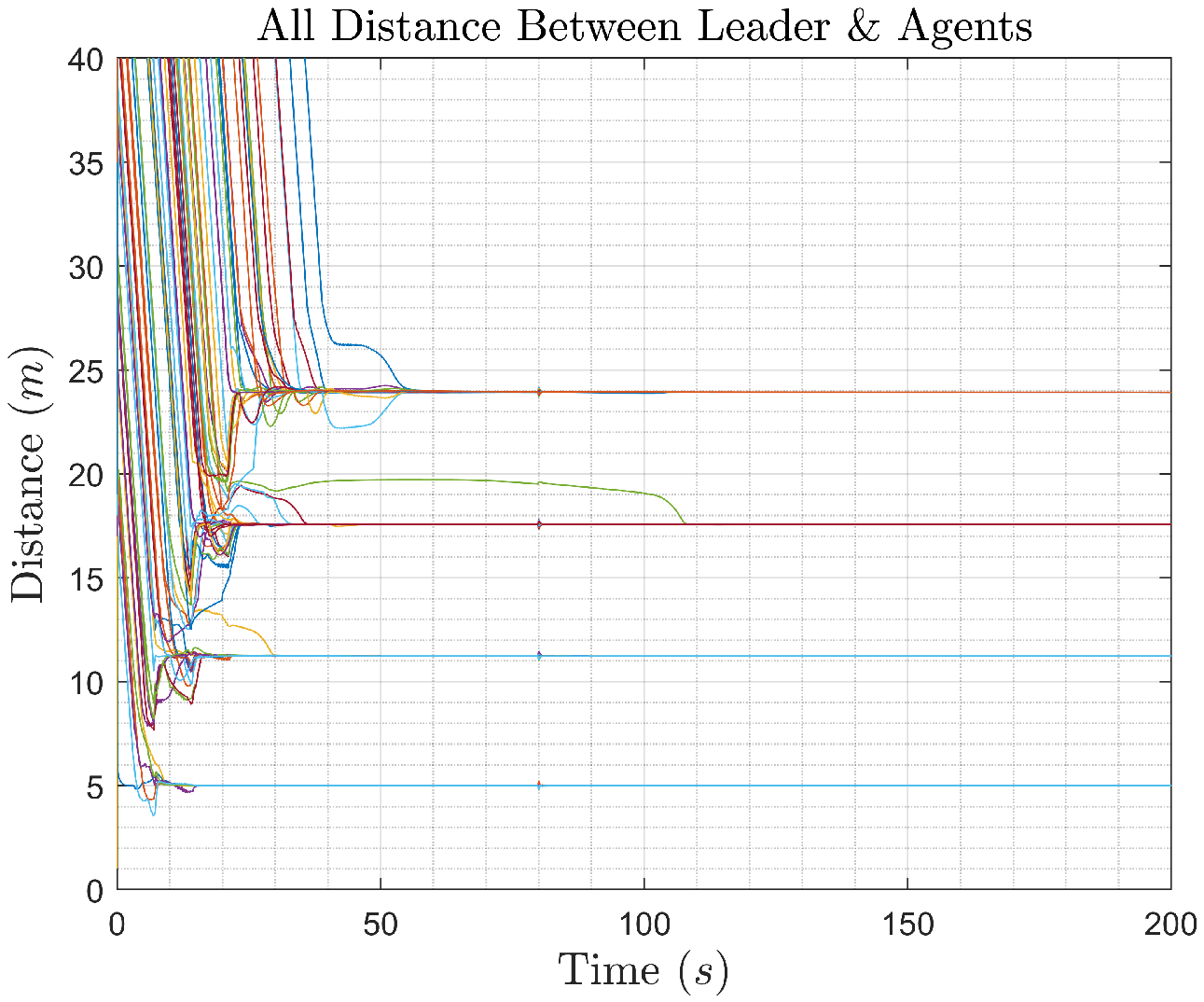}
						\caption{Hexagon}
						\label{fig:mlt6}
					\end{subfigure}\\ % <-- added
					\begin{subfigure}{0.33\textwidth}
						\includegraphics[width=\linewidth, height=0.2\textheight]{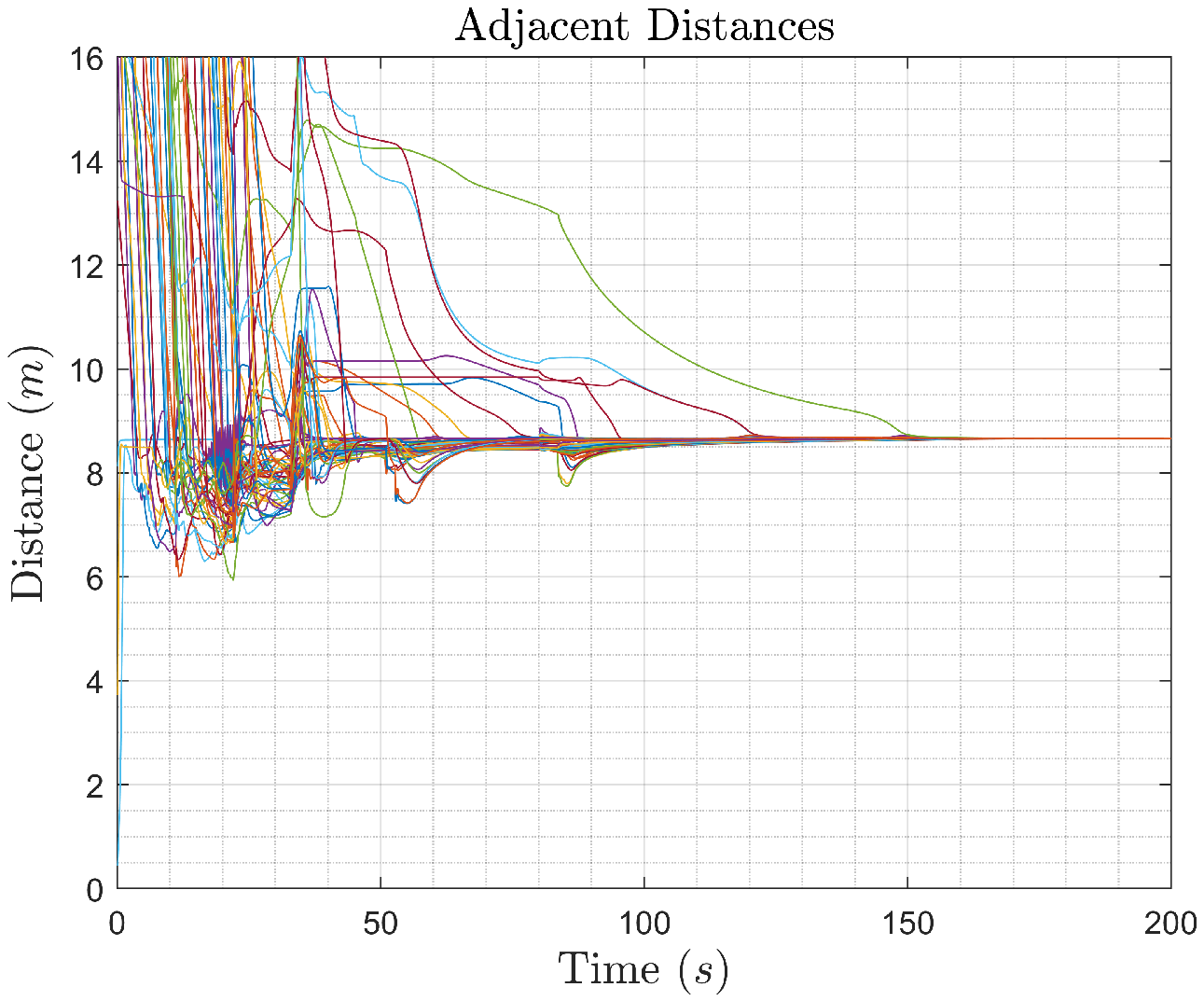}
						\caption{Triangle}
						\label{fig:mlt7}
					\end{subfigure}\hfil % <-- added
					\begin{subfigure}{0.33\textwidth}
						\includegraphics[width=\linewidth, height=0.2\textheight]{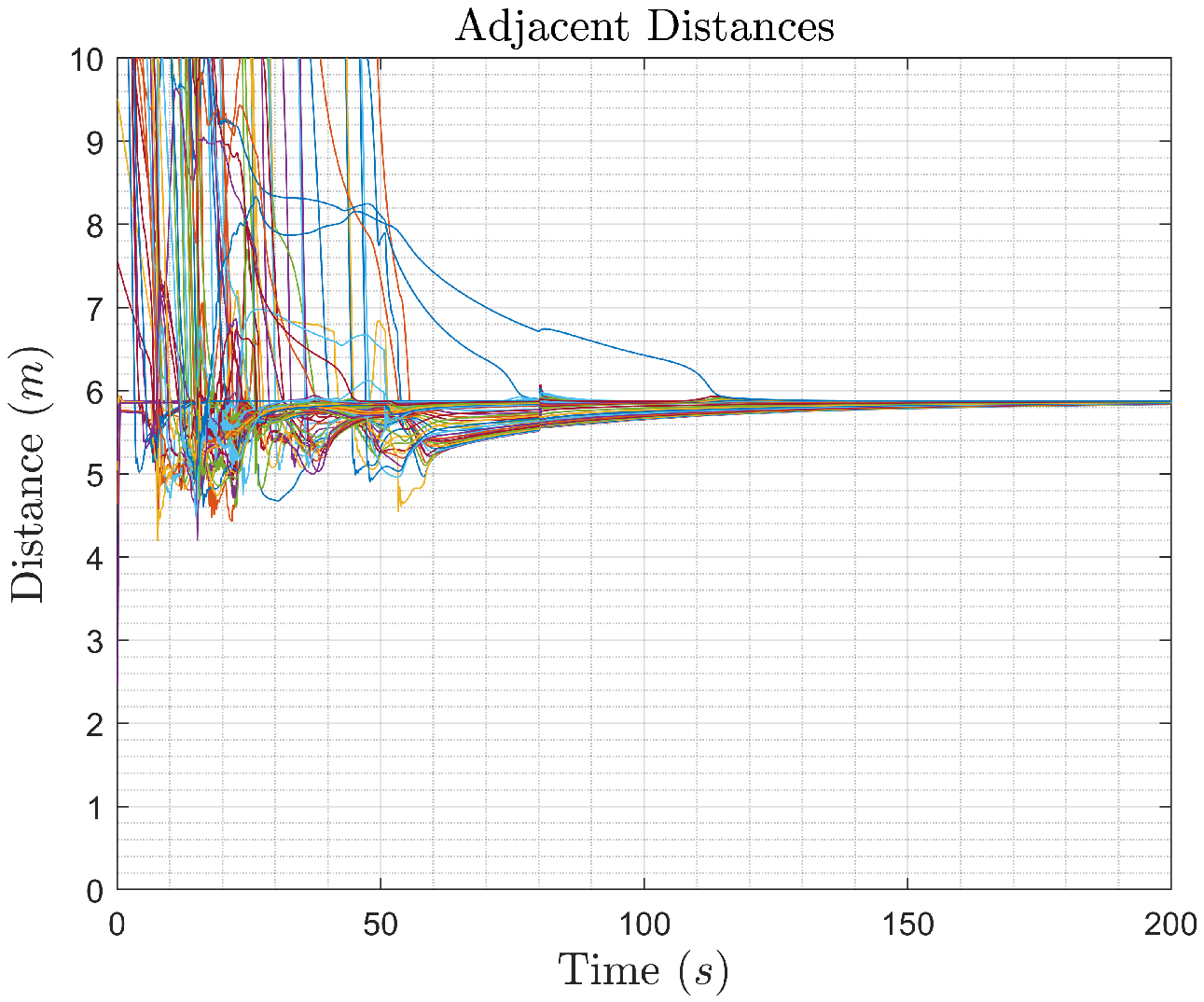}
						\caption{Pentagon}
						\label{fig:mlt8}
					\end{subfigure}\hfil % <-- added
					\begin{subfigure}{0.33\textwidth}
						\includegraphics[width=\linewidth, height=0.2\textheight]{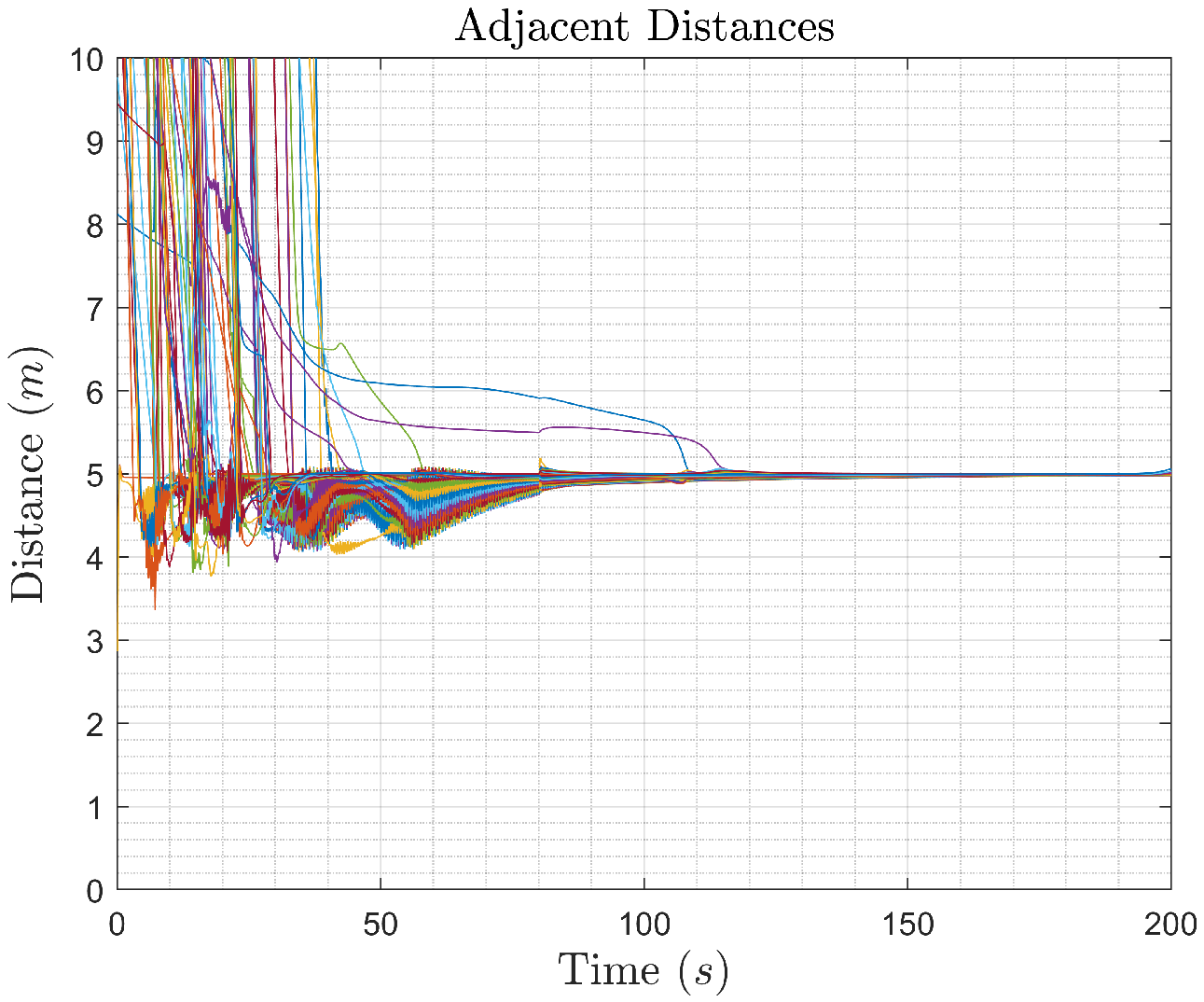}
						\caption{Hexagon}
						\label{fig:mlt9}
					\end{subfigure} % <-- added
					\caption{Multi-circles formations }
					\label{fig:fig11}
				\end{figure*}
				The radius of the circles formation and cut-off are determined according to Assumption \ref{asum:4} and  (\ref{eq12}), respectively.
				To determine the radius of circles, firstly, the distance between agents is calculated according to the radius and number of agents in the first circle by  (\ref{eq8}). Then, according to the desired number of agents in other circles, i.e., the second circle to the last one, their radius is achieved by (\ref{eq8}) and is checked by Assumption \ref{asum:4}.
				For example, with three agents and a radius of 5 in the first circle, the distance between the agents is 8.66. The desired number of agents in the second, third and fourth circles is considered to be 11, 18, and 26 agents, respectively, so the radius of the circles is obtained 15.37, 24.93, and 35.92 according to the number of agents.
				The simulation time is 200, and the initial conditions of the agents are chosen as normally distributed random numbers with a mean of zero and a standard deviation of 60.
				Also, in  (\ref{eq10}), $a_1=1, a_2=1.5$, and $a_3=3$ are considered. 				
				The leader is stationary in the beginning of the simulation until four circles are formed. Then, after 80 seconds, the leader starts moving at a constant speed. The rest of the parameters and settings are shown in table \ref{Table3}.
				The four-circle formations for 3, 5, and 6 agents in the first circle i.e. geometric patterns of triangle, pentagon, and hexagon, are shown in Fig. \ref{fig:mlt1}, \ref{fig:mlt2}, and \ref{fig:mlt3}, respectively. Also, Fig. \ref{fig:mlt4}, \ref{fig:mlt5}, and \ref{fig:mlt6} illustrate the distance between agents and the leader, and Fig. \ref{fig:mlt7}, \ref{fig:mlt8}, and \ref{fig:mlt9} demonstrate the distance between adjacent agents for relevant formations, respectively.  It can be seen in Fig. \ref{fig:mlt4}, \ref{fig:mlt5}, and \ref{fig:mlt6} all agents are finally placed at the desired distances of the leader. In Fig. \ref{fig:mlt7}, \ref{fig:mlt8}, and \ref{fig:mlt9}, the minimum adjacent distance of agents in bad conditions is approximately 70 percent of the desired distance, so collision avoidance is guaranteed.
				\begin{table*}[ht]
					\centering
					\caption{Algorithm parameters}
					\begin{tabular}{|c|c|c|c|c|c|c|c|c|c|c|c|c|} \hline 
						\label{Table3}
						& $c_1^{\alpha}$ & $c_2^{\alpha}$ & $c_1^{\gamma}$ & $c_2^{\gamma}$ & $d$ & $k$ & $d_{L_2}$ & $d_{L_3}$ & $d_{L_4}$ & $d_{\varepsilon}$ & SW & $N$\\  
						\hline
						Triangle  & 6.6 & 2.4 & 4.3 & 11.2 & 8.6603 & 1.07 & 15.3696 & 24.9362 & 35.9237 & 1.5 & 110 & 58  \\\hline
						Pentagon  & 6.1 & 2.8 & 4.9 & 12.7 & 5.8779 & 1.09 & 13.2074 & 20.6509 & 29.0499 & 0.9 & 75 & 72 \\\hline
						Hexagon  & 5.9 & 2.3 & 5.2 & 13.3 & 5 & 1.1& 11.2349 & 17.5667 & 23.9169 & 1 & 70 & 72 \\\hline
					\end{tabular}
				\end{table*} 
				\section{Conclusion}\label{section4}
				This paper investigated the dynamic circular formations with a leader in the center by a flocking approach. Polygon formations were achieved when some agents failed. Obstacle avoidance was guaranteed, and size scaling was proposed for better passing through obstacles. For the multi-circle formation, the switching piecewise potential function was proposed. Finally, the parameters of the flocking algorithm were optimized by optimization algorithms with different scenarios.
				\bibliographystyle{IEEEtranN}
				\bibliography{mybib.bib}

% Generated by IEEEtranN.bst, version: 1.14 (2015/08/26)
\begin{thebibliography}{24}
\providecommand{\natexlab}[1]{#1}
\providecommand{\url}[1]{#1}
\csname url@samestyle\endcsname
\providecommand{\newblock}{\relax}
\providecommand{\bibinfo}[2]{#2}
\providecommand{\BIBentrySTDinterwordspacing}{\spaceskip=0pt\relax}
\providecommand{\BIBentryALTinterwordstretchfactor}{4}
\providecommand{\BIBentryALTinterwordspacing}{\spaceskip=\fontdimen2\font plus
\BIBentryALTinterwordstretchfactor\fontdimen3\font minus
  \fontdimen4\font\relax}
\providecommand{\BIBforeignlanguage}[2]{{%
\expandafter\ifx\csname l@#1\endcsname\relax
\typeout{** WARNING: IEEEtranN.bst: No hyphenation pattern has been}%
\typeout{** loaded for the language `#1'. Using the pattern for}%
\typeout{** the default language instead.}%
\else
\language=\csname l@#1\endcsname
\fi
#2}}
\providecommand{\BIBdecl}{\relax}
\BIBdecl

\bibitem[Reynolds(1987)]{reynolds1987flocks}
C.~W. Reynolds, ``Flocks, herds and schools: A distributed behavioral model,''
  in \emph{Proceedings of the 14th annual conference on Computer graphics and
  interactive techniques}, 1987, pp. 25--34.

\bibitem[Olfati-Saber(2006)]{olfati2006flocking}
R.~Olfati-Saber, ``Flocking for multi-agent dynamic systems: Algorithms and
  theory,'' \emph{IEEE Transactions on automatic control}, vol.~51, no.~3, pp.
  401--420, 2006.

\bibitem[Chen et~al.(2018)Chen, Pei, Lai, and Yan]{chen2018multitarget}
S.~Chen, H.~Pei, Q.~Lai, and H.~Yan, ``Multitarget tracking control for coupled
  heterogeneous inertial agents systems based on flocking behavior,''
  \emph{IEEE Transactions on Systems, Man, and Cybernetics: Systems}, vol.~49,
  no.~12, pp. 2605--2611, 2018.

\bibitem[Manzoor et~al.(2017)Manzoor, Lee, and Choi]{manzoor2017coordinated}
S.~Manzoor, S.~Lee, and Y.~Choi, ``A coordinated navigation strategy for
  multi-robots to capture a target moving with unknown speed,'' \emph{Journal
  of Intelligent \& Robotic Systems}, vol.~87, no.~3, pp. 627--641, 2017.

\bibitem[Ma et~al.(2018)Ma, Yao, Dai, Lu, Xiao, and Zheng]{ma2018cooperative}
J.~Ma, W.~Yao, W.~Dai, H.~Lu, J.~Xiao, and Z.~Zheng, ``Cooperative encirclement
  control for a group of targets by decentralized robots with collision
  avoidance,'' in \emph{2018 37th Chinese Control Conference (CCC)}.\hskip 1em
  plus 0.5em minus 0.4em\relax IEEE, 2018, pp. 6848--6853.

\bibitem[Zhang et~al.(2019)Zhang, Ling, and Mo]{zhang2019distributed}
T.~Zhang, J.~Ling, and L.~Mo, ``Distributed finite-time rotating encirclement
  control of multiagent systems with nonconvex input constraints,'' \emph{IEEE
  access}, vol.~7, pp. 102\,477--102\,486, 2019.

\bibitem[Ma and Sun(2017)]{ma2017finite}
D.~Ma and Y.~Sun, ``Finite-time circle surrounding control for multi-agent
  systems,'' \emph{International Journal of Control, Automation and Systems},
  vol.~15, no.~4, pp. 1536--1543, 2017.

\bibitem[Wu et~al.(2021)Wu, Yu, Ma, Wu, Han, Shi, and Gao]{wu2021autonomous}
J.~Wu, Y.~Yu, J.~Ma, J.~Wu, G.~Han, J.~Shi, and L.~Gao, ``Autonomous
  cooperative flocking for heterogeneous unmanned aerial vehicle group,''
  \emph{IEEE Transactions on Vehicular Technology}, vol.~70, no.~12, pp.
  12\,477--12\,490, 2021.

\bibitem[Brust et~al.(2017)Brust, Danoy, Bouvry, Gashi, Pathak, and
  Gon{\c{c}}alves]{brust2017defending}
M.~R. Brust, G.~Danoy, P.~Bouvry, D.~Gashi, H.~Pathak, and M.~P.
  Gon{\c{c}}alves, ``Defending against intrusion of malicious uavs with
  networked uav defense swarms,'' in \emph{2017 IEEE 42nd conference on local
  computer networks workshops (LCN workshops)}.\hskip 1em plus 0.5em minus
  0.4em\relax IEEE, 2017, pp. 103--111.

\bibitem[Wang et~al.(2020)Wang, Shen, Song, and Zhang]{wang2020circle}
Y.~Wang, T.~Shen, C.~Song, and Y.~Zhang, ``Circle formation control of
  second-order multi-agent systems with bounded measurement errors,''
  \emph{Neurocomputing}, vol. 397, pp. 160--167, 2020.

\bibitem[Song et~al.(2018)Song, Liu, and Xu]{song2018circle}
C.~Song, L.~Liu, and S.~Xu, ``Circle formation control of mobile agents with
  limited interaction range,'' \emph{IEEE Transactions on Automatic Control},
  vol.~64, no.~5, pp. 2115--2121, 2018.

\bibitem[Wang and Xie(2017)]{wang2017limit}
C.~Wang and G.~Xie, ``Limit-cycle-based decoupled design of circle formation
  control with collision avoidance for anonymous agents in a plane,''
  \emph{IEEE Transactions on Automatic Control}, vol.~62, no.~12, pp.
  6560--6567, 2017.

\bibitem[Jin et~al.(2018)Jin, Yu, and Ren]{jin2018circular}
L.~Jin, S.~Yu, and D.~Ren, ``Circular formation control of multiagent systems
  with any preset phase arrangement,'' \emph{Journal of Control Science and
  Engineering}, vol. 2018, 2018.

\bibitem[Wen et~al.(2018)Wen, Wang, and Xie]{wen2018asynchronous}
J.~Wen, C.~Wang, and G.~Xie, ``Asynchronous distributed event-triggered circle
  formation of multi-agent systems,'' \emph{Neurocomputing}, vol. 295, pp.
  118--126, 2018.

\bibitem[Wen et~al.(2019)Wen, Xu, Wang, Xie, and Gao]{wen2019distributed}
J.~Wen, P.~Xu, C.~Wang, G.~Xie, and Y.~Gao, ``Distributed event-triggered
  circle formation control for multi-agent systems with limited communication
  bandwidth,'' \emph{Neurocomputing}, vol. 358, pp. 211--221, 2019.

\bibitem[Yang et~al.(2022)Yang, Yan, Zeng, Zhan, and Shi]{yang2022decoupled}
B.~Yang, H.~Yan, L.~Zeng, X.~Zhan, and K.~Shi, ``Decoupled design of
  distributed event-triggered circle formation control for multi-agent
  system,'' \emph{International Journal of Systems Science}, pp. 1--10, 2022.

\bibitem[Yu et~al.(2018)Yu, Wang, Xie, and Jin]{yu2018event}
M.~Yu, H.~Wang, G.~Xie, and K.~Jin, ``Event-triggered circle formation control
  for second-order-agent system,'' \emph{Neurocomputing}, vol. 275, pp.
  462--469, 2018.

\bibitem[Muslimov and Munasypov(2020)]{muslimov2020adaptive}
T.~Z. Muslimov and R.~A. Munasypov, ``Adaptive decentralized flocking control
  of multi-uav circular formations based on vector fields and backstepping,''
  \emph{ISA transactions}, vol. 107, pp. 143--159, 2020.

\bibitem[Chen et~al.(2019)Chen, Yu, Zhang, and Liu]{chen2019circular}
Y.~Chen, R.~Yu, Y.~Zhang, and C.~Liu, ``Circular formation flight control for
  unmanned aerial vehicles with directed network and external disturbance,''
  \emph{IEEE/CAA Journal of Automatica Sinica}, vol.~7, no.~2, pp. 505--516,
  2019.

\bibitem[Berlinger et~al.(2021)Berlinger, Gauci, and
  Nagpal]{berlinger2021implicit}
F.~Berlinger, M.~Gauci, and R.~Nagpal, ``Implicit coordination for 3d
  underwater collective behaviors in a fish-inspired robot swarm,''
  \emph{Science Robotics}, vol.~6, no.~50, p. eabd8668, 2021.

\bibitem[Han et~al.(2015)Han, Wang, Lin, and Zheng]{han2015formation}
Z.~Han, L.~Wang, Z.~Lin, and R.~Zheng, ``Formation control with size scaling
  via a complex laplacian-based approach,'' \emph{IEEE transactions on
  cybernetics}, vol.~46, no.~10, pp. 2348--2359, 2015.

\bibitem[Mitchell(1998)]{mitchell1998introduction}
M.~Mitchell, \emph{An introduction to genetic algorithms}.\hskip 1em plus 0.5em
  minus 0.4em\relax MIT press, 1998.

\bibitem[Kennedy and Eberhart(1995)]{kennedy1995particle}
J.~Kennedy and R.~Eberhart, ``Particle swarm optimization,'' in
  \emph{Proceedings of ICNN'95-international conference on neural networks},
  vol.~4.\hskip 1em plus 0.5em minus 0.4em\relax IEEE, 1995, pp. 1942--1948.

\bibitem[Mirjalili et~al.(2014)Mirjalili, Mirjalili, and
  Lewis]{mirjalili2014grey}
S.~Mirjalili, S.~M. Mirjalili, and A.~Lewis, ``Grey wolf optimizer,''
  \emph{Advances in engineering software}, vol.~69, pp. 46--61, 2014.

\end{thebibliography}
				
			\end{document}